\documentclass[twocolumn]{aastex63}

\usepackage[utf8]{inputenc}
\usepackage{natbib}
\usepackage[scaled]{helvet} 
\usepackage{graphicx}
\usepackage{txfonts} 
\usepackage{epsfig}
\usepackage{longtable}
\bibpunct{(}{)}{;}{a}{}{,}
\usepackage[normalem]{ulem} 
\newcommand{\Bcep}{$\beta$ Cephei~}
\newcommand{\Dsct}{$\delta$ Scuti~}
\newcommand{\Gdor}{$\gamma$ Doradus~}
\newcommand{\cd}{d$^{-1}$\,}
\newcommand{\kms}{km\,s$^{-1}$\,}

\submitjournal{AAS Journals}

\shorttitle{New Beta Cephei stars with KELT}
\shortauthors{Labadie-Bartz et al.}

\begin{document}

\title{New Beta Cephei stars from the KELT project}

\correspondingauthor{Jonathan Labadie-Bartz}
\email{jbartz@usp.br}

\author[0000-0002-2919-6786]{Jonathan Labadie-Bartz}
\affiliation{Instituto de Astronomia, Geof\'{i}sica e Ci\^{e}ncias Atmosf\'{e}ricas, Universidade de S\~{a}o Paulo, Rua do Mat\~{a}o 1226, Cidade Universit\'{a}ria, 05508-900 S\~{a}o Paulo, SP, Brazil}
\affiliation{Department of Physics \& Astronomy, University of Delaware, Newark, DE 19716, USA}
\affiliation{Department of Physics, Lehigh University, 16 Memorial Drive East, Bethlehem, PA 18015, USA}

\author[0000-0001-7756-1568]{Gerald Handler}
\affiliation{Nicolaus Copernicus Astronomical Center, Bartycka 18, 00-716 Warsaw, Poland}

\author[0000-0002-3827-8417]{Joshua Pepper}
\affiliation{Department of Physics, Lehigh University, 16 Memorial Drive East, Bethlehem, PA 18015, USA}

\author[0000-0003-1241-9792]{Luis Balona}
\affiliation{South African Astronomical Observatory, PO Box 9, Observatory, 7935, Cape Town, South Africa}

\author[0000-0001-5419-2042]{Peter De Cat}
\affiliation{Royal Observatory of Belgium, Av Circulaire 3-Ringlaan 3, B-1180 Brussels, Belgium}

\author[0000-0002-5951-8328]{Daniel J.\ Stevens}
\affiliation{Center for Exoplanets and Habitable Worlds, The Pennsylvania State University, 525 Davey Lab, University Park, PA 16802, USA}
\affiliation{Department of Astronomy \& Astrophysics, The Pennsylvania State University, 525 Davey Lab, University Park, PA 16802, USA}

\author[0000-0003-2527-1598]{Michael B.\ Lund}
\affiliation{Caltech IPAC – NASA Exoplanet Science Institute 1200 E. California Ave, Pasadena, CA 91125, USA}

\author[0000-0002-3481-9052]{Keivan G.\ Stassun}
\affiliation{Department of Physics and Astronomy, Vanderbilt University, Nashville, TN 37235, USA}
\affiliation{Department of Physics, Fisk University, 1000 17th Avenue North, Nashville, TN 37208, USA}

\author[0000-0001-8812-0565]{Joseph E.\ Rodriguez}
\affiliation{Center for Astrophysics \textbar \ Harvard \& Smithsonian, 60 Garden St, Cambridge, MA 02138, USA}

\author[0000-0001-5016-3359]{Robert J.\ Siverd}
\affiliation{Department of Physics and Astronomy, Vanderbilt University, Nashville, TN 37235, USA}

\author[0000-0001-5160-4486]{David J. James}
\affiliation{Center for Astrophysics \textbar \ Harvard \& Smithsonian, 60 Garden St, Cambridge, MA 02138, USA}
\affiliation{Black Hole Initiative at Harvard University, 20 Garden Street, Cambridge, MA 02138, USA}

\author[0000-0002-4236-9020]{Rudolf B.\ Kuhn}
\affiliation{South African Astronomical Observatory, PO Box 9, Observatory, 7935, Cape Town, South Africa}
\affiliation{Southern African Large Telescope, PO Box 9, Observatory, 7935, Cape Town, South Africa}

\begin{abstract}

We present the results of a search for Galactic \Bcep stars, which are massive pulsating stars with both pressure modes and mixed modes. Thus, these stars can serve as benchmarks for seismological studies of the interiors of massive stars. We conducted the search by performing a frequency analysis on the optical light curves of known O- and B-type stars with data from the KELT exoplanet survey. We identify 113 \Bcep stars, of which 86 are new discoveries, which altogether represents a ~70\% increase in the number presently known. An additional 96 candidates are identified. Among our targets, we find five new eclipsing binaries and 22 stars with equal frequency spacings suggestive of rotational splitting of nonradial pulsation modes. Candidates for runaway stars among our targets and a number of interesting individual objects are discussed.  Most of the known and newly discovered \Bcep stars will be observed by the TESS mission, providing by far the most comprehensive observational data set of massive main sequence pulsating stars of sufficient quality for detailed asteroseismic studies. Future analysis of these light curves has the potential to dramatically increase our understanding of the structure of stellar interiors and the physical processes taking place therein.

\end{abstract}


\section{Introduction} \label{sec:intro}

Despite recent advances, there remain uncertainties regarding the evolution and structure of massive stars. 
Currently, the role of rotation, internal angular momentum distribution and transport, and internal mixing are not satisfactorily understood in the context of stellar evolution. Mixing of material into the hydrogen-burning stellar core (``convective overshooting'') considerably affects the main sequence lifetime of massive stars \citep[\textit{e.g.}][]{Mowlavi1994}, and causes surface abundances to change. Rotation influences this process, but the details remain poorly constrained \citep[\textit{e.g.}][]{Maeder1987}. Differential rotation is sometimes measured in massive stars, but there are significant uncertainties regarding the coupling between the stellar core and envelope, and the degree to which angular momentum is transported from the core outward \citep{Aerts2017}. To make further progress in understanding the nature of massive-star interiors, studies of massive stars that are amenable to seismology are needed.

As a class, \Bcep stars are massive non-supergiant variable stars with spectral type O or B with photometric, radial velocity, and/or line profile variations caused by low-order pressure and gravity mode pulsations \citep{Stankov2005}. Most of them are early B-type stars (roughly spanning spectral types B0 -- B2.5) with masses between 8 --17 M$_{\odot}$. They are characterized by their relatively high-frequency pulsations (with typical periods between 2 -- 7 hours) driven by the $\kappa$ mechanism \citep{Moskalik1992,Dziembowski1993}.

The pulsational properties of \Bcep stars make them particularly well-suited for detailed asteroseismic studies. They tend to oscillate in several nonradial modes, and sometimes have both \textit{p}- and \textit{g}-mode pulsations \citep[\textit{e.g.}][]{Handler2009}. Because the frequency of each oscillation mode is determined by the physical conditions in the region in which it propagates, measuring these frequencies (and knowing their geometries on the stellar surface and the interior) translates to constraints on the physical conditions in the stellar interior. Seismic modelling of a small number of \Bcep stars has already yielded significant progress. Quantitative estimates of the core overshooting parameter have been derived for the \Bcep star HD 129929 \citep{Aerts2003}, which is also found to undergo non-rigid internal rotation \citep{Dupret2004}. Similar analyses have been done for $\beta$ CMa \citep{Mazumdar2006}, $\delta$ Ceti \citep{Aerts2006}, 12 Lac \citep{Handler2006,Dziembowski2008}, V2052 Oph \citep{Briquet2012}, and a few others. Asteroseismology can also be used to measure stellar mass and radius, the surface rotation rate, and the evolutionary stage. Given that these stars are fairly massive, they are correspondingly rare: only about 250 such pulsators are known in our Galaxy \citep{Stankov2005,Pigulski2008,Kahraman2016}.

The technique of asteroseismology relies on having a long enough time baseline and a high enough cadence and precision to identify the periodic signals of as many pulsation modes as possible. Space-based time-series photometry provides excellent data for these purposes. Although current and past space observatories have been successfully used for asteroseismology, only a small number of the high-mass \Bcep type stars have been observed, and subsequently modeled, to date in this capacity (\textit{e.g.} HD~180642 with CoRoT; \citealt{Aerts2011}, $\delta$~Ceti with MOST; \citealt{Aerts2006}, $\beta$~Cen with BRITE; \citealt{Pigulski2016}) -- there have been no comprehensive asteroseismic studies of this class of star. 

The recently launched NASA Transiting Exoplanet Survey Satellite (TESS) mission \citep{Ricker2014} is a nearly all-sky photometric survey with a photometric precision similar to that of the Kepler mission, but designed primarily to target stars some five magnitudes brighter. The photometric precision for a 10th magnitude star in one hour of TESS observations is estimated to be 200 ppm. While the primary science goal of the TESS mission is to discover transiting exoplanets, TESS light curves will be leveraged for a host of ancillary science efforts. Among these is the use of TESS data for an asteroseismic study of a large sample of \Bcep stars. To maximize the scientific yield,  we are mining ground-based time-series photometric data to identify as many \Bcep stars as possible, to then be targeted by TESS for the purpose of asteroseismology. The TESS light curves of these stars will allow for detailed modeling of an unprecedented number of massive stars across a wide range of parameter space, and will result in important constraints on theories regarding stellar structure and evolution.

In Section~\ref{sec:data} we introduce the data used for our search, and the cuts used to narrow down the number of light curves to be analyzed. Section~\ref{sec:analysis} describes the methods by which the data are analyzed. Our results are presented in Section~\ref{sec:results}, and a summary of our main conclusions are given in Section~\ref{sec:conclusions}.

\section{Data} \label{sec:data}
The source of data used for our analysis is the Kilodegree Extremely Little Telescope (KELT) survey \citep{Pepper2007,Pepper2012}. KELT is a photometric survey designed to discover transiting exoplanets orbiting stars in the magnitude range of 7 $\lesssim$ V $\lesssim$ 12, with light curves for $\sim$4.9 million objects across $\sim$70\% of the sky. The KELT survey employs two small-aperture (42 mm) wide-field (26$^{\circ}$ x 26$^{\circ}$) telescopes, with a northern location at Winer Observatory in Arizona in the United States, and a southern location at the South African Astronomical Observatory near Sutherland, South Africa. The effective passband is roughly equivalent to a broad R-band filter. With time baselines of up to 10 years, a typical cadence of 30 minutes, and the precision to detect periodic signals down to approximately one to a few mmag, KELT light curves are well-suited to the detection of the periodic variability exhibited by \Bcep stars. The utility of KELT light curves to detect periodic oscillations in brightness was demonstrated by \citet{Cargile2014}, \citet{Wisniewski2015}, and \citet{Labadie-Bartz2017}. 

The goal of this work is the identification of new and candidate \Bcep stars. We begin with a star catalog that is used internally by the TESS target selection committees and maintained by one of us (LB, \url{http://redcliffcottage.co.za/tess/}), which contains stars brighter than 12th magnitude, and includes literature spectral types (when available), and photometric magnitudes in multiple passbands. All stars with a spectral type between O4 -- B7 (determined either by literature spectral type, or by photometric color) were selected. These 16682 stars were then cross-matched to the KELT catalog of all fields reduced until February 2016, resulting in 5840 matches. The KELT light curves of these 5840 O4 -- B7 stars are then analyzed for the signatures of \Bcep pulsation. The range of stellar brightness covered by KELT coincides very well with that of a large fraction of TESS targets, which makes KELT an extremely valuable source to pre-select targets to be later observed from space.

\section{Analysis} \label{sec:analysis}

For each of the 5840 light curves for stars with spectral types between O4 -- B7, a Fourier periodogram was computed in the range of 0 -- 20 \cd. To perform a pre-selection of candidates, each periodogram that had the strongest peak in the range of known \Bcep star pulsation frequencies ($f$ $\gtrsim$ 3 \cd) was visually inspected, as were the light curves phased to twice the recovered period. On the basis of these plots, and in doubtful cases based on additional frequency analyses, all stars that had either untrustworthy data, showed no significant variability, exhibited obvious binary or rotational light curves, and those whose periodograms could be explained solely by reduction imperfections (residual differential color extinction) were rejected. Those that were not rejected were scrutinized, and a few more obvious non-pulsators were removed. The remaining stars were preliminarily classified into \Bcep stars and candidate \Bcep stars.

Objects pre-classified as \Bcep stars either showed periodic variability at a single frequency between 4 -- 14 \cd and have spectral types between B0 -- B2, or showed periodic variability at multiple frequencies in the range 4 -- 14 \cd (suggesting they are multi-mode pulsators), but without further spectral type cuts. The choice of the low-frequency limit near 4 \cd is justified by this being approximately the expected value of the radial fundamental mode frequency at the end of the main sequence in the \Bcep mass range, but also as a reasonable cutoff to discriminate against close binaries, rotational variables or other pulsators.

The group pre-classified as candidates contains stars that do not fall in the previous category, but have primary frequencies in the range 3 -- 22 \cd (sometimes with multiple significant frequencies), and stars where only one frequency is detected (if their spectral type is outside the range of B0 -- B2) and the light curve shape suggests a pulsational origin for the variability. This group likely contains genuine \Bcep stars, but is certainly contaminated by other types of variable stars. 

Other types of variable objects are found near \Bcep stars in the HR diagram. Slowly pulsating B (SPB) stars tend to have spectral types between approximately B2 -- B9, and pulsate in relatively lower frequency g-modes compared to \Bcep stars \citep{DeCat2002}. However, very rapid rotation can lead some of these g modes to attain frequencies in the \Bcep domain \citep{Salmon2014}. There are also ``hybrid'' pulsators, oscillating in low-frequency g-modes and high-frequency p-modes simultaneously \citep{Handler2002,DeCat2007,Pigulski2008,Handler2009}. Classical Be stars span the entire spectral range of B-type stars (even extending into the O and A types), and can also be pulsators, often with multiple modes. They are very rapidly rotating as a class, on average rotating at about 80\% of their critical (or breakup) velocity, and occasionally eject matter to form a circumstellar disk \citep{Rivinius2013}. The rapid rotation of classical Be stars can complicate the observed frequency spectrum \citep[\textit{e.g.}][]{Kurtz2015}, as can variability in the circumstellar environment \citep{Stefl1998,Rivinius2016}. Whereas some of the hotter Be stars can also be \Bcep pulsators, some of their gravity modes may also attain sufficiently high frequency to constitute a source of confusion.

The pixel scale of KELT is large (23$^{\prime\prime}$), and some stars in our sample lie in relatively crowded fields near the Galactic plane. As a result, light from other sources will sometimes be blended with the target star in the aperture used by KELT. If a neighboring star is blended with the target star and is variable, then this variability can appear in the light curve for the target. This contamination issue is addressed by first inspecting images of high spatial resolution (DSS and 2MASS images) that show a patch of sky in the vicinity of the target, and identifying any neighboring sources that are close enough to the target to be blended in KELT photometry. Then, difference images of the pixels in the KELT images are analyzed to determine precisely which pixels are the source of the detected variability. This blending analysis can robustly identify contaminating sources further than two KELT pixels away from the target, and is effective at determining which source is variable in somewhat crowded fields. Through this analysis, we reject 3 stars that would have otherwise been classified as candidates. Additionally, three stars that were pre-selected as \Bcep were found to have their signal originating in a neighboring star that could possibly be of a spectral type consistent with \Bcep pulsation, and this variable source is included in our list of candidates. Another consequence of blending is that the measured amplitude of variability will be diluted proportional to the amount of flux from neighboring sources that leaks into the aperture used to extract the target star light curve. Therefore, all photometric amplitudes quoted here should be understood as a lower limit, although this effect is small in practice for the majority of stars in our samples. Since it is primarily the frequency (and spectral type or number detected of modes) that was used to pre-classify the stars in this work, and not the photometric amplitude, the amplitude suppression is not a significant concern for the purposes of classification. Of course, there can be cases where the amplitude suppression is so severe that we would miss the stellar signal completely. Another important cause of amplitude suppression is CCD saturation for bright stars. In this case, only non-saturated pixels would carry the stellar signal, but the total flux would include the saturated ones, resulting in a net decrease in amplitude in units of magnitude.

Keeping in mind the considerations above, the stars that survived the pre-selection (148 objects preliminarily identified as \Bcep stars, and an additional 90 candidates) were frequency-analyzed by hand. To this end we used the {\sc Period04} software \citep{Lenz2005}. This package applies single-frequency Fourier analysis and simultaneous multi-frequency sine-wave least-squares fitting. It also includes advanced options such as the calculation of optimal light-curve fits for multiperiodic signals including harmonic, combination, and equally spaced frequencies. 

It is important to discuss under which conditions the detection of a variability signal is considered significant. The classical approach is to evaluate the signal-to-noise ratio of a peak in the Fourier amplitude spectrum. We computed the noise level as the average amplitude in a 2\cd interval centered on the frequency of interest, i.e. considering possible additional stellar signals as noise at each prewhitening step. For ground-based campaign observations, $S/N>4$ is usually considered a safe choice \citep{Breger1993}. On the other hand, in their search for \Bcep stars in ASAS data whose basic characteristics should be similar to that of our data sets, \citet{Pigulski2008} adopted a more conservative criterion of $S/N>5$. After careful examination of all our ``borderline'' cases that have the strongest signal within $4<S/N<5$ we decided to select $S/N>5$ as the threshold for a star to be classified as a \Bcep pulsator, and a more relaxed $S/N>4$ for candidates. At this stage, two stars were rejected because they were found to have met the $S/N$ criterion only because of uncorrected artifacts in the data. Two further stars were rejected because of heavy crowding in the center of a cluster and likely data artifacts. Two more stars separated by 89" showed the same variability frequencies, suggesting the same source of origin of the variations. Our blending analysis unambiguously shows this signal as coming from the brighter of this pair, and so this brighter source (ALS 7011) is selected as a candidate while the other is not. 

Following that, the astronomical literature was checked for each of the individual stars. The position of these stars in the HR Diagram was determined, or at least estimated. In most cases, a luminosity estimate from the Gaia DR2 parallax \citep{Gaia2016,Gaia2018,Luri2018} and Galactic reddening determinations \citep{Chen1998} could be obtained, but for some stars more information such as accurate multicolor photometry was available, mostly in the Str\"omgren system \citep{Paunzen2015}; calibrations thereof \citep{Napiwotzki1993} were applied to estimate $T_{\rm eff}$ and log\,$g$. These criteria allowed us to identify stars with frequency spectra similar to \Bcep pulsators but with different stellar properties (such as \Dsct and sdB pulsators). Based on these examinations, a few stars were moved in between the two groups (\Bcep stars and candidates), and 22 more were rejected. We finally classify 113 stars as \Bcep and 96 as candidates; 27 and three of those, respectively, have been reported in the literature before.

\section{Results} \label{sec:results}

\subsection{New and Candidate \Bcep Systems} \label{sec:gen_results}

\begin{figure}[!ht]
\centering\epsfig{file=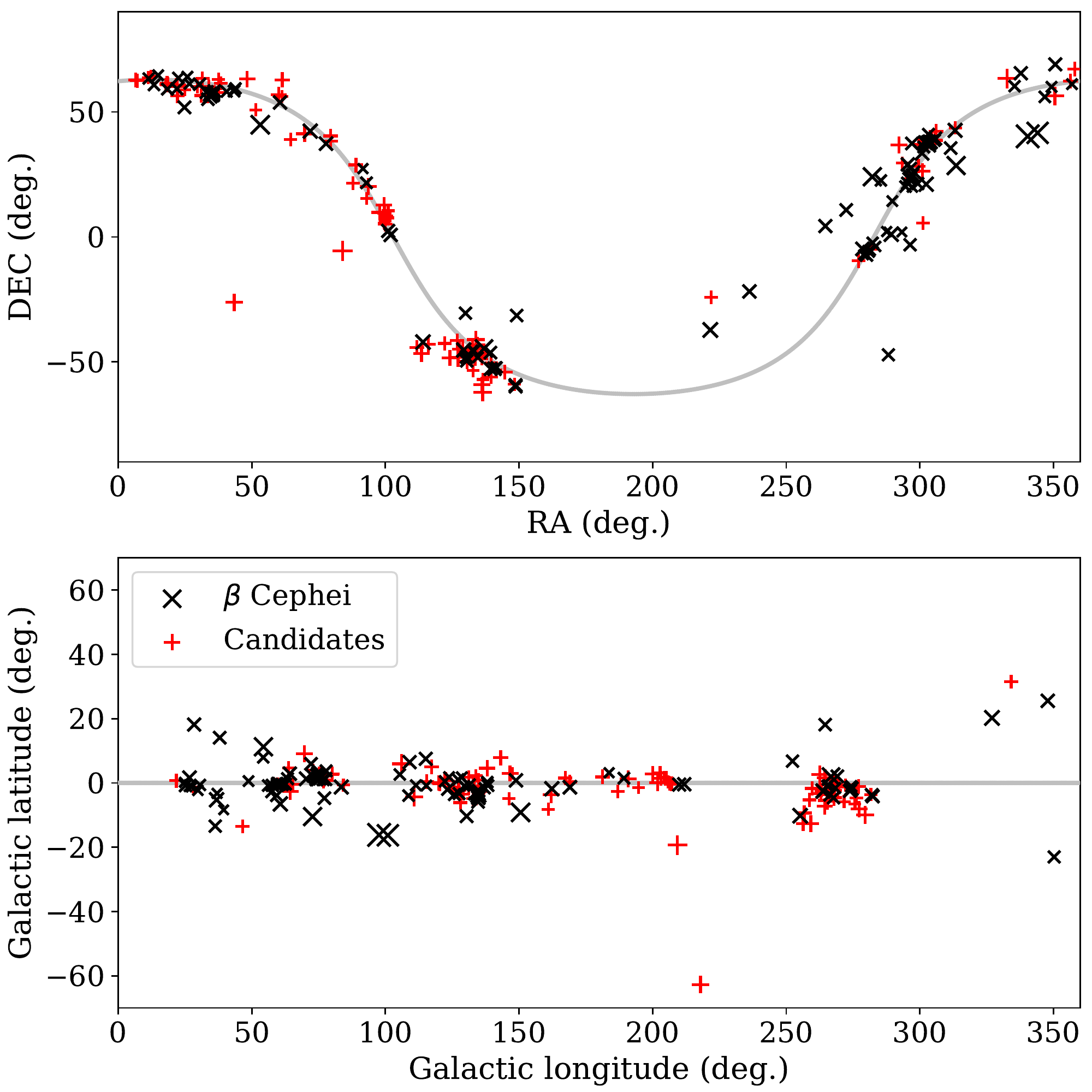,clip=,width=0.99\linewidth}
\caption{Distribution of \Bcep and candidate stars according to their RA and Dec, and Galactic longitude and latitude (epoch J2000). Marker size is proportional to V-band magnitude.}
\label{fig:coords}
\end{figure}

Basic information on the stars investigated is listed in Tables~\ref{tbl:Beta-Ceph-table},~\ref{tbl:candidate-table}, and~\ref{tbl:rejected-table}, sorted by increasing TESS Input Catalog (TIC, \citealt{Stassun2018}) identifier. In addition to a primary identifier, the primary frequency and its photometric amplitude, the number of independent modes in the \Bcep domain detected, coordinates, the V-band magnitude, and the spectral type are listed. All reported amplitudes are the semi-amplitudes. The next column indicates if the star is previously known to belong to the \Bcep class (showing the appropriate reference), followed by a column that indicates whether the star is a known cluster member. Remarks are listed next, which are especially useful if the target has close visual companions (which is important input for TESS target selection). The amplitudes listed in Tables ~\ref{tbl:Beta-Ceph-table} and ~\ref{tbl:candidate-table} should be viewed as lower limits, since light from neighboring sources can blend with the target star, acting to dilute the signal, as discussed in \ref{sec:analysis}. This dilution is strongest in crowded fields. The sky positions of the \Bcep and candidate stars identified in this work are shown in Fig.~\ref{fig:coords}. KELT does not uniformly observe the sky. Some regions have no data, and there can be a large range in the number of observations from field to field. The distributions in Fig.~\ref{fig:coords} are some convolution of KELT sampling and the intrinsic distribution of \Bcep and candidate stars across the sky.

\begin{figure}[!ht]
\centering\epsfig{file=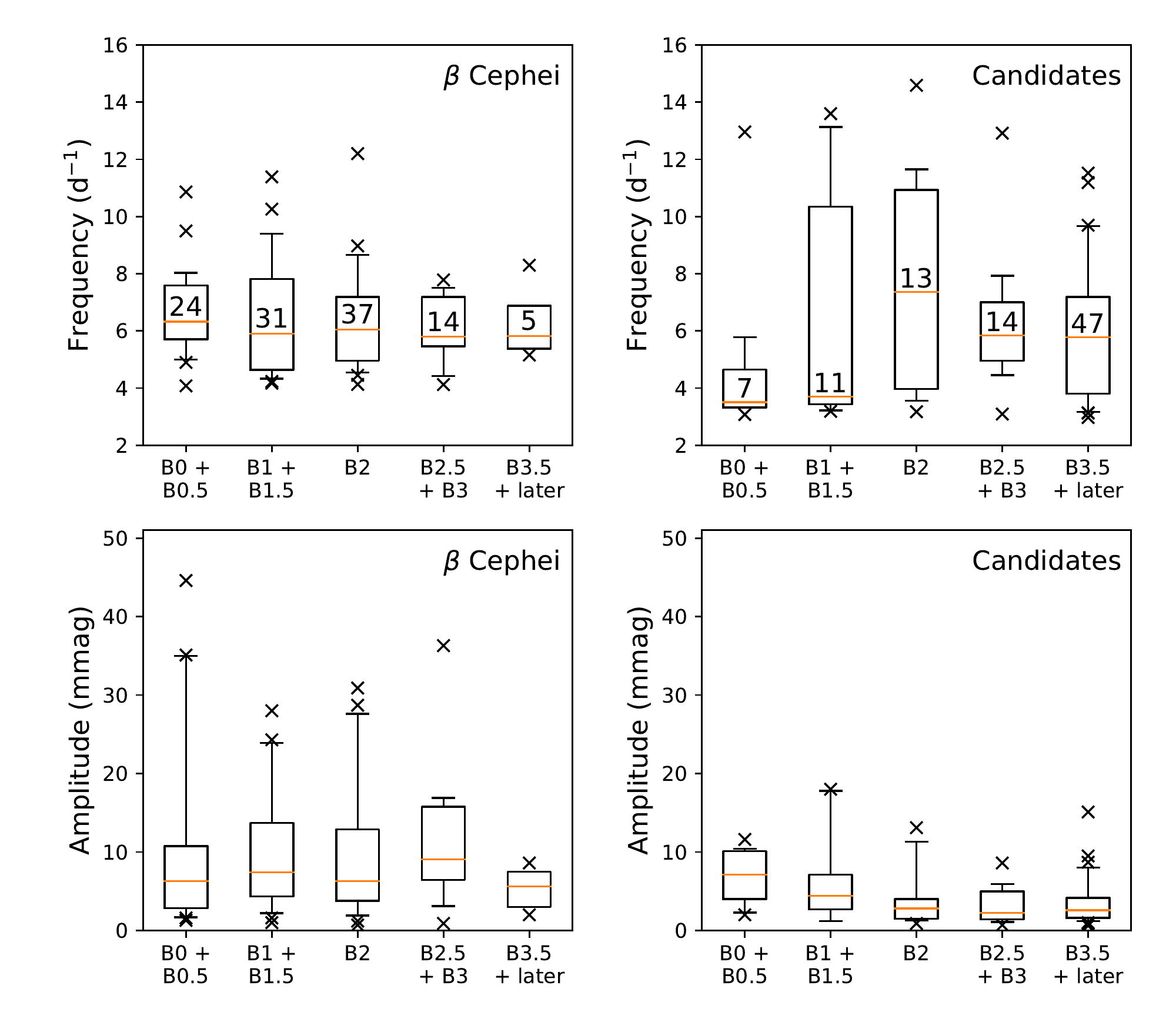,clip=,width=0.99\linewidth}
\caption{Boxplot showing the frequency (top row) and the amplitude (bottom row) distributions for the \Bcep stars (left column) and the candidates (right column). These are split into bins according to their spectral types, as reported in the literature. The numbers in the boxes in the top row indicate the number of objects in that bin. The corresponding bins in the lower row contain the same number of stars. The middle red line in each box is the median, the top and bottom of the boxes mark the 25th and 75th percentile, and the lower and upper whiskers denote the 5th and 95th percentile. Outliers are shown as x's. The \Bcep star HD 190336 has a spectral type of B0.7, and candidate CPD-45 2977 has a spectral type of O9.5, both of which are in the first bin in their respective categories.}
\label{fig:boxplot}
\end{figure}

\begin{figure}[!ht]
\centering\epsfig{file=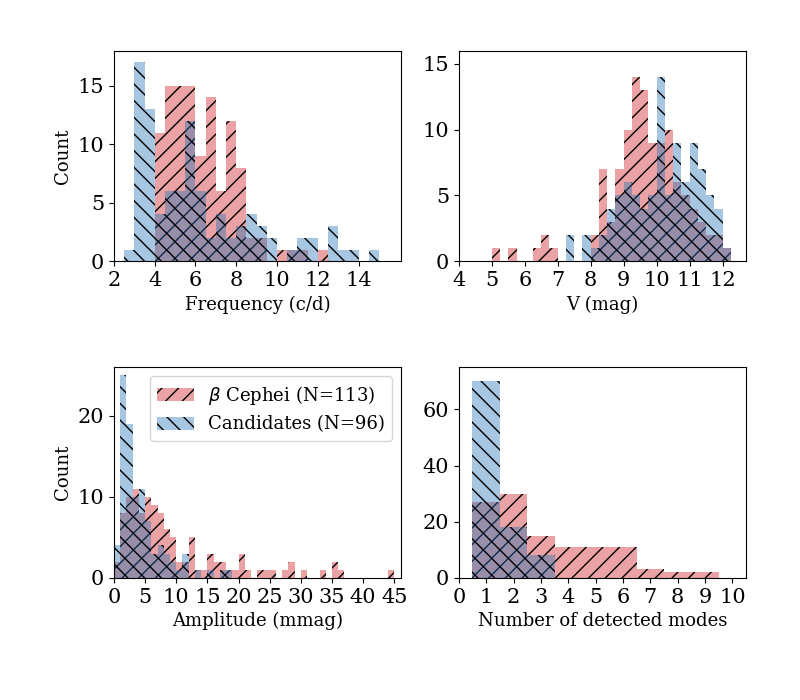,clip=,width=0.99\linewidth}
\caption{Histogram of the primary frequency (top left) and the corresponding amplitude (bottom left) recovered, the V-band magnitude (top right), and the number of detected modes (bottom right) for the \Bcep and candidate stars.}
\label{fig:Hist_fig}
\end{figure}

In Fig.~\ref{fig:boxplot} we show the distribution of the primary detected frequency and its photometric amplitude, separated by spectral type bins. The overall median primary frequency is 6.04 \cd for the \Bcep group and 5.53 \cd for the candidates, and the median amplitudes are 6.5 mmag and 3.0 mmag, respectively. Fig.~\ref{fig:Hist_fig} shows histograms of the frequency, amplitude, number of modes, and $V$-band brightness for the \Bcep and candidate stars.

\begin{figure}[!ht]
\centering\epsfig{file=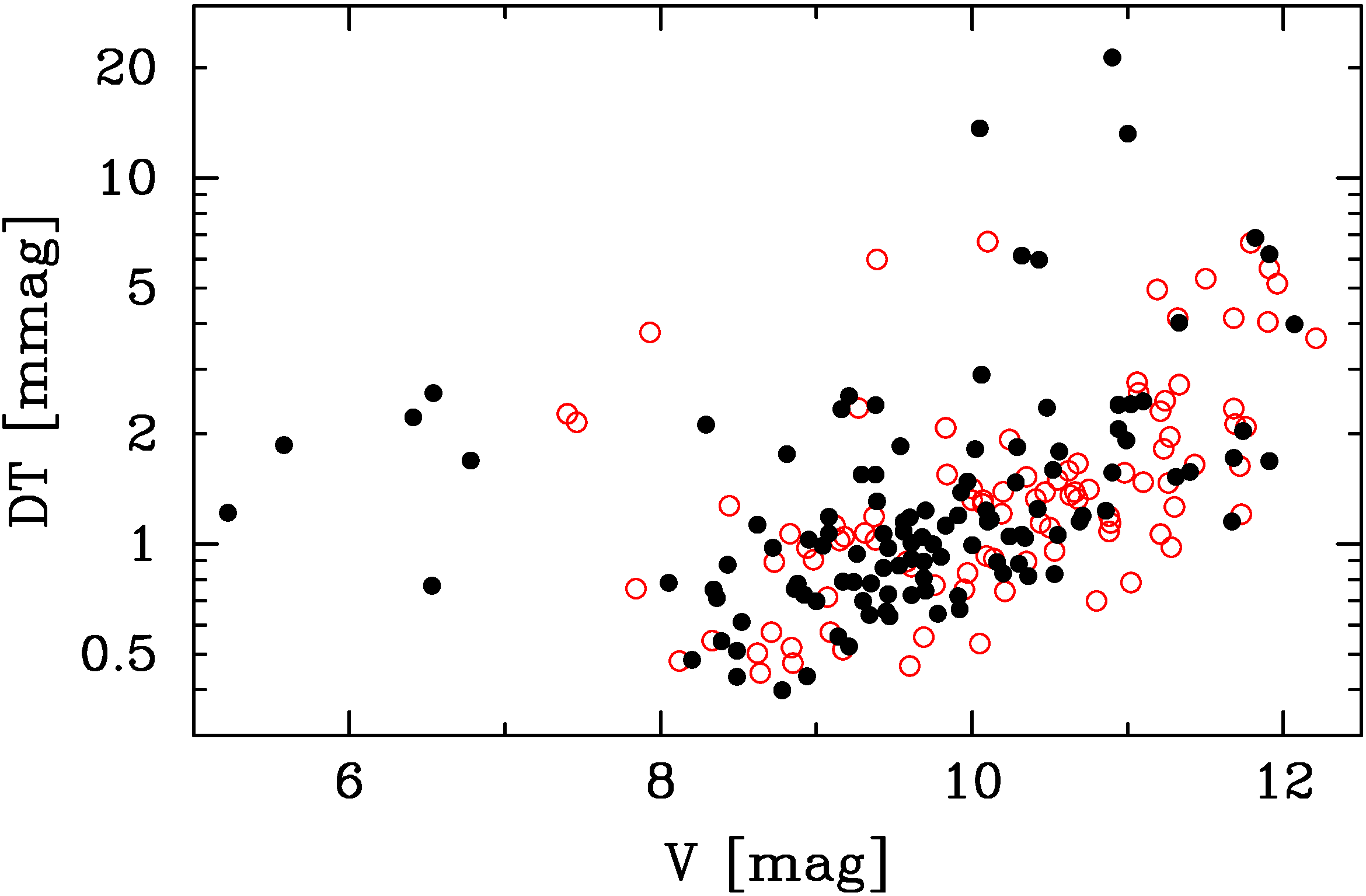,clip=,width=0.99\linewidth}
\caption{Detection level of periodic variability in the \Bcep domain for stars classified as \Bcep pulsators (filled black circles) and candidates (open red circles). Note the logarithmic ordinate scale.}
\label{fig:mag_det}
\end{figure}

In Fig.~\ref{fig:mag_det} we show the detection level ($S/N=4)$ for pulsations in the \Bcep domain for the stars with detections. Despite a few outliers (mostly blended faint stars) it is satisfactory, with a median detection level of 1.15 mmag and mean value of 1.8 mmag. This is approximately a factor of 3 lower than that of \citet{Pigulski2008}, largely due to higher photometric precision and greater number of observations. As can be expected, there is a dependence of detection level on stellar magnitude, with median levels of 0.7 mmag for $8 < V < 9$, 1.0 mmag for $9 < V < 10$, 1.3 mmag for $10 < V < 11$, and 2.3 mmag for $V>11$. The brightest stars we could detect variability for are the known \Bcep pulsators 12 and 16 Lac, the faint end lies at $V=12.1$, which is however more of a limitation in the availability of spectral classification (the major historical star catalogues have a magnitude limit of $V\approx 10$) than of the KELT time series data.

\begin{figure}[!ht]
\centering\epsfig{file=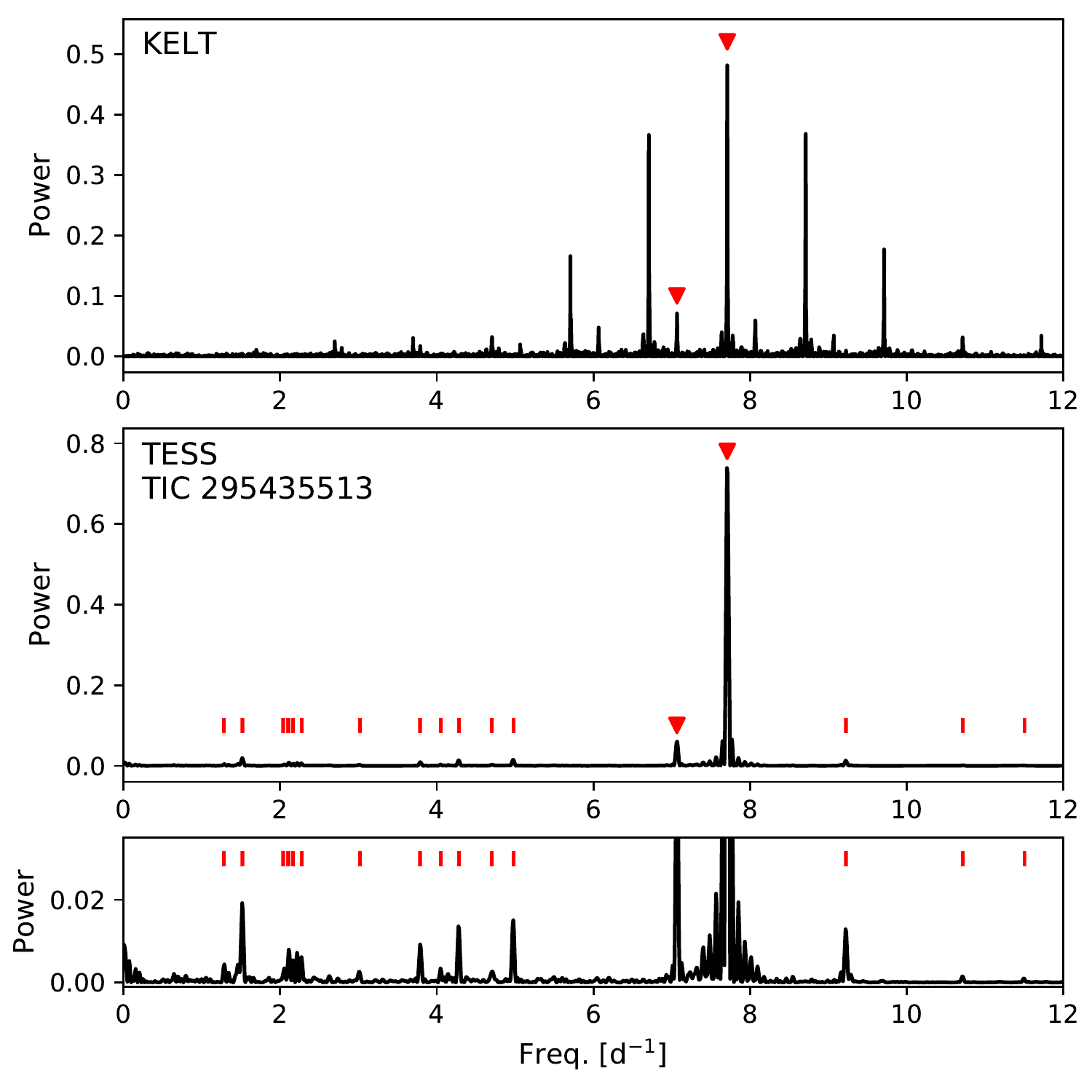,clip=,width=0.99\linewidth}
\caption{Frequency spectrum of the \Bcep star TIC 295435513 computed from KELT (top) and TESS (middle) data. The two frequencies identified in this work from KELT data are marked with red triangles. Additional frequencies detected in the TESS data are indicated by red tick marks. The lower panel shows these frequencies more clearly. The non-marked peaks in the top panel are aliases of the marked frequencies.}
\label{fig:KT_comp}
\end{figure}

To demonstrate how this work can be applied in the context of new and upcoming data from the NASA TESS mission, Fig.~\ref{fig:KT_comp} shows the frequency spectrum of a typical and arbitrarily-chosen \Bcep star from this work, with the upper panel displaying the spectrum computed from KELT data, and the lower panels showing the results from TESS 2-minute cadence data. The two significant frequencies identified in this work from the KELT data are indicated with red triangles. These are also the most significant frequencies in the TESS data. Additional significant frequencies found in the TESS data are indicated with red tick marks. The lower-most panel zooms in on these peaks. A comprehensive analysis of available TESS data for the stars in our sample is beyond the scope of this work.

\subsection{Interesting Systems} \label{sec:interesting}
During the process of manual inspection for each new and candidate \Bcep systems, a number of systems show interesting features in addition to the periodic photometric signals characteristic of \Bcep stars. Here we identify and briefly discuss these cases. 

\subsubsection{Eclipsing Binaries} \label{sec:EBs}

\begin{figure}[!ht]
\centering\epsfig{file=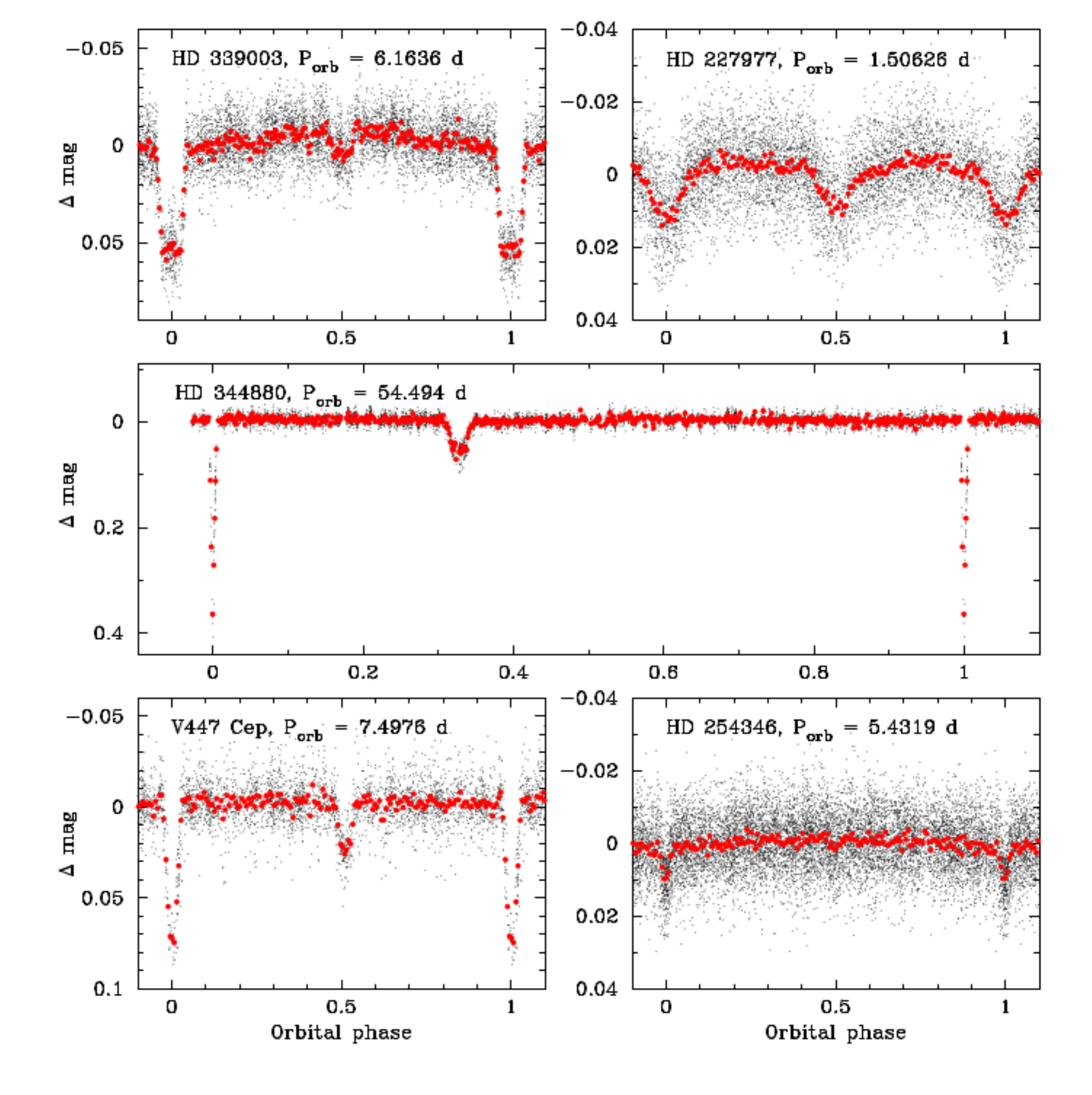,clip=,width=0.99\linewidth}
\caption{Phase-folded light curves of the eclipsing binaries discovered in this work; the pulsational variability has been removed for clarity. Black points show the KELT data and red points show the binned data to illustrate the basic light curve shape.}
\label{fig:eclipse}
\end{figure}

Approximately two thirds of all stars with spectral types that would put them into the \Bcep domain are located in binary systems \citep[\textit{e.g.}][]{Chini2012}. Therefore it can be expected that some of our targets show eclipses. Indeed, we found five objects that can confidently be classified as eclipsing binaries; others may also have binary-induced light variations and are mentioned in Appendix A. We show phase-folded light curves of the pulsation-removed light curves of the eclipsing binaries in Fig.~\ref{fig:eclipse}. We give ephemerides for the times of primary minimum in Table~\ref{tbl:eclephem}, in the form
\begin{displaymath}
min_I=T_0+E*P_{\rm orb}
\end{displaymath}
It should be pointed out that we classified only the first three stars in Table~\ref{tbl:eclephem} as definite \Bcep pulsators; V447 Cep is among the candidates and HD 254346 among the rejected stars (see discussion in Sect. A.3).

\begin{deluxetable}{lcr}
\tablenum{1}
\tablecaption{Ephemerides of the eclipsing binaries containing a \Bcep or candidate star. Error estimates are given in parentheses after the last significant digit.\label{tbl:eclephem}}
\tablewidth{0pt}
\tablehead{
\colhead{ID} & \colhead{$P_{\rm orb}$} & \colhead{$T_0$}\\
\colhead{ } & \colhead{(d)} & \colhead{(HJD)}
}
\decimals
\startdata
TIC 10891640 = HD 339003 & 6.1636(1) & 2456003.90(1)\\
TIC 42365645 = HD 227977 & 1.50626(1) & 2456001.131(6)\\
TIC 451932686 = HD 344880 & 54.494(2) & 2456050.2(1)\\
TIC 335484090 = V447 Cep & 7.4976(7) & 2456006.06(2)\\
TIC 426520557 = HD 254346 & 5.4319(3) & 2456002.15(4)\\
\enddata
\end{deluxetable}

Inspecting Fig.~\ref{fig:eclipse}, one notices a reflection effect in the light curve of HD 339003 in addition to the total eclipses. Given this light curve, the companion star must be of grossly different surface temperature and smaller than the \Bcep pulsator. This means that it could either be an evolved star or belongs to the recently proposed class of binaries by \citet{Moe2015}, containing a hotter main sequence star and a cooler pre-main sequence companion. HD 227977 is the shortest-period eclipsing system we have discovered. There is some evidence for geometrical (ellipsoidal) distortion of the primary. The star also has a rich pulsation spectrum which is discussed in Sect.~\ref{sec:splittings}. HD 344880, on the other hand, is the longest-period system detected. Its orbit is obviously eccentric; secondary eclipse lasts approximately four times as long as primary eclipse. At $V=7.46$, V447 Cep is the brightest new eclipsing system in our sample. The light curve appears to have a short phase of totality. Finally, HD 254346 also shows some evidence for a reflection effect, suggesting the presence of a secondary similar to that of HD 339003. Follow-up observations of all these stars are underway.

\subsubsection{Stillstand Phenomenon} \label{sec:stillstand}
Some \Bcep stars show the ``stillstand phenomenon,'' where the phased light curve shows a marked departure from a pure sinusoid. This rare phenomenon was first observed in a \Bcep star in BW Vul \citep{Sterken1986}. As the brightness is increasing, the brightness stalls for some time, and then continues to increase towards its maximum value \citep{Sterken1987}. The ensuing decrease in brightness is relatively steep. Hydrodynamic models of BW Vul suggest that shocks generated below the photosphere can give rise to the stillstand phenomenon \citep{Mathias1998,Fokin2004}. 

\begin{figure}[!ht]
\centering\epsfig{file=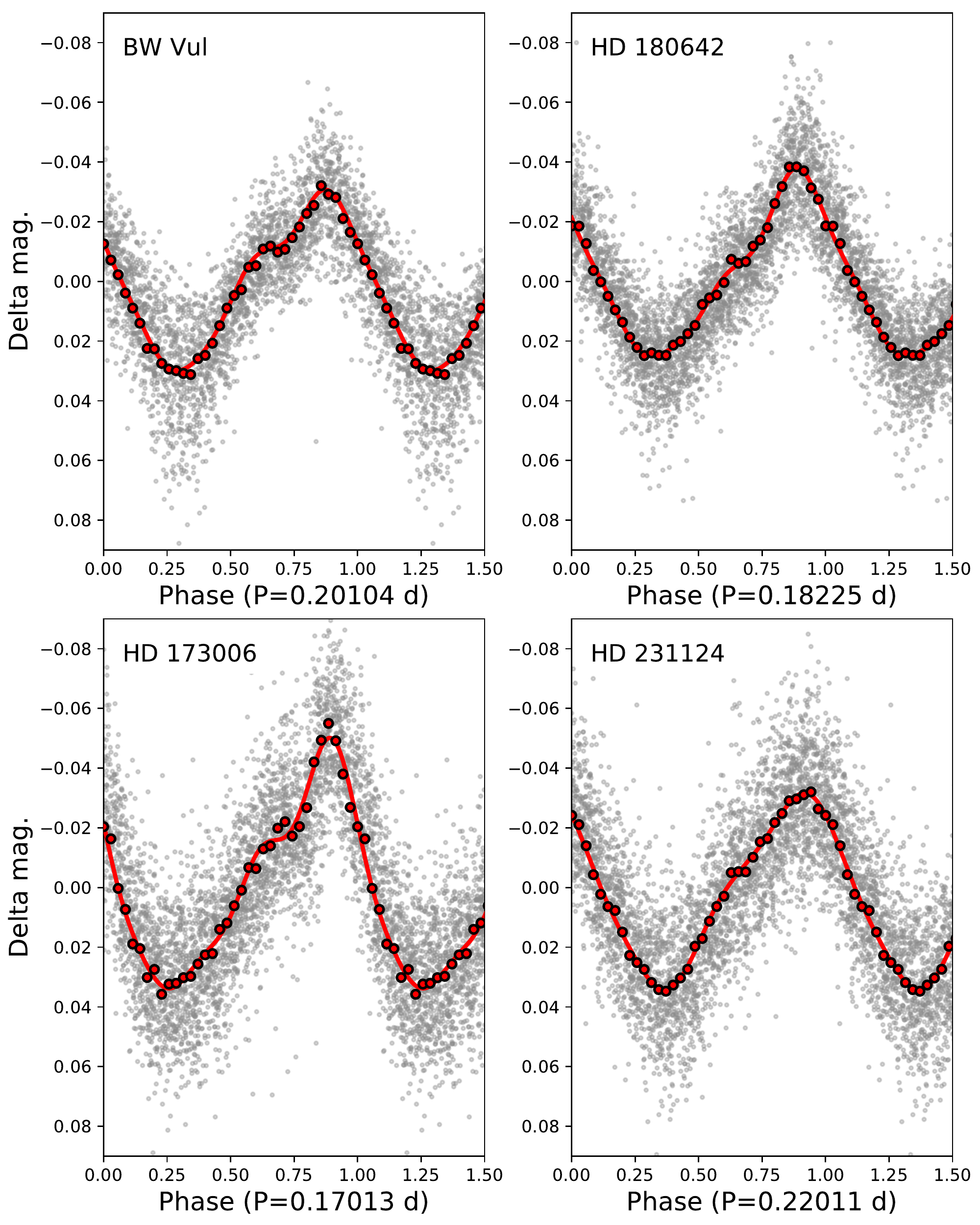,clip=,width=0.99\linewidth}
\caption{Phased light curves of the four \Bcep stars whose light curves show the ``stillstand phenomenon,'' when phased to their primary period. Black points show the KELT data, red points show the binned data, with 35 bins in phase, and the red curve shows a four-term sinusoidal fit to the binned data. Of these, HD 231124 is the only newly-reported \Bcep star. The other three are known \Bcep pulsators where the stillstand phenomenon has previously been reported.}
\label{fig:stillstand}
\end{figure}

Only a handful of \Bcep stars have been found to exhibit the stillstand phenomenon to date \citep[\textit{e.g.}][]{Sterken1986,Pigulski2008,Degroote2009}. Of the four \Bcep stars in the present sample that show the stillstand phenomenon, three have been previously reported as \Bcep stars that have this characteristic (BW Vul, HD 173006, HD 180642). The fourth, HD 231124, is identified here for the first time, having been so far not known to pulsate. Light curves for these four stars are shown in Fig.~\ref{fig:stillstand}, and they are marked in Table~\ref{tbl:Beta-Ceph-table}. 

\subsubsection{Other Non-sinusoidal and Unusual Behavior} \label{sec:non-sinusoidal}

\begin{figure}[!ht]
\centering\epsfig{file=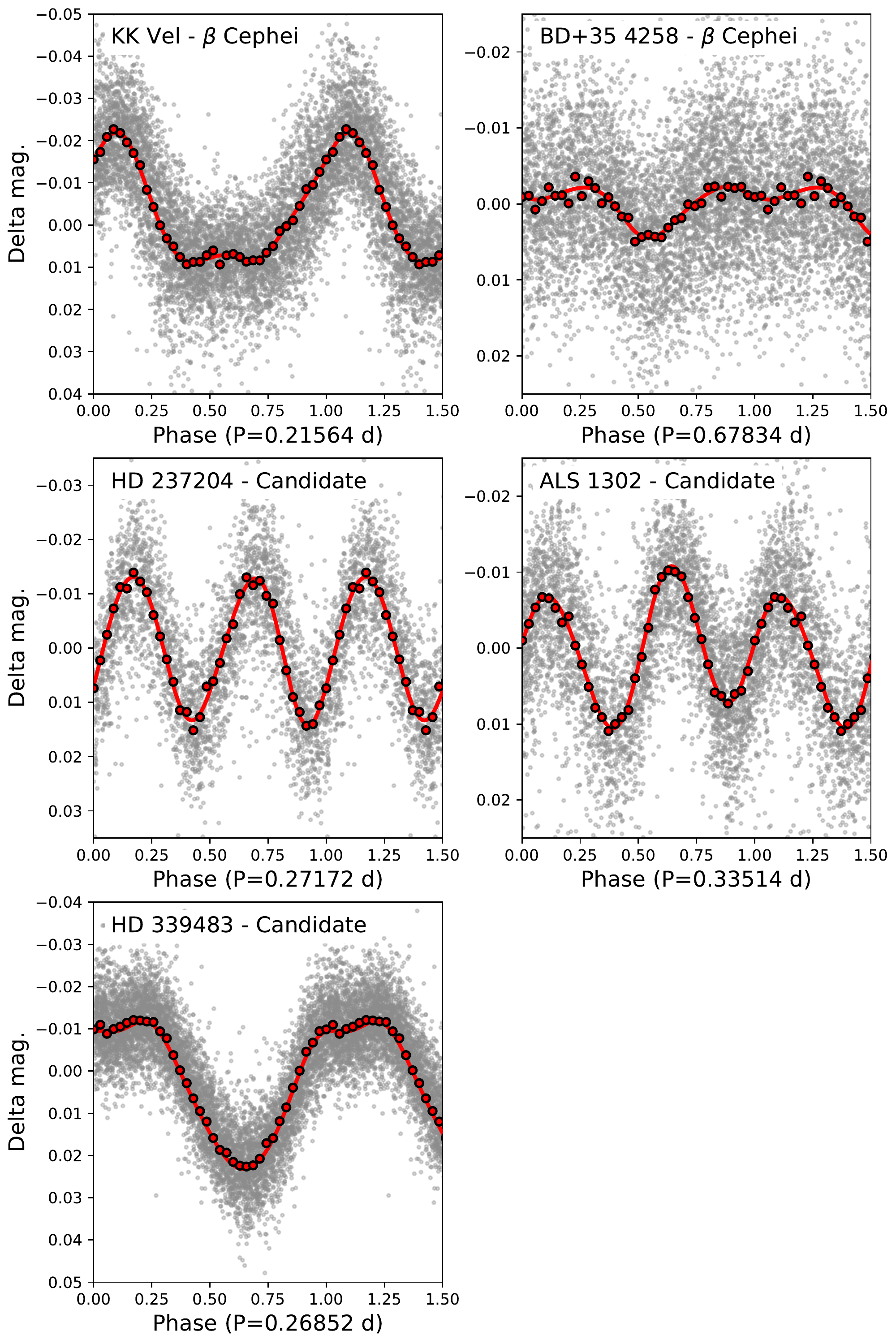,clip=,width=0.99\linewidth}
\caption{Phased light curves for systems with non-sinusoidal signals. Symbols are the same as in Figure~\ref{fig:stillstand}.}
\label{fig:nonsinusoidal}
\end{figure}

Besides the four systems showing the stillstand phenomenon, there are two \Bcep stars and three candidates with unusual or non-sinusoidal behavior that we briefly discuss below. These are shown in Fig.~\ref{fig:nonsinusoidal}.

In the course of the discovery of \Bcep pulsations in KK Vel, \citet{Cousins1982} noticed the non-sinusoidal light curve. \citet{Aerts1994} verified the presence of an additional frequency at twice the primary frequency, noticing differences in the color behavior between these frequencies. The primary frequency of KK Vel is somewhat variable. This was already pointed out by \citet{Cousins1982} who could not phase his three seasons of data (with photometry from 1980 to 1982) with a single frequency. He reported the main frequency to vary between $f=4.6351\pm0.0001$ and $f=4.6361\pm0.0001$\,\cd. \citet{Heynderickx1992} finds $f=4.63637 \pm 0.00003 \,d^{-1}$ with observations in March 1987 and March 1988 with Walraven five-color observations, and November 22 1988 -- January 25, 1989. Also in this work, we see evidence for a variable main pulsation frequency - and its amplitude. In the 2009/2010 observing season, the frequency was $f=4.6342\pm0.0002 \,d^{-1}$ with an amplitude of $A=13.1\pm0.2$\,mmag, in 2012/13 $f=4.63717\pm0.00004 \,d^{-1}$ and $A=21.2\pm0.4$\,mmag, and in 2013/14 $f=4.63760\pm0.00004 \,d^{-1}$, $A=16.6\pm0.4$\,mmag. The amplitude and frequency variations are likely present on time scales shorter than a single observing season. Unfortunately, our data and the previous data are too sparsely sampled in time to facilitate a deeper analysis, although TESS observations may prove valuable in this regard. 

We have classified BD+35 4258 as a \Bcep star. Apart from the single pulsation frequency (at P = 0.13174\,d), the shape of the light curve suggests binary-induced or rotational variability with a base period of 0.67834\,d (Fig.~\ref{fig:nonsinusoidal}, top right), which appears rather short for an early B-type star. Vilnius photometry \citep{Sudzius1992} and its calibration \citep{Straizys1993} indicates $T_{\rm eff} \approx 26000$\,K, log $g \approx 3.9$. From the Gaia DR2 parallax and Galactic reddening one can estimate $M_v<-3.6$. With bolometric corrections and the spectral type-temperature calibration by \citet{Pecaut2013} one derives $M_{bol}<-6.2$, hence $R>7.6$\,R$_{\sun}$ for the star. A rotational modulation with a period of 0.67834\,d hence requires $v_{rot}>560$\,\kms, which is close to breakup speed. Estimating a mass of 13\,M$_{\sun}$ for BD+35 4258 by placing it into a theoretical HR Diagram \citep[e.g.,][]{Martins2013} and assuming a binary scenario with a massless companion and an orbital period of 0.67834\,d requires a semimajor axis of $R=7.6$\,R$_{\sun}$, which is similar to the primary's radius. Alternative hypotheses would be SPB-type pulsation with a non-sinusoidal light curve \citep{Kurtz2015} or a base period twice as long. Another possible scenario is that this signal arises from a close binary system with an orbital period of 0.67834\,d which is not spatially resolved with respect to the B-type star in which we detect the \Bcep pulsation. The light curve shape and period is suggestive of a close binary where both stellar surfaces are gravitationally deformed (\textit{i.e.} an ellipsoidal variable).

The remaining stars discussed in this section are all classified as candidate \Bcep pulsators. The light curve of HD 237204 (Fig.~\ref{fig:nonsinusoidal}, middle left) can be decomposed into two frequencies related exactly by 2:3, suggesting binary-induced or rotational variability with a base period of 0.27172\,d (Fig.~\ref{fig:nonsinusoidal} middle left). Similar to the previously discussed star one can estimate $T_{\rm eff} \approx 29000$\,K, $M_v<-3.2$ from Vilnius photometry \citep{Zdanav2005} and the Gaia DR2 parallax. The bolometric corrections and the spectral type-temperature calibration by \citet{Pecaut2013} then yields $M_{bol}<-6.0$, hence $R>5.6$\,R$_{\sun}$ for HD 237204. A rotational modulation with a 0.27172\,d period thus requires $v_{rot}>1000$\,\kms, far above breakup speed. Estimating a mass of 15\,M$_{\sun}$ for HD 237204 in a possible binary scenario with a massless companion and an orbital period of 0.27172\,d requires a semimajor axis of $R=4.3$\,R$_{\sun}$, smaller than the primary's radius. For these reasons we favor an interpretation in terms of pulsation, although the light curve shape with a descending branch steeper than the rising branch would be remarkable in this case.

ALS 1302 has two variability frequencies that are harmonically related and give rise to a light curve shape (Fig.~\ref{fig:nonsinusoidal}, middle right) reminiscent of rotational or binary-induced variation. However, the base period ($P=0.33514$\,d) would again be too short for such explanations. We appear to be left with an interpretation in terms of pulsation with an unusual light curve shape (\textit{e.g.}, see Fig. 6 of \citealt{Handler2006}).

The light curve of HD 339483 (Fig.~\ref{fig:nonsinusoidal} bottom left) when phase-folded with respect to the strongest variability signal shows an unusual double-humped maximum. Just as for HD 237204, the short period of this signal ($P=0.26852$\,d) effectively rules out a binary or rotational origin. The observation that the rising branch of the phase-folded light curve is steeper than the descending branch suggests that pulsation could be responsible for this variability; perhaps we are seeing a stillstand phenomenon shortly before light maximum, or a curious superposition of a base and a harmonic frequency (e.g., Fig. 1 by \citealt{Kurtz2015}). Although HD 339483 is classified as a classical Be star \citep{Jaschek1982}, it is not clear that there is evidence to support this designation. None of the spectra that are published in the literature or are publicly available show evidence of the circumstellar disk characteristic of classical Be stars. This includes five optical spectra in the Be Star Spectra (BeSS) database\footnote{http://basebe.obspm.fr} that span the dates 2015 September 05 through 2018 September 12, five optical spectra with one observation taken 1998 August 3 and the next four taken between 2009 September and 2010 November \citep{Barnsley2013}, an optical spectrum taken 1998 August 2 \citep{Steele1999}, a K-band spectrum taken 1996 June 28 \citep{Clark2000}, and an H-band spectrum taken 1996 June 29 \citep{Steele2001}. There is, however, a somewhat nebulous feature that is readily apparent in the far-IR WISE W4-band photometry in the vicinity of this object \citep{Wright2010} that may be responsible for the excess at long wavelengths. This star is classified as a \Bcep star in \citet{Pigulski2008}, and while they find the same frequency with ASAS and Hipparcos data as we find in the KELT data, the unusual shape of the phased data is not reported.

\subsubsection{Stars with Frequency Splittings} \label{sec:splittings}

\begin{figure*}[!ht]
\centering\epsfig{file=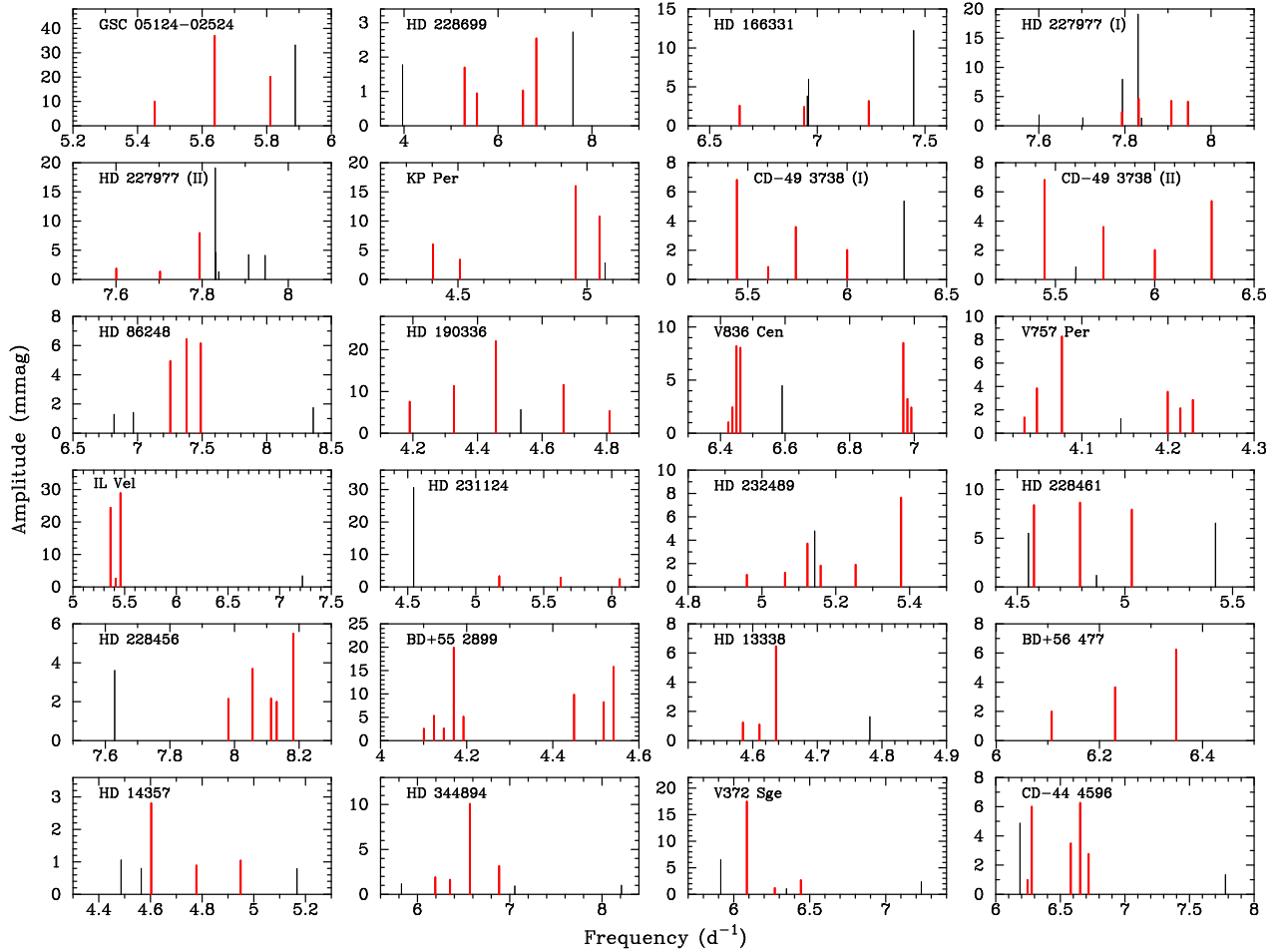,clip=,width=1.0\linewidth}
\caption{Schematic \Bcep mode spectra for stars with equal frequency splittings. The signals suspected to form multiplets are indicated with thick red lines, other oscillation frequencies with thin black lines. Note that some of these possible multiplets are incomplete, that some stars have more than one splitting with the same spacing (that may sometimes even overlap in frequency), and that for two stars, HD 227977 and CD-49 3738, there are two possibilities for the base splitting frequency; see Table~\protect\ref{tbl:split} for more information.}
\label{fig:splittings}
\end{figure*}

\Bcep stars often show nonradial pulsations. Stellar rotation introduces frequency splittings into the pulsation spectra, which can be used to determine their interior rotation \citep[e.g.,][]{Dupret2004,Pamyatnykh2004}. Thus stars that show frequency splittings are highly interesting for asteroseismology, although it needs to be cautioned that such splittings may also occur by coincidence (see \citealt{Handler2006} for an example). We have examined the targets classified as \Bcep stars for frequency splittings, keeping in mind that second-order effects of rotation make them somewhat uneven, in the sense that within a given pulsation mode multiplet the frequency separations of consecutive signals slightly decrease with increasing frequency \citep{Dziembowski1992}, and that some frequency multiplets may be incomplete. We found a total of 22 stars with frequency splittings (Fig.~\ref{fig:splittings}, Table~\ref{tbl:split}) that could reveal their rotation rate. The splittings of two of those stars, IL Vel and V836 Cen, were already known; those of the 20 other stars are new discoveries and await confirmation and/or resolution of ambiguities via pulsational mode identification. Pinpointing possible rotational frequency splittings in this way is expected to be successful only in cases of slow to moderate rotation. For fast rotators (50\% of breakup speed or more), the asymmetry of the rotationally split multiplets hampers their identification from the underlying frequency pattern only \citep{Deupree2010}.

\begin{deluxetable}{lr}  
\tablenum{2}
\tablecaption{Stars with detected frequency spacings and estimates of the first-order rotational splitting. For objects where several possible splittings were found, there is more than one solution.\label{tbl:split}}  
\tablehead{ 
\colhead{ID} & \colhead{$\Delta f$} \\
\colhead{ } & \colhead{(d$^{-1}$)}
}
\startdata
TIC 5076425 = GSC 05124-02524  & 0.0895 or 0.1790\\
TIC 11696250 = HD 228699  & 0.068 or 0.136 or 0.273\\
TIC 25070410 = HD 166331  & 0.1500 or 0.2999\\
TIC 42365645 = HD 227977  & 0.0383 or 0.0484 or 0.0968 \\
TIC 65166720 = KP Per  & 0.025 or 0.049 or 0.098\\
TIC 93730538 = CD-49 3738  & 0.073 or 0.1388 or 0.146 or 0.291\\
TIC 101423289 = HD 86248  & 0.059 or 0.118\\
TIC 105517114 = HD 190336  & 0.070 or 0.140\\
TIC 159932751 = V836 Cen  & 0.0124\\
TIC 264613619 = V757 Per  & 0.0145\\
TIC 293680998 = IL Vel  & 0.0483\\
TIC 299821534 = HD 231124  & 0.221 or 0.443\\
TIC 308954763 = HD 232489  & 0.11 or 0.23\\
TIC 312626970 = HD 228461  & 0.114 or 0.227\\
TIC 312637783 = HD 228456  & 0.036 or 0.072\\
TIC 314833456 = BD+55 2899  & 0.0460\\
TIC 347486043 = HD 13338  & 0.0128 or 0.0257\\
TIC 348137274 = BD+56 477  & 0.0603 or 0.1206\\
TIC 348506748 = HD 14357  & 0.0865 or 0.1730\\
TIC 361324132 = HD 344894  & 0.079 or 0.159\\
TIC 393662110 = V372 Sge  & 0.0889 or 0.1779\\
TIC 461607866 = CD-44 4596  &  0.0346\\
\enddata
\end{deluxetable}

Our results for V836 Cen deserve separate discussion. This star has been among the first to be successfully modeled asteroseismically \citep{Dupret2004}. \citet{Aerts2003} identified a radial oscillation, a rotationally split triplet of dipole modes and two components of a quadrupole mode. Our analysis adds two more components (one of which already tentatively reported by \citealt{Aerts2004}) to the quadrupole mode quintuplet, of which four consecutive components are detected now (see Table in the Appendix). This reduces the previously four possibilities for which is the axisymmetric component of the quadrupole mode to two, and rules out the conjectured identification of this axisymmetric mode by \cite{Aerts2004}. Furthermore, the relative amplitudes of the oscillations in the two nonradial mode multiplets have changed in the $\sim$ two decades that lie between the two data sets. Also, whereas \citet{Aerts2003} find a first-order rotational splitting of 0.0121295\,\cd for the triplet of dipole modes, we obtain a slightly, but statistically significantly larger value of $0.01238\pm0.00002$\,\cd. This suggests that the rotation rate in the resonant cavity of the dipole mode of V836 Cen has {\it increased} in the meantime, counter-intuitive to what would be expected from evolutionary stellar expansion.

\subsubsection{Galactic Distribution} \label{sec:high-lat}

Because \Bcep stars are massive main sequence pulsators, they are relatively young stars. In most cases they should not have had the time to move away from the Galactic plane where they were formed. However, some interactions with other stars, such as a supernova disrupting a binary system \citep{Zwicky1957} or ejection of a star from a young open cluster \citep{Poveda1967} are able to move a young star away from the Galactic plane. Given the interesting evolutionary histories such stars will have, they are of increased interest for asteroseismic investigations.

\begin{figure}[!ht]
\centering\epsfig{file=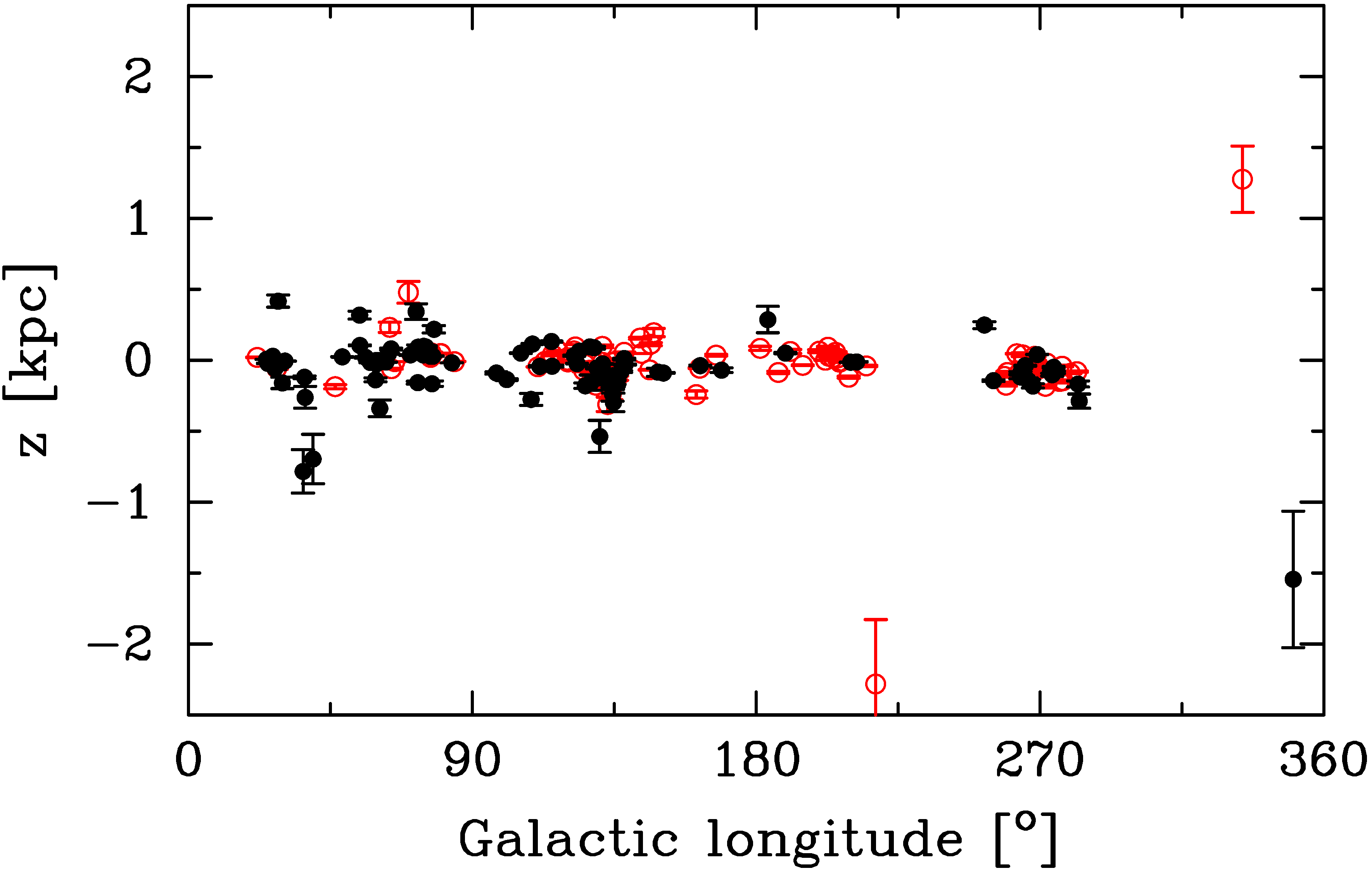,clip=,width=0.99\linewidth}
\caption{Distribution of \Bcep (filled black circles) and candidate stars (open red circles) according to their Galactic latitude and distance from the Galactic plane.}
\label{fig:gal_plane}
\end{figure}

We have therefore computed the distance from the Galactic plane of our targets with parallaxes with a relative error below 25\% and show the result with respect to Galactic longitude in Fig.~\ref{fig:gal_plane}; Table \ref{tbl:gal-table} lists all stars that are located more than 400 pc off the Galactic plane. We also quote their radial velocities if available. Some of these stars have rather large radial velocities, adding evidence for a possible runaway nature.

\begin{deluxetable}{lcr}  
\tablenum{3}
\tablecaption{Stars located more than 400 pc away from the Galactic plane.\label{tbl:gal-table}}  
\tablehead{ 
\colhead{ID} & \colhead{z} & \colhead{RV}\\
\colhead{ } & \colhead{(kpc)} & \colhead{(\kms)}
}
\startdata
\multicolumn{3}{c}{\Bcep stars}\\
TIC 61516388 = HD 178987 & $-1.5\pm0.5$ & \\
TIC 251637508 = HD 186610 & $-0.8\pm0.2$ & +24.4 \\
TIC 308954763 = HD 232489 & $-0.5\pm0.1$ & $-32.0$ \\
TIC 349332755 = HD 160233 & $+0.42\pm0.04$ & +14.0 \\
TIC 392965683 = ALS 10332 &  $-0.7\pm0.2$ & \\
\hline
\multicolumn{3}{c}{\Bcep candidates}\\
TIC 92193 = HD 130195 &  $+1.3\pm0.2$ & \\
TIC  65516748 = HD 18100 &  $-2.3\pm0.5$ & +76.0 \\
TIC 137489662 = HD 183535 & $+0.48\pm0.08$ & +14.0\\
\enddata
\end{deluxetable}

We noticed that there are several stars that failed our criterion for the precision of their parallaxes, but that are rather faint (hence distant in any case) and that are located at relatively high Galactic latitude. These are therefore also good candidates for runaway stars. Among the \Bcep pulsators, these are TIC 25070410 = HD 166331 ($b=-14.1^o, RV=+33$\,\kms), TIC 101423289 = HD 86248 ($b=-18.1^o, RV=+73$\,\kms), and TIC 287467101 = HD 140543 ($b=-25.5^o$); among the candidates, TIC 384540878 = CD$-$56 2603 ($b=-6.6^o$) stands out in this respect. Finally, we remark that V836 Cen ($b=+20.2^o, RV=+66$\,\kms) has already been suspected to be far away from the Galactic plane \citep{Waelkens1983,Aerts2004}. There is no reliable and precise parallax for the star, but its luminosity has been derived asteroseismically \citep{Dupret2004}. Using these results and reddening estimates from \citet{Chen1998} and Str\"omgren photometry, we find $z\approx430$\,pc for this star.

\subsubsection{Extreme Helium Stars} \label{sec:extreme-He}

Extreme Helium stars are rare objects that are not main sequence B-type stars such as the \Bcep pulsators. They are believed to be more likely the product of a merger of two white dwarfs than post-AGB objects that underwent a final Helium flash \citep[see][]{Jeffery2014}. Nevertheless, two of these stars, V652 Her (a known pulsator) and HD 144941, survived the pre-selection criteria for \Bcep pulsators or candidates.

HD 144941 was initially identified as having a tentative low-amplitude signal within the frequency range of the \Bcep stars. However, further analysis suggests this detection was spurious. \citet{Jeffery2018} have analyzed $K2$ photometry and find somewhat complex light variations with a period of 13.9 days and a full amplitude of 4 parts per thousand. This variability is attributed to rotation of an inhomogeneous surface. We do not recover this signal in the noisier KELT data. Further, the high-frequency signal that was tentatively detected in our original analysis (Table\,\ref{tbl:rejected-table}) is not found in the publicly available $K2$ data (within a limit of 40\,$\mu$mag). The lack of pulsation in HD 144941, despite it lying in the instability strip, may be attributed to its low metallicity, as suggested by \citet{Jeffery2018}, or the possible presence of a large organized magnetic field (which the implied surface inhomogeneities may be associated with) could act to suppress pulsation. 

\begin{figure}[!ht]
\centering\epsfig{file=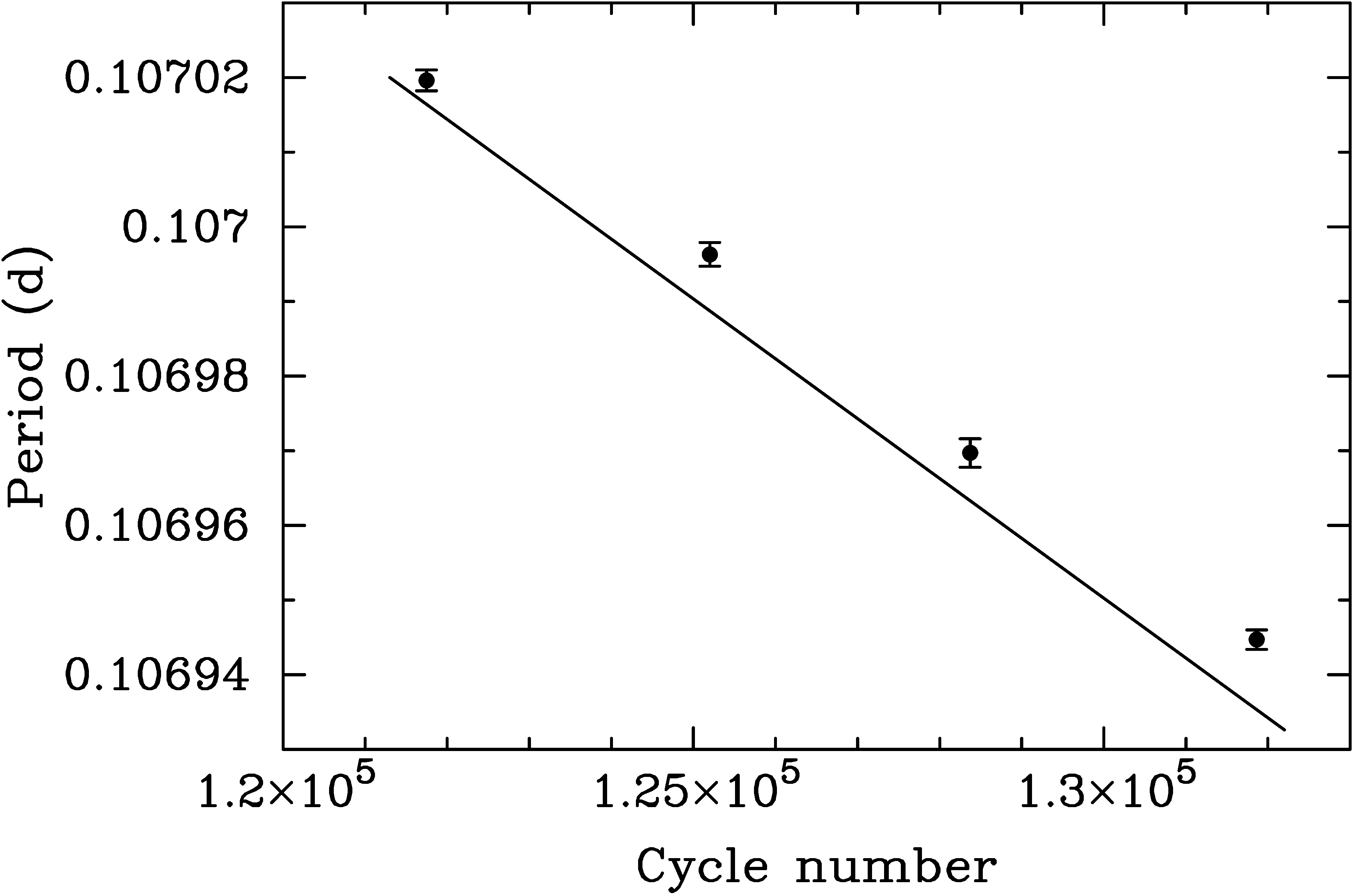,clip=,width=0.99\linewidth}
\caption{Yearly values of the pulsation period of V652 Her (dots with error bars) compared to the fit to earlier data by \citet{Kilkenny2005}.}
\label{fig:v652her}
\end{figure}

V652 Her was readily detected in our analysis, but its dominating frequency is the first harmonic of the actual pulsation frequency due to the unusual light curve shape of this star \citep[see, e.g.][]{Kilkenny1999}. Also, its rapid period change \citep{Kilkenny2005} is evident in our data that cannot be phased with a constant period. We show a comparison of the ``instantaneous" period for our yearly data sets of V652 Her with the fit obtained by \citet{Kilkenny2005} to their earlier period measurements in Fig.~\ref{fig:v652her}. Given the $\approx 6$-year gap between the latest measurements in this paper and the first ones in ours, the general trend of a decreasing pulsation period is still well preserved. However, systematic deviations are also visible, likely a consequence of the non-linear nature of the period change of this star.

\section{Discussion and Conclusions} \label{sec:conclusions}
Through a periodicity analysis of light curves from the KELT survey for O- and B-type stars, we identify 113 \Bcep pulsators, of which 89 are new discoveries. We identify a further 96 stars as \Bcep candidates, a group that likely contains a mix of genuine \Bcep stars, plus other O- and B-type variables. Additionally, we identify 27 stars that meet the pre-selection criteria, but are rejected upon a more scrupulous investigation. 

Our pre-selection was based on a visual inspection of the Fourier spectra of the KELT light curves. Based on our experience with these data, we derived some pre-selection criteria that can be used to facilitate more automatic identifications of \Bcep pulsators and candidates from similar datasets. For stars to be pre-classified as \Bcep pulsators, we required variability frequencies between 4 and 13\,\cd, a signal-to-noise ratio of the strongest signal in the amplitude spectrum to exceed 5, a light curve shape not indicative of binary-induced variability, and a spectral type between O9 and B5 (allowing for inaccurate spectral classifications). \Bcep candidates would be stars not classified as \Bcep that have to have frequencies between $3-15$\,\cd, $S/N>4$ and spectral types O$-$B7. This pre-selection resulted in 148 \Bcep identifications plus 90 candidates. 

The previously described criteria are necessary to be fulfilled, but not sufficient: after a careful literature search and taking into account their position in the HR Diagram, we had to reject seven stars (5\%) pre-classified as \Bcep stars as well as 18 candidates (20\%; see Appendix A.3 for short discussions of each individual star). In addition, we moved 27 (18\%) stars from \Bcep to candidate status to err on the side of caution, but also ``upgraded" three from candidate to \Bcep because reliable spectral classifications not available in our input catalogue were found in the literature. Further, three stars selected as \Bcep stars were found to be blended with a variable neighbor, and in all cases the neighbor is possibly of the correct spectral type and is so classified as a candidate. An additional three stars identified as candidates were rejected in this fashion. We conclude that a mistake-free selection of \Bcep stars is not possible from analyzing light curves and spectral type information alone, but a reasonably clean sample for deeper investigation can be extracted using the criteria above. 

Among the 113 stars finally classified as \Bcep, we found 22 (19\%) with regular frequency spacings suggestive of rotational splitting of nonradial pulsation modes. In comparison, \citet{Pigulski2008} detected only seven such stars in their total sample of 103 (i.e., 7\%). This is probably a consequence of the lower noise level in our data, only about one third of that by \citet[][cf. Sect. \ref{sec:gen_results}]{Pigulski2008}. The average number of pulsation modes detected per star is 3.14 in our data, whereas it is 1.93 in the work by \citet{Pigulski2008}. On the other hand, the number of eclipsing binaries amongst the \Bcep pulsators is three in this work and four in \citet{Pigulski2008}. It is somewhat unfortunate that there is not yet a \Bcep pulsator with more than three oscillation modes that is part of an eclipsing SB2 system.

Aside from the eclipsing binaries, we found some unusual light curve shapes for some of the pulsators. Whereas some of them can be associated with the stillstand phenomenon, others are more difficult to explain. The shapes of some of these light curves phenomenologically resemble those of close binary stars, but the associated periods are too short to be explained that way. Time-resolved spectroscopy of these objects would be valuable to understand their true nature.

Some of the pulsating stars or candidates reported here are located at rather large distances from the Galactic plane and are therefore candidate runaway stars. This is interesting both from the stellar evolutionary and asteroseismic point of view because the flight time of a runaway star provides a limit on the stellar age that can be imposed on seismic models \citep{Handler2019}. Having such a constraint at hand could be of similar importance as basic stellar parameters from pulsators in eclipsing binaries.

These new discoveries (about a 70\% increase over the currently known sample), plus previously known \Bcep stars, will be targeted by the TESS mission. The high-quality TESS light curves will then be used to perform asteroseismic studies on this population, which will reveal valuable information about the interior structure and evolution of massive stars.

\clearpage

\startlongtable
\begin{deluxetable*}{p{2.5cm}lcrcccccccccc} 
\tabletypesize{\footnotesize}
\tablenum{4}
\tablecaption{\Bcep Stars \label{tbl:Beta-Ceph-table}}  
\tablehead{ 
\colhead{ID} & \colhead{TIC} & \colhead{Freq.}    & \colhead{Amp.} &   \colhead{num.}   & \colhead{RA (2000)} & \colhead{Dec (2000)} & \colhead{V}     & \colhead{SP} & \colhead{known?} & \colhead{cluster?} & \colhead{Rem.}  \\
\colhead{}   & \colhead{ID}  & \colhead{(d$^{-1}$)} & \colhead{(mmag)} &  \colhead{modes} & \colhead{}          & \colhead{}           & \colhead{(mag)} & \colhead{}   & \colhead{ }      & \colhead{}         & \colhead{ }  
} 
\startdata 
GSC 05124-02524 & 5076425 & 5.6387 & 36.3 & 4 & 18 35 53.63 & $-$07 09 53.6 & 11.91 & B3 & & & M \\
HD 339003 & 10891640 & 6.7997 & 12.7 & 2 & 19 51 02.86 & +25 57 15.4 & 9.93 & B0.5III & & & M \\
HD 228699 & 11696250 & 7.5964 & 2.7 & 6 & 20 16 36.74 & +37 41 12.8 & 9.46 & B0.5III & & IC 4996 & M \\
HD 228690 & 11698190 & 5.8721 & 9.3 & 4 & 20 16 29.22 & +37 55 21.2 & 9.29 & B0.5V & & & A \\
HD 229085 & 13332837 & 6.8744 & 5.5 & 3 & 20 21 35.13 & +38 36 47.8 & 9.8 & B0V & & NGC 6913 &  \\
HD 194205 & 13967727 & 4.4455 & 11.5 & 2 & 20 23 01.50 & +39 20 40.5 & 9.08 & B2III & & & A \\
HD 166331 & 25070410 & 7.4476 & 12.6 & 6 & 18 09 50.41 & +10 46 26.5 & 9.39 & B1.5III & & & A, M  \\
CD-44 4876 & 29123576 & 7.7826 & 16.9 & 2 & 08 50 16.09 & $-$45 23 02.5 & 10.94 & B3/5 & & & A \\
CD-45 4663 & 29585482 & 8.0717 & 5.5 & 1 & 08 52 29.74 & $-$45 37 03.9 & 11.31 & B1/4 & & &  \\
KK Vel & 31921921 & 4.6373 & 14.8 & 1 & 09 07 42.52 & $-$44 37 56.8 & 6.78 & B1.5II & SH05 & & M \\
HD 227728 & 41327931 & 8.9711 & 1.2 & 2 & 20 06 49.54 & +38 01 39.4 & 9.91 & B2V & & &  \\
HD 227977 & 42365645 & 7.8305 & 20.1 & 9 & 20 09 17.22 & +37 30 07.9 & 9.68 & B2III & & & A, M \\
HD 228101 & 42940133 & 6.2998 & 4.1 & 6 & 20 10 36.69 & +37 27 30.6 & 8.49 & B1IV & & & A \\
HD 228290 & 43881045 & 7.1226 & 7.5 & 1 & 20 12 20.66 & +38 00 07.6 & 9.47 & B1II & & &  \\
HD 171305 & 44980675 & 5.1586 & 8.3 & 2 & 18 34 15.85 & $-$04 48 48.8 & 8.72 & B3III & PP08 & & A \\
HD 253021 & 45799839 & 7.5753 & 3.7 & 2 & 06 11 42.41 & +21 37 58.7 & 10.16 & B2 & & &  \\
ALS 10035 & 55702566 & 6.5490 & 6.3 & 4 & 18 53 57.85 & $-$03 48 49.0 & 11.4 & B0.5III & & & A \\
HD 178987 & 61516388 & 7.1466 & 7.6 & 2 & 19 12 59.64 & $-$47 09 39.5 & 9.83 & B2II & PP08 & & A, M \\
KP Per & 65166720 & 4.9558 & 15.5\dag & 5 & 03 32 38.98 & +44 51 20.7 & 6.41 & B2IV & SH05 & & A, M \\
HD 80279 & 75745359 & 7.2591 & 3.1 & 2 & 09 17 14.93 & $-$46 16 30.2 & 9.45 & B3II/III & & & A \\
ALS 8706 & 80897625 & 4.4060 & 4.6 & 2 & 06 06 28.15 & +27 18 32.3 & 11.68 & B1IIIe & & & A \\
TYC 2682-73-1 & 89756665 & 8.4179 & 3.8 & 2 & 20 05 57.78 & +35 57 13.7 & 10.09 & B1Vn & & &  \\
HD 74339 & 93723398 & 5.2201 & 25.1 & 5 & 08 41 32.97 & $-$48 01 30.7 & 9.38 & B2III & PP08 & IC 2395 & A \\
CD-49 3738 & 93730538 & 5.4451 & 7.2 & 5 & 08 41 38.55 & $-$49 35 52.9 & 9.69 & B3 & & & M \\
HD 86248 & 101423289 & 7.4897 & 6.2 & 6 & 09 56 33.26 & $-$31 26 31.0 & 9.56 & B3II & PP08 & & M \\
HD 190336 & 105517114 & 4.4571 & 21.7 & 6 & 20 03 18.68 & +33 26 59.7 & 8.62 & B0.7II-III & JSH09 & & M \\
HD 61193 & 123211175 & 7.1893 & 0.7 & 1 & 07 36 14.33 & $-$42 04 56.2 & 8.2 & B2Vn & & &  \\
12 Lac & 128821888 & 5.1789 & 7.6\dag & 4 & 22 41 28.65 & +40 13 31.6 & 5.22 & B2III & SH05 & & A \\
16 Lac & 129538133 & 5.8503 & 4.4\dag & 3 & 22 56 23.63 & +41 36 13.9 & 5.58 & B2IV & SH05 & & A \\
HD 73568 & 141292310 & 4.5453 & 3.9 & 4 & 08 37 19.48 & $-$45 12 26.0 & 8.36 & B2III/IV & PP08 & & A \\
V836 Cen & 159932751 & 6.9661 & 8.2\dag & 8 & 14 46 25.76 & $-$37 13 20.1 & 8.05 & B3V & SH05 & & A, M \\
HD 225884 & 168996597 & 5.8220 & 2.9 & 5 & 19 48 15.12 & +37 21 59.2 & 9.43 & B5 & & & A \\
HD 180642 & 175760664 & 5.4869 & 28\dag & 1 & 19 17 14.80 & +01 03 33.9 & 8.29 & B1.5II & SH05 & & A \\
ALS 10186 & 180952390 & 6.6275 & 6.9 & 2 & 19 10 27.87 & +02 07 32.3 & 11.67 & B0.5V & & &  \\
BD-02 4752 & 182714198 & 6.9779 & 7.0 & 2 & 18 49 25.06 & -02 21 09.8 & 10.48 & B0.5V & & & A \\
HD 173006 & 184874234 & 5.8779 & 35.1 & 1 & 18 43 26.26 & $-$05 46 47.7 & 10.06 & B0.5IV & PP08 & & A, M \\
HD 73918 & 185476952 & 8.2949 & 2 & 5 & 08 39 53.96 & $-$30 29 59.9 & 9.7 & B5III & & &  \\
HD 33308 & 187483462 & 6.7355 & 0.9 & 1 & 05 11 07.96 & +37 18 06.8 & 8.78 & B3 & & &  \\
HD 172367 & 197332002 & 4.6882 & 8.5 & 2 & 18 40 09.71 & $-$07 15 02.0 & 9.54 & B2III & & & A \\
HD 332408 & 216976987 & 6.7778 & 2.5 & 1 & 19 42 07.98 & +28 59 45.6 & 8.94 & B2IV & & & A \\
ALS 9955 & 225687900 & 5.4431 & 23.9 & 2 & 18 45 34.43 & $-$05 21 59.0 & 11.02 & B1.5II & & & blend \\
BD+35 4258 & 232846315 & 7.5905 & 2.9 & 1 & 20 46 12.66 & +35 32 25.6 & 9.46 & B0.5Vn & & & M \\
HD 48553 & 234068267 & 5.5980 & 9.8 & 3 & 06 44 09.94 & +02 23 29.6 & 9.08 & B3 & PP08 & &  \\
HD 49330 & 234230792 & 10.8591 & 1.7 & 2 & 06 47 57.27 & +00 46 34.0 & 8.95 & B0nnep & H09 & & A \\
BD+57 614 & 245719692 & 6.9250 & 2.4 & 2 & 02 42 19.47 & +58 05 30.0 & 10.69 & B2III & & & A \\
BD+57 655 & 251196433 & 6.3730 & 6.5 & 3 & 02 53 28.41 & +58 19 32.9 & 10.12 & B2III & & IC 1848 &  \\
ALS 7546 & 251250184 & 4.1204 & 16.3 & 5 & 02 55 36.68 & +59 24 40.1 & 10.52 & B3III & & IC 1848 &  \\
ALS 7541 & 251250634 & 12.2005 & 1.9 & 2 & 02 54 57.47 & +59 15 57.6 & 10.71 & B2II & & IC 1848 &  \\
HD 186610 & 251637508 & 5.4999 & 15.4 & 5 & 19 45 27.32 & $-$03 09 06.6 & 9.7 & B3n & PP08 & & M \\
ALS 6426 & 255974332 & 4.8242 & 10.7 & 5 & 00 59 49.44 & +64 39 37.4 & 10.99 & B2III & & & A \\
BD+56 488 & 264612228 & 6.2736 & 5.3 & 4 & 02 18 13.52 & +57 21 30.9 & 10.1 & B & & NGC 869 & M \\
V611 Per & 264613043 & 5.8246 & 3.2 & 1 & 02 18 29.83 & +57 09 03.1 & 9.35 & B0.5I/V & SH05 & NGC 869 & A \\
V757 Per & 264613619 & 4.0769 & 7.9 & 7 & 02 18 23.05 & +57 00 36.7 & 8.43 & B0.5III & & NGC 869 & A, M \\
HD 14014 & 264615882 & 4.8970 & 6.3 & 6 & 02 18 00.02 & +56 13 57.3 & 8.86 & B0.5V & & &  \\
V665 Per & 264731122 & 4.1275 & 20.4 & 5 & 02 18 48.02 & +57 17 07.9 & 9.38 & B2V & SH05 & NGC 869 &  \\
HD 232874 & 266338052 & 5.7400 & 5.6 & 3 & 04 02 15.74 & +53 45 11.8 & 8.92 & B2III & & & A \\
ALS 12866 & 269228628 & 4.9919 & 44.6 & 3 & 23 17 13.70 & +60 00 27.9 & 10.9 & B0.5V & & & blend \\
ALS 13180 & 272922818 & 8.0279 & 35 & 1 & 23 47 51.68 & +61 05 45.9 & 11 & B0III & & &  \\
HD 345370 & 279236955 & 8.0541 & 18.8 & 4 & 19 56 58.66 & +21 19 49.7 & 9.75 & B2III & & &  \\
BD+68 1373 & 279659875 & 5.8301 & 3.8 & 2 & 23 22 50.65 & +69 00 34.7 & 9.14 & B2III & & & A \\
HD 343642 & 282529703 & 7.5007 & 11.2 & 2 & 19 01 46.01 & +22 34 17.8 & 10.42 & B3 & & &  \\
ALS 6392 & 283136444 & 7.9518 & 13.5 & 1 & 00 53 35.57 & +60 47 08.7 & 10.32 & B2IVnn & & &  \\
HD 140543 & 287467101 & 5.6086 & 2.0 & 1 & 15 44 56.66 & $-$21 48 53.9 & 8.88 & B0.5IIIn & & & A, M \\
HD 339039 & 287690192 & 8.2695 & 9.5 & 5 & 19 48 46.69 & +24 48 21.0 & 9.69 & B1.5V & & & A \\
IL Vel & 293680998 & 5.4598 & 28.7 & 4 & 09 17 31.15 & $-$52 50 19.5 & 9.16 & B2III & SH05 & & A, M \\
HD 81370 & 295435513 & 7.7088 & 10.1 & 2 & 09 23 17.73 & $-$52 44 52.3 & 8.81 & B0.5IVn & & &  \\
HD 298411 & 296570221 & 6.0909 & 5.4 & 5 & 09 26 27.12 & $-$52 09 33.0 & 10.53 & B2/5 & & & blend \\
HD 199021 & 297259536 & 11.3847 & 2.2 & 3 & 20 52 53.21 & +42 36 27.9 & 8.49 & B1IV & & & A \\
HD 231124 & 299821534 & 4.5432 & 30.9 & 4 & 19 18 52.34 & +14 19 41.0 & 11.1 & B2III & & & A, M \\
HD 232489 & 308954763 & 5.3769 & 7.5 & 7 & 01 39 12.85 & +51 49 19.1 & 9.26 & B5 & & & A, M \\
HD 228365 & 311943795 & 8.2886 & 4.8 & 4 & 20 13 01.17 & +41 01 42.1 & 9.97 & B1V & & & A \\
HD 228461 & 312626970 & 4.7912 & 8.5 & 6 & 20 14 06.42 & +38 14 38.4 & 9.56 & B2II & & & M \\
HD 228463 & 312630206 & 6.0847 & 17.3 & 12 & 20 14 03.30 & +37 45 30.1 & 9.6 & B1V & & & A \\
HD 228456 & 312637783 & 8.1827 & 5.5 & 6 & 20 14 02.31 & +36 48 07.0 & 10.2 & B2IV & & & M \\
HD 228450 & 312639883 & 5.9532 & 4.7 & 5 & 20 13 59.02 & +36 32 37.9 & 9.24 & B0.5p & & & blend \\
BD+55 2899 & 314833456 & 4.1698 & 20.4 & 9 & 23 07 08.78 & +56 00 21.1 & 10.29 & B1IIIp & & & M \\
HD 236664 & 331782797 & 7.9873 & 33.9 & 1 & 01 13 41.79 & +59 05 57.4 & 10.05 & B0.5V & & & \\
ALS 12345 & 337105253 & 5.9461 & 15.9 & 2 & 22 21 51.49 & +60 17 09.7 & 10.43 & B3V & & &  \\
BD+64 1677 & 338076483 & 4.7785 & 2.8 & 2 & 22 31 13.58 & +65 27 58.5 & 9 & B2III-IV & & & A \\
HD 13338 & 347486043 & 4.6360 & 6.3 & 4 & 02 12 19.17 & +57 56 27.1 & 9.17 & B1III & & & A, M \\
BD+54 490 & 347684339 & 5.8959 & 3.5 & 2 & 02 14 20.11 & +55 03 33.6 & 9.53 & B1V & & &  \\
BD+56 477 & 348137274 & 6.3488 & 6.1 & 3 & 02 17 04.49 & +56 58 07.2 & 10.02 & B & & NGC 869 & M \\
BD+56 537A & 348231245 & 6.8061 & 2.3 & 1 & 02 19 39.13 & +57 16 13.2 & 10.34 & B2V & & NGC 869 & blend \\
BD+56 540 & 348232246 & 7.7903 & 3.2 & 1 & 02 19 42.66 & +56 58 45.8 & 10.28 & B0.6V & & NGC 869 & blend \\
BD+56 560 & 348443241 & 5.1899 & 4.4 & 1 & 02 21 32.75 & +57 34 07.1 & 10.24 & B2III & & NGC 884 & A \\
HD 14357 & 348506748 & 4.6030 & 2.5 & 6 & 02 21 10.44 & +56 51 56.4 & 8.52 & B2II/III & & NGC 884 & A, M \\
BD+56 584 & 348607991 & 4.5816 & 3.8 & 3 & 02 22 29.86 & +57 12 28.8 & 9.61 & B0.7I & S13 & NGC 884 &  \\
ALS 7146 & 348668924 & 5.5486 & 6.2 & 2 & 02 23 14.38 & +58 09 49.5 & 10.56 & B1V & & &  \\
NGC 884 2579 & 348671634 & 4.4147 & 3.9 & 2 & 02 22 50.28 & +57 08 50.7 & 11.91 & B3e & S13 & NGC 884 &  \\
HD 160233 & 349332755 & 7.6411 & 4.1 & 1 & 17 38 40.64 & +04 20 09.8 & 9.04 & B2IV/V & & & M \\
HD 174298 & 357853776 & 10.2571 & 5.3 & 1 & 18 48 55.63 & +24 03 21.1 & 6.53 & B1.5IV & KE02 & &  \\
HD 344775 & 360063836 & 5.1399 & 7.4 & 6 & 19 43 09.76 & +23 26 15.7 & 10.36 & B1III & & NGC 6823 & A \\
HD 344894 & 361324132 & 6.5681 & 9.5 & 8 & 19 45 48.21 & +23 11 42.7 & 9.61 & B2IIIn & & & A, M \\
BD+58 241 & 370128780 & 4.8896 & 1.6 & 1 & 01 27 54.88 & +59 14 08.9 & 9.91 & B1V & & & A \\
V1143 Cas & 372724051 & 5.2636 & 19.8 & 3 & 01 43 35.58 & +64 02 06.8 & 10.86 & B1 & HM11 & NGC 637 & A \\
BD+60 416 & 374038418 & 5.7341 & 1.6 & 1 & 02 01 44.21 & +61 03 13.0 & 9.61 & B0.5III & & &  \\
HD 350202 & 377088966 & 4.2299 & 9.5 & 3 & 19 38 41.60 & +20 07 46.8 & 10.3 & B1.5III & & &  \\
HD 344842 & 378429048 & 6.5255 & 6.3 & 1 & 19 41 44.28 & +21 29 03.5 & 9.78 & B2III & & &  \\
BD+62 258A & 389532846 & 4.3293 & 8.3 & 2 & 01 30 06.14 & +63 34 57.2 & 10.32 & B1IV & & &  \\
ALS 10538 & 391082033 & 9.3965 & 24.3 & 1 & 19 48 53.68 & +19 58 07.0 & 11.33 & B1V & & &  \\
HD 30209 & 391836737 & 6.1933 & 12.2 & 2 & 04 47 30.26 & +42 19 11.8 & 8.39 & B1.5V & & & A \\
ALS 10332 & 392965683 & 6.0443 & 12.9 & 3 & 19 33 11.45 & +01 56 44.2 & 12.07 & B2 & & & M \\
V372 Sge & 393662110 & 6.0859 & 17.5 & 6 & 20 09 39.59 & +21 04 43.6 & 8.34 & B0.5IIIe & H05 & & A, M \\
NGC 663 4 & 399436828 & 5.1536 & 5.6\dag & 1 & 01 46 39.00 & +61 14 06.1 & 11.06 & B5 & SH05 & NGC 663 & A, blend \\
CD-47 4562 & 400852783 & 6.8803 & 8.6 & 3 & 08 58 23.73 & $-$48 11 08.7 & 10.9 & B5V & & &  \\
BW Vul & 419354107 & 4.9741 & 27.6\dag & 2 & 20 54 22.40 & +28 31 19.2 & 6.54 & B2III & SH05 & & A, M \\
ALS 6331 & 421117635 & 5.2183 & 4.5 & 2 & 00 45 35.06 & +63 21 07.4 & 10.55 & B0.5V & & &  \\
HD 14645 & 445619007 & 5.2131 & 1.7 & 1 & 02 24 01.13 & +58 19 25.9 & 9.43 & B0IVnn & & & A \\
HD 344880 & 451932686 & 9.4929 & 1.3 & 1 & 19 45 42.31 & +23 59 04.0 & 9.34 & B0.5IIInn & & Roslund 2 & A, M \\
HD 338862 & 452018494 & 8.6542 & 4.0 & 1 & 19 46 18.34 & +27 27 37.4 & 9.92 & B2V & & &  \\
CD-44 4596 & 461607866 & 6.6530 & 5.6 & 7 & 08 36 59.62 & $-$45 17 23.1 & 9.3 & B1III & & & M \\
HD 86214 & 469221047 & 4.4426 & 12.4 & 2 & 09 54 58.43 & $-$59 49 46.7 & 9.21 & B1III & PP08 & NGC 3114 & A \\
HD 86162 & 469223889 & 7.8416 & 1.1 & 3 & 09 54 47.75 & $-$59 16 03.3 & 9.21 & B01/IV & & & A \\
\enddata
\tablecomments{Table of \Bcep stars identified in this work, including the ID, TESS Input Catalog (TIC) number, information about the primary frequency and number of modes, coordinates, V-band magnitude, and spectral type, a citation if the star is known to be a \Bcep star, a note if the star is a known cluster member, and information about remarks. Stars labeled ``M" in this column are discussed in the main paper text, stars labeled ``A" are commented upon in the Appendix, while ``blend'' indicates at there are bright nearby sources contributing flux in the aperture used by KELT (and likewise with TESS). The citations for known \Bcep stars are as follows: SH05 = \citet{Stankov2005}, H05 = \citet{Handler2005}, PP08 = \citet{Pigulski2008}, H09 = \citet{Huat09}, JSH09 = \citet{Jurcsik2009}, HM11 = \citet{Handler2011}, S13 = \citet{Saesen2013}. \dag Amplitudes may be largely suppressed.} 
\end{deluxetable*}

\startlongtable
\begin{deluxetable*}{p{2.5cm}lcrcccccccccc} 
\tabletypesize{\footnotesize}
\tablenum{5}
\tablecaption{Candidate \Bcep Stars\label{tbl:candidate-table}}  
\tablehead{ 
\colhead{ID} & \colhead{TIC} & \colhead{Freq.}    & \colhead{Amp.} &   \colhead{num.}   & \colhead{RA (2000)} & \colhead{Dec (2000)} & \colhead{V}     & \colhead{SP} & \colhead{known?} & \colhead{cluster?} & \colhead{Rem.}  \\
\colhead{}   & \colhead{ID}  & \colhead{(d$^{-1}$)} & \colhead{(mmag)} &  \colhead{modes} & \colhead{}          & \colhead{}           & \colhead{(mag)} & \colhead{}   & \colhead{ }      & \colhead{}         & \colhead{ }  
} 
\startdata 
HD 130195 & 92193 & 5.9765 & 1.6 & 1 & 14 47 31.92 & $-$24 12 05.2 & 10.66 & B6II & & & M\\
TYC 4032-93-1 & 11158867 & 12.9147 & 2.0 & 1 & 01 58 38.10 & +60 05 55.7 & 10.68 & B3 & & & A \\
TYC 3151-109-1 & 12249117 & 3.2891 & 3.0 & 1 & 20 17 43.65 & +39 20 36.2 & 11.02 & B5V & & & \\
ALS 7011 & 12647534 & 3.5118 & 5.7 & 1 & 02 15 24.53 & +60 08 21.4 & 10.63 & B0III & & IC 1805 & A\\
TYC 4050-1949-1 & 13839931 & 4.9542 & 3.9 & 1 & 02 30 19.98 & +63 00 23.4 & 11.21 & B7V & & IC 1805 & \\
TYC 3152-1307-1 & 13971092 & 9.6946 & 1.8 & 1 & 20 23 06.70 & +38 49 15.6 & 10.62 & B5 & & NGC 6913 & A\\
HD 229171 & 13973539 & 3.5030 & 9.8 & 2 & 20 23 02.88 & +38 27 20.8 & 9.38 & B0.5IIIne & & NGC 6913 & A \\
CD-44 4871 & 28949811 & 13.127 & 1.2 & 1 & 08 50 09.76 & $-$44 37 22.5 & 10.44 & B1/3V & & & A\\
CD-46 4639 & 29036690 & 7.9231 & 0.7 & 1 & 08 49 39.68 & $-$46 50 53.2 & 10.05 & B3He & & & A\\
HD 76307 & 29598925 & 4.0110 & 4.9 & 1 & 08 53 28.33 & $-$47 31 07.6 & 9.27 & B2/3V & & & A \\
HD 76967 & 30569481 & 5.5781 & 5.2 & 2 & 08 57 54.00 & $-$43 09 12.7 & 9.07 & B3/5V & & & A \\
HD 192003 & 43301361 & 8.8495 & 3.0 & 2 & 20 11 16.85 & +38 13 48.0 & 8.85 & B2IV & & & A \\
ALS 7310 & 49721269 & 11.6465 & 1.3 & 2 & 02 33 43.00 & +61 26 12.2 & 10.89 & B2III & & IC 1805 & blend \\
TYC 4030-800-1 & 53709049 & 6.6171 & 7.3 & 2 & 01 11 12.15 & +61 06 06.2 & 11.96 & B5 & & & A \\
BD+60 185 & 53961302 & 9.3790 & 1.5 & 1 & 01 13 05.69 & +61 24 44.9 & 10.07 & B7V & & & \\
BD+60 192 & 53968977 & 6.0136 & 9.5 & 1 & 01 14 30.29 & +60 53 28.8 & 9.39 & B5 & & & A\\
HD 59259 & 60245596 & 6.4427 & 1.5 & 2 & 07 27 03.03 & $-$44 10 40.2 & 9.95 & B7/8V & & & \\
HD 18100 & 65516748 & 8.5481 & 1.6 & 1 & 02 53 40.81 & $-$26 09 20.4 & 8.44 & B5II/III & & & M \\
HD 237204 & 72696935 & 7.3604 & 13.1 & 1 & 04 00 23.28 & +56 54 05.7 & 9.18 & B2III & & & M \\
ALS 7879 & 72944678 & 3.3264 & 10.4 & 1 & 04 05 03.97 & +56 13 06.3 & 11.79 & B0p & & & \\
CD-45 4896 & 74197071 & 2.9677 & 4.7 & 3 & 09 08 50.23 & $-$46 25 58.9 & 10.75 & B7IV & & & A\\
HD 249179 & 78499882 & 3.7216 & 2.8 & 1 & 05 55 55.05 & +28 47 06.4 & 10 & B5ne & & & A\\
HD 67980 & 80814494 & 5.4045 & 2.3 & 1 & 08 08 44.04 & $-$42 37 07.6 & 10.53 & B7II & & & A\\
HD 74533 & 93923487 & 3.4462 & 1.2 & 1 & 08 42 28.52 & $-$49 45 53.1 & 9.17 & B5IV & & & \\
HD 74581 & 94000461 & 4.5450 & 15.1 & 1 & 08 42 47.92 & $-$48 13 31.1 & 9.12 & B6/8V & & IC 2395 & A \\
HD 248434 & 114249288 & 3.1336 & 4.0 & 1 & 05 51 38.52 & +21 32 28.1 & 10.68 & B5ne & & & \\
HD 60794 & 123036723 & 4.7814 & 4.3 & 1 & 07 34 06.86 & $-$46 38 37.7 & 8.73 & B3/5IIIe & & & \\
HD 62894 & 123828144 & 5.5254 & 1.0 & 1 & 07 44 15.56 & $-$43 01 04.2 & 9.6 & B7/9e & & & A \\
HD 183535 & 137489662 & 11.5196 & 1.6 & 1 & 19 28 38.16 & +36 46 45.1 & 8.64 & B5 & & & A, M \\
CD-44 4484 & 140309502 & 7.1391 & 3.2 & 1 & 08 30 53.51 & $-$44 58 33.2 & 9.76 & B5 & & & A \\
HD 72090 & 140429699 & 6.1756 & 1.6 & 1 & 08 28 52.24 & $-$48 11 25.6 & 7.84 & B6V & & & \\
CD-44 4571 & 141190209 & 4.7460 & 1.5 & 1 & 08 35 52.40 & $-$44 39 23.5 & 10.88 & B5V & & & \\
CD-46 4437 & 141903641 & 5.2911 & 5.8\dag & 2 & 08 40 35.29 & $-$47 14 08.4 & 10.24 & --- & & & A \\ 
HD 279639 & 143530557 & 3.2317 & 4.1 & 3 & 04 18 03.37 & +38 57 47.0 & 11.06 & B7 & & & A\\
CPD-52 1713 & 145672806 & 3.1628 & 1.8 & 1 & 08 51 15.55 & $-$53 27 03.0 & 11.3 & B7e & & & A \\
HD 29332 & 155977294 & 4.4491 & 1.5 & 1 & 04 39 04.89 & +41 15 00.0 & 8.71 & B3ne & & & \\
TYC 2682-3173-1 & 172560369 & 8.9049 & 2.4 & 1 & 19 58 39.75 & +37 00 23.3 & 11.72 & B5 & & & \\
CD-40 4269 & 184226481 & 3.7008 & 4.4 & 1 & 08 27 31.64 & $-$41 24 48.4 & 10.35 & B1/3 & & & \\
HD 76554 & 190783321 & 15.5906 & 1.0 & 3 & 08 55 25.94 & $-$41 04 43.9 & 8.33 & B2Vne & & & A\\
BD+41 3731 & 193573033 & 3.5563 & 2.8 & 1 & 20 24 15.72 & +42 18 01.4 & 9.84 & B2/3ne & & NGC 6910 & \\
BD+41 3736 & 193608395 & 6.9397 & 2.6 & 2 & 20 24 33.90 & +42 14 15.5 & 10.55 & B6 & & NGC 6910 & \\
HD 258853 & 207045768 & 6.1398 & 2.5 & 1 & 06 31 14.87 & +09 47 25.0 & 8.83 & B3Vnn & & & A\\
HD 261172 & 220238856 & 3.7193 & 11.3 & 3 & 06 38 31.56 & +09 25 12.2 & 10.1 & B2III & & NGC 2264 & \\
HD 261589 & 220298408 & 6.0890 & 5.1 & 1 & 06 39 44.90 & +06 27 35.6 & 11.43 & B7 & & NGC 2244 & \\
ALS 9974 & 226144867 & 12.676 & 6.4 & 1 & 18 47 15.69 & $-$05 00 57.5 & 12.21 & B1V & & & \\
TYC 746-578-1 & 231149746 & 9.0529 & 2.9 & 2 & 06 41 53.08 & +08 24 17.7 & 11.68 & B6/9 & & & A\\
HD 262595 & 231177417 & 5.9053 & 5.9 & 1 & 06 43 06.31 & +07 36 03.8 & 11.19 & B3/5 & & & \\
TYC 750-467-1 & 231236273 & 9.6644 & 2.3 & 1 & 06 44 16.00 & +10 28 11.8 & 11.26 & B6/9 & & NGC 2264 & \\
HD 339483 & 244327575 & 3.7241 & 17.8 & 1 & 20 04 00.75 & +26 16 16.8 & 8.98 & B1IIIe & KE02 & & M \\
TYC 3324-92-1 & 252864874 & 5.3848 & 7.1 & 3 & 03 26 04.74 & +50 49 46.6 & 11.33 & B7 & & Melotte 20 & A\\
BD+61 2515 & 272625912 & 3.3052 & 11.6 & 1 & 23 45 43.89 & +62 17 31.2 & 10 & B0.5V & & & \\
GSC 00155-00374 & 281533292 & 5.5399 & 7.9 & 1 & 06 38 13.40 & +05 33 20.0 & 11.9 & B7V & & NGC 2244 & \\
HD 333172 & 282737511 & 3.2152 & 6.9 & 1 & 19 57 53.88 & +28 19 51.2 & 10.35 & B1II & & & \\
ALS 14570 & 285122116 & 7.9203 & 1.1 & 1 & 00 49 11.38 & +64 11 21.9 & 11.28 & B3IV & & & \\
ALS 6915 & 285684866 & 3.0730 & 7.1 & 1 & 02 06 08.56 & +63 22 11.8 & 10.24 & B0.5:ep & & & \\
HD 69824 & 286352720 & 6.0635 & 3.6 & 2 & 08 16 38.69 & $-$48 26 11.2 & 9.09 & B4/6V & PP08 & & \\
HD 78507 & 290300414 & 4.3680 & 0.9 & 2 & 09 05 39.03 & $-$62 06 12.6 & 8.12 & B6V & & & A\\
GSC 01314-00792 & 294306184 & 7.2264 & 5.4 & 1 & 06 12 02.90 & +15 23 10.0 & 11.5 & B5 & & & \\
ALS 1302 & 296449179 & 5.9676 & 8.6 & 1 & 09 25 42.12 & $-$53 14 19.5 & 11.73 & B3 & & & M \\
ALS 11602 & 297264447 & 8.7851 & 1.3 & 1 & 20 53 04.97 & +43 37 13.2 & 11.21 & B2Vn & & & \\
HD 260858 & 307944768 & 5.8902 & 1.2 & 1 & 06 37 46.71 & +12 46 05.1 & 9.15 & B6He & & & A\\
HD 220300 & 319302209 & 3.3830 & 8.7 & 1 & 23 22 10.46 & +56 20 53.6 & 7.93 & B6IVne & & & A \\
V447 Cep & 335484090 & 3.1845 & 18.1 & 1 & 22 10 59.56 & +63 23 58.5 & 7.46 & B1Vk & & & A, M \\
BD+05 4404 & 345652701 & 3.1828 & 2.4 & 1 & 20 04 42.72 & +05 33 19.6 & 10.47 & B5 & & & \\
V352 Per & 347585038 & 3.6438 & 2.9 & 1 & 02 13 37.02 & +56 34 14.3 & 9.31 & B1III & & & A \\
BD+56 579 & 348609224 & 3.9560 & 1.5 & 1 & 02 22 19.23 & +57 37 12.9 & 10.88 & B7IVe & & & A \\
LS II +23 36 & 360749347 & 4.8883 & 11.8\dag & 1 & 19 44 26.15 & $+$23 17 54.1 & 10.68 & OB & & & A \\ 
CD-49 4294 & 364146227 & 3.0901 & 5.7 & 1 & 09 18 19.05 & $-$50 18 46.9 & 11.24 & B3/5 & & & \\
ALS 6216 & 366108295 & 12.9568 & 2.0 & 1 & 00 28 37.71 & +62 29 17.7 & 10.2 & B0.5V & & & A \\
BD+66 1651 & 368237682 & 7.2884 & 1.8 & 2 & 23 52 12.14 & +67 10 07.5 & 9.97 & B3Ve & & & A\\
HD 261630 & 369708912 & 3.2593 & 2.6 & 1 & 06 39 49.73 & +05 04 37.7 & 10.07 & B5 & & NGC 2244 & \\
BD+55 334 & 370269139 & 5.9102 & 1.5 & 1 & 01 28 39.05 & +56 21 04.3 & 10.41 & B2e & & & A \\
BD+57 579 & 372115570 & 3.9621 & 4.0 & 1 & 02 30 14.05 & +57 40 30.3 & 10.09 & B2III & & & A \\
HD 236939 & 374696382 & 5.7791 & 3.6 & 1 & 02 04 33.68 & +56 33 00.6 & 10.19 & B5 & & & \\
BD+29 3644 & 379932247 & 8.3034 & 2.4 & 1 & 19 33 49.91 & +29 29 56.3 & 11.27 & B5 & & & blend \\
HD 78206 & 384257658 & 3.6997 & 1.8 & 2 & 09 04 22.00 & $-$59 09 24.9 & 8.84 & B7/8V & & & A\\
CD-56 2603 & 384540878 & 3.5513 & 3.6 & 1 & 09 05 41.99 & $-$57 00 02.9 & 11.76 & B1III/Ve & & & M \\
CPD-55 2071 & 386513392 & 3.1218 & 1.6 & 1 & 09 18 11.17 & $-$56 01 33.3 & 10.8 & B7e & & & \\
HD 42896 & 386693012 & 13.597 & 7.3 & 3 & 06 14 06.19 & +20 10 10.9 & 8.62 & B1Vnn & & & A\\
BD-09 4742 & 387153140 & 11.3216 & 2.1 & 2 & 18 28 20.13 & $-$09 35 05.2 & 10.5 & B2V & & & A\\
CD-47 4494 & 400080677 & 9.0207 & 0.6 & 1 & 08 54 20.85 & $-$48 25 49.5 & 9.69 & B5 & & & \\
TYC 3683-1328-1 & 400533469 & 11.1849 & 4.5 & 1 & 01 39 47.92 & +58 45 24.4 & 11.68 & B5 & & & \\
CD-48 4390 & 401847213 & 5.1564 & 4.9 & 1 & 09 04 23.69 & $-$48 46 25.1 & 11.32 & B4/6 & & & \\
HD 280753 & 408941991 & 4.6168 & 1.4 & 1 & 05 18 05.96 & +38 17 40.6 & 10.21 & B3 & & & \\
HD 277933 & 408991617 & 5.7713 & 1.4 & 1 & 05 17 51.12 & +40 23 27.0 & 10.14 & B3 & & & \\
BD+61 77 & 419246605 & 8.0205 & 1.2 & 1 & 00 26 23.21 & +62 45 38.9 & 9.61 & B1IV & & & A \\
ALS 6330 & 420545430 & 3.3131 & 2.5 & 1 & 00 45 17.23 & +63 42 36.8 & 11.1 & B1III & & & \\
TYC 4031-1324-1 & 421691939 & 6.0165 & 3.5 & 2 & 01 27 04.47 & +60 49 08.1 & 11.07 & B5 & & & \\
BD+59 254 & 422533344 & 5.3948 & 3.8\dag & 2 & 01 28 15.06 & $+$60 13 45.8 & 10.07 & A2 & & & A \\ 
HD 37115 & 427396133 & 5.9099 & 5.8 & 2 & 05 35 54.08 & $-$05 37 42.3 & 7.4 & B7Ve & & & A\\
CD-45 4501 & 430625041 & 5.7728 & 2.3\dag & 1 & 08 45 19.49 & $-$45 57 34.4 & 9.08 &  OB & & & A \\ 
HD 298610 & 440971753 & 3.1712 & 4.1 & 2 & 09 39 00.81 & $-$54 03 45.3 & 9.83 & B2/4e & & & \\
HD 77769 & 447933173 & 5.4681 & 3.9 & 3 & 09 02 48.21 & $-$46 57 48.9 & 9.37 & B3IV & PP08 & & A\\
HD 19635 & 458911894 & 3.8763 & 2.4 & 3 & 03 12 56.84 & +63 11 12.4 & 8.94 & B4 & & & A\\
BD+60 770 & 459014234 & 11.7894 & 10.8 & 5 & 04 00 49.57 & +61 17 26.5 & 9.8 & B5 & & & A \\ 
BD+62 647 & 459105076 & 10.9253 & 2.3 & 1 & 04 05 34.01 & +62 47 28.6 & 9.58 & B2V & & NGC 1502 & \\
TYC 8610-895-1 & 469095496 & 8.0080 & 4.3 & 1 & 09 53 23.95 & $-$58 57 36.2 & 11.23 & B5/7 & & & blend \\
\enddata
\tablecomments{Table of Candidate stars identified in this work, including the ID, TESS Input Catalog (TIC) number, information about the primary frequency and number of modes, coordinates, V-band magnitude, and spectral type, a citation if the star is known to be a \Bcep star, a note if the star is a known cluster member, and information about remarks. Stars labeled ``M" in this column are discussed in the main paper text, stars labeled ``A" are commented upon in the Appendix, while ``blend'' indicates at there are bright nearby sources contributing flux in the aperture used by KELT (and likewise with TESS). The citations for candidate \Bcep stars are as follows: PP08 = \citet{Pigulski2008}, KE02 = \citet{Koen02}; some of these were already listed as candidates by \citet{Stankov2005}. \dag Amplitudes may be largely suppressed.}
\end{deluxetable*}

\startlongtable
\begin{deluxetable*}{p{2.5cm}lcrcccccccccc} 
\tabletypesize{\footnotesize}
\tablenum{6}
\tablecaption{Rejected stars\label{tbl:rejected-table}}  
\tablehead{ 
\colhead{ID} & \colhead{TIC} & \colhead{Freq.}    & \colhead{Amp.} &   \colhead{num.}   & \colhead{RA (2000)} & \colhead{Dec (2000)} & \colhead{V}     & \colhead{SP} & \colhead{known?} & \colhead{cluster?} & \colhead{Rem.}  \\
\colhead{}   & \colhead{ID}  & \colhead{(d$^{-1}$)} & \colhead{(mmag)} &  \colhead{modes} & \colhead{}          & \colhead{}           & \colhead{(mag)} & \colhead{}   & \colhead{ }      & \colhead{}         & \colhead{ }  
} 
\startdata 
TYC 4033-2268-1 & 12647620 & 3.5118 & 4.6 & 1 & 02 15 28.61 & +60 09 44.6 & 11.69 & B3 & & IC 1805 & A\\ 
HD 221991 & 26283875 & 14.2427 & 0.8 & 1 & 23 36 32.16 & +52 37 12.6 & 9.84 & B5 & & & A  \\
HD 75290 & 28709025 & 5.4869 & 1.9 & 3 & 08 47 33.74 & $-$42 29 07.0 & 8.09 & B3/5V & & Trumpler 10 & A \\
HD 190088 & 40102236 & 5.2399 & 3.0 & 1 & 20 01 45.46 & +38 44 06.9 & 7.86 & B5 & & & A \\
HD 59446 & 60320306 & 7.0191 & 0.8 & 1 & 07 27 42.78 & $-$47 24 50.4 & 7.59 & B6II/III & & & A \\
HD 144941 & 67985749 & 6.344 & 1.5 & 1 & 16 09 24.55 & $-$27 13 38.2 & 10.02 & B1/2II & & & A, M \\
HD 72539 & 90134626 & 2.9852 & 1.6 & 2 & 08 31 22.31 & $-$48 44 56.9 & 7.97 & B5V & & & A \\
o Vel & 93549165 & 3.9746 & 0.7 & 1 & 08 40 17.59 & $-$52 55 18.8 & 3.63 & B3/5V & & & A \\
HD 331621 & 102161004 & 3.3436 & 1.2 & 2 & 19 56 59.97 & +31 17 16.0 & 9.98 & B7 & & & A \\
HD 29450 & 118680798 & 4.6172 & 0.7 & 1 & 04 39 13.54 & +22 39 08.1 & 8.57 & B7V & & & A \\
HD 62755 & 123754451 & 3.4328 & 0.8 & 5 & 07 43 23.09 & $-$47 02 48.0 & 7.85 & B5V & & & A \\
HD 290564 & 138905907 & 8.6575 & 12.2 & 4 & 05 32 08.73 & +00 07 36.8 & 11.2 & B5 & & & A \\
CD-46 4432 & 141903541 & 5.2911 & 5.8 & 2 & 08 40 30.48 & $-$47 12 35.2 & 10 & B5 & & & A \\  
HD 151654 & 157535787 & 12.5893 & 6.4 & 1 & 16 48 49.41 & $-$03 36 39.4 & 8.6 & B0.5V & & & A \\
HD 181124 & 162012064 & 22.1863 & 2.2 & 3 & 19 19 18.34 & $-$01 22 50.4 & 9.62 & B5 & & & A  \\
HD 35612 & 264485563 & 9.0839 & 0.8 & 1 & 05 26 06.00 & +00 50 02.4 & 8.3 & B7Vn & & & A \\
TYC 4804-1086-1 & 281803267 & 3.9139 & 18.5 & 1 & 06 50 31.54 & $-$02 19 45.9 & 11.91 & B5 & & & \\ 
HD 59325 & 291556636 & 3.7152 & 3.9 & 3 & 07 26 56.07 & $-$51 11 07.0 & 10.55 & B7V & & & A \\
HD 190066 & 352529679 & 7.0184 & 15.6 & 0 & 20 02 22.10 & +22 09 05.2 & 6.6 & B1Iab & & & A  \\
ALS 10464 & 360661624 & 4.8883 & 11.8 & 1 & 19 44 21.08 & +23 17 05.9 & 11.82 & B0.5V & & & A \\ 
V652 Her & 377498419 & 18.6905 & 7.6 & 1 & 16 48 04.69 & +13 15 42.4 & 10.51 & B2He & & & A, M  \\
TYC 3315-1807-1 & 384992041 & 3.7614 & 44.1 & 1 & 03 21 39.63 & +47 27 18.8 & 11.73 & B7 & & Melotte 20 & A \\
HD 196035 & 417438983 & 4.0083 & 2.8 & 1 & 20 34 09.98 & +20 59 06.7 & 6.47 & B3IV & & & A \\
TYC 4031-1770-1 & 422533347 & 5.3948 & 3.8 & 2 & 01 28 21.97 & +60 14 43.9 & 10.94 & B5 & & & A \\  
HD 350990 & 424032547 & 5.0251 & 3.4 & 1 & 19 58 49.85 & +20 31 29.9 & 10.32 & B7II/III & & Roslund 3 & A  \\
TYC 1624-299-1 & 424032634 & 7.0305 & 1.4 & 1 & 19 58 41.30 & +20 31 48.6 & 12.04 & B7IVnnp & & Roslund 3 & A  \\
HD 254346 & 426520557 & 17.1232 & 4.2 & 4 & 06 16 57.32 & +22 11 42.0 & 9.74 & B2/3III & & & A, M \\
CPD-45 2977 & 430625174 & 5.7728 & 2.3 & 1 & 08 45 10.44 & $-$45 58 54.7 & 10.98 & O9.5II & & & \\ 
TYC 4269-482-1 & 434178307 & 5.9868 & 1.9 & 2 & 22 50 26.49 & +62 42 03.0 & 10.78 & B5/8V & & & A \\
HD 282433 & 446041643 & 4.4838 & 5.2 & 3 & 04 45 28.76 & +30 16 54.4 & 9.52 & B5 & & & A \\
\enddata
\tablecomments{Table of Candidate stars rejected in this work, including the ID, TESS Input Catalog (TIC) number, information about the primary frequency and number of modes, coordinates, V-band magnitude, and spectral type, a citation if the star is known to be a \Bcep star, a note if the star is a known cluster member, and information about remarks. Stars labeled ``M" in this column are discussed in the main paper text, stars labeled ``A" are commented upon in the Appendix.} 
\end{deluxetable*}


\acknowledgments*

J.L.-B. acknowledges support from FAPESP (grant 2017/23731-1). GH has been supported by the Polish NCN grant 2015/18/A/ST9/00578. LB acknowledges support from the NRF (South Africa) and thanks Mr Trevor Robinson for database assistance. DJS acknowledges support from the Pennsylvania State University's Eberly Research Fellowship. DJJ was supported through the Black Hole Initiative at Harvard University, through a grant (60477) from the John Templeton Foundation and by National Science Foundation award AST-1440254. This project makes use of data from the KELT survey, including support from The Ohio State University, Vanderbilt University, and Lehigh University, along with the KELT follow-up collaboration. This work has made use of data from the European Space Agency (ESA) mission {\it Gaia} (\url{https://www.cosmos.esa.int/gaia}), processed by the {\it Gaia} Data Processing and Analysis Consortium (DPAC, \url{https://www.cosmos.esa.int/web/gaia/dpac/consortium}). Funding for the DPAC has been provided by national institutions, in particular the institutions participating in the {\it Gaia} Multilateral Agreement. This research has made use of NASA's Astrophysics Data System. This research has made use of the SIMBAD database, operated at CDS, Strasbourg, France. This work has made use of the BeSS database, operated at LESIA, Observatoire de Meudon, France: http://basebe.obspm.fr.

\vspace{5mm}
\facilities{TESS, Gaia, KELT}

\software{astropy \citep{Astropy2013},  
          Period04 \citep{Lenz2005} 
          }

\clearpage 

\appendix

\section{Notes on individual stars}

\subsection{\Bcep stars}

{\it TIC 11698190 = HD 228690:} There is a slow drift in the light curve with a time scale of about 280 d that appears to be intrinsic to the star. There is also evidence for more pulsation frequencies than detected here.

{\it TIC 13967727 = HD 194205:} The frequency spectrum is complicated, suggesting the presence of more \Bcep pulsation frequencies, but also binarity or rotational variability, or SPB-type pulsations, or any combination thereof.

{\it TIC 25070410 = HD 166331:} This star is listed as a spectroscopic binary by \citet{Braganca2012}. There are equally pulsation frequencies; see main paper text.

{\it TIC 29123576 = CD-44 4876:} The strongest pulsation frequency is likely to be variable in amplitude.

{\it TIC 42940133 = HD 228101:} Possible \Bcep/SPB ``hybrid" pulsator and/or rotational modulation/binarity.

{\it TIC 44980675 = HD 171305:} Possible \Bcep/SPB ``hybrid" pulsator.

{\it TIC 55702566 = ALS 10035:} More frequencies are present, but the periodogram is too complicated to push the analysis further. The low frequency detected indicates a possible \Bcep/SPB ``hybrid" pulsator, or rotational modulation/binarity.

{\it TIC 61516388 = HD 178987:} This is likely a ``hybrid" pulsator.

{\it TIC 65166720 = KP Per:} The pulsation amplitudes for this star are probably underestimated because of its brightness. There are regular frequency spacings; see main paper text.

{\it TIC 75745359 = HD 80279:} Possible \Bcep/SPB ``hybrid" pulsator or rotational modulation/binarity.

{\it TIC 80897625 = ALS 8706:} The time scales of the temporal modulation of the variability amplitudes and phases of this Be star are different, which argues against a simple multifrequency beating phenomenon.

{\it TIC 93723398 = HD 74339:} There is a suspected $F6=5.222037$\,\cd, but unresolved from the strongest mode $F1$ in our data set (Table~\ref{tbl:bcep-freqtable}); \citet{Pigulski2008} detected this mode.

{\it TIC 128821888 = 12 Lac and TIC 129538133 = 16 Lac:} The pulsation amplitudes of these bright famous \Bcep stars are strongly suppressed in our data as simultaneous observations from other telescopes available to one of us (GH) show.

{\it TIC 159932751 = V836 Cen:} Perhaps the pulsation amplitudes we determined here are somewhat suppressed due to some saturation of this bright star. See Sect. \ref{sec:splittings} for a detailed discussion of the pulsation spectrum.

{\it TIC 168996597 = HD 225884:} This is probably a \Bcep/SPB ``hybrid" pulsator, the pulsation frequency $F4$ (Table~\ref{tbl:bcep-freqtable}) has variable amplitude.

{\it TIC 175760664 = HD 180642:} The pulsation amplitudes of this well-studied bright \Bcep star are probably somewhat suppressed in our data.

{\it TIC 182714198 = BD-02 4752:} This object shows binary induced or rotational modulation with a base period of 2.629\,d.

{\it TIC 184874234 = HD 173006:} This star shows the stillstand phenomenon \citep{Pigulski2008}. We detect more harmonics of the pulsation frequency than these authors, and also a very low frequency that appears to be real.

{\it TIC 197332002 = HD 172367:} This rapidly rotating star ($v \sin i = 240$\,\kms, \citealt{Daflon2007}) is possibly a \Bcep/SPB ``hybrid" pulsator.

{\it TIC 216976987 = HD 332408:} Possible shallow ($\approx 3$ mmag) eclipses with a 8.5862-d period.

{\it TIC 234230792 = HD 49330:} The CoRoT data of this Be star have been analysed in detail by \citet{Huat09}; an alias of the strongest pulsation frequency in the latter paper is the strongest signal in our data.

{\it TIC 245719692 = BD+57 614:} Possible \Bcep/SPB ``hybrid" pulsator or rotational modulation/binarity.

{\it TIC 255974332 = ALS 6426:} This heavily reddened star could be a \Bcep/SPB ``hybrid" pulsator or show the effects of rotational modulation/binarity.

{\it TIC 264613043 = V611 Per:} This star could be a binary or rotationally modulated with a short period of 1.3277\,d.

{\it TIC 264613619 = V757 Per:} This star was among the \Bcep candidates of \citet{Stankov2005} and is confirmed here. It also has regular frequency spacings; see main paper text.

{\it TIC 266338052 = HD 232874:} This star could be a binary or an ellipsoidal variable or undergo rotational modulation with a period of 3.0358\,d.

{\it TIC 279659875 = BD+68 1373:} This star shows some evidence for binary- or rotationally induced variability with a period of 3.2719\,d.

{\it TIC 287467101 = HD 140543:} This high Galactic latitude, high luminosity star (Sect. \ref{sec:high-lat}) may show ``hybrid" \Bcep/SPB pulsations. \citet{Martin2006} suggested that it could have formed in the Galactic halo and noted ``irregularly shaped lines" in his spectra of HD 140543, perhaps an effect of nonradial pulsation.

{\it TIC 287690192 = HD 339039:} There appear to be many more pulsation modes of \Bcep type present in this star, yet daily aliasing and possible amplitude variations precluded their reliable detection. Some g-mode pulsations of the SPB type may also be present.

{\it TIC 293680998 = IL Vel:} We find a fourth signal in addition to the known mode triplet of the star \citep{Handler2003}; the presence of more pulsation frequencies is also indicated.

{\it TIC 297259536 = HD 199021:} Possible \Bcep/SPB ``hybrid" pulsator or rotational modulation/binarity.

{\it TIC 299821534 = HD 231124:} Possible \Bcep/SPB ``hybrid" pulsator or rotational modulation/binarity. There are regular frequency spacings; see main paper text.

{\it TIC 308954763 = HD 232489:} Str\"omgren photometry \citep{Westin1982} places this possible \Bcep/SPB ``hybrid" pulsator or binary into the \Bcep domain in the HR Diagram, unlike its commonly quoted literature spectral type of B5 \citep{Skiff2014}. There are regular frequency spacings; see main paper text.

{\it TIC 311943795 = HD 228365:} The frequency spectrum of the star contains two signals between the common \Bcep and SPB domains, perhaps an effect of rapid rotation. Our frequency analysis was hampered by some aliasing problems.

{\it TIC 312630206 = HD 228463:} We were able to detect as many as twelve independent mode frequencies for this star (Table ~\ref{tbl:bcep-freqtable}), but the density of the frequency spectrum and some aliasing ambiguities hampered the identification of a possible rotational frequency splitting.

{\it TIC 338076483 = BD+64 1677:} Possible \Bcep/SPB ``hybrid" pulsator.

{\it TIC 347486043 = HD 13338:} There are regular frequency spacings; see main paper text. We also find a binary or rotationally induced variation with a period of 1.7404\,d.

{\it TIC 348443241 = BD+56 560:} Possible \Bcep/SPB ``hybrid" pulsator.

{\it TIC 348506748 = HD 14357:} Possible \Bcep/SPB ``hybrid" pulsator or rotational modulation/binarity. There are regular frequency spacings; see Sect. \ref{sec:splittings}.

{\it TIC 360063836 = HD 344775:} The signal $F3$ in Table~\ref{tbl:bcep-freqtable} is likely an artifact from amplitude/frequency variations of $F1$.

{\it TIC 361324132 = HD 344894:} The signal $F5$ in Table~\ref{tbl:bcep-freqtable} is likely an artifact from amplitude/frequency variations of $F1$. There are regular frequency spacings; see Sect. \ref{sec:splittings}.

{\it TIC 370128780 = BD+58 241:} Possible \Bcep/SPB ``hybrid" pulsator or rotational modulation/binarity.

{\it TIC 372724051 = V1143 Cas:} The signals listed for this object are the combination of the star's \Bcep pulsations and that of a nearby ellipsoidal variable, V1142 Cas (cf. \citealt{Handler2011}).

{\it TIC 391836737 = HD 30209:} There are probably more pulsation frequencies in the \Bcep domain, but apparent amplitude/frequency variations of the dominant signal hamper their reliable detection.

{\it TIC 393662110 = V372 Sge:} Again, there is evidence for several more pulsation frequencies in the \Bcep domain, but apparent amplitude/frequency variations of the strongest signals hamper their reliable detection. There are also several possibilities for repetitive frequency splittings (Sect. \ref{sec:splittings}).

{\it TIC 399436828 = NGC 663 4:} Whereas our algorithms originally identified this variable with the open cluster member NGC 663 2, a search in the surroundings revealed that in reality we recovered the pulsations of the known \Bcep star NGC 663 4. NGC 663 2 is thus rejected.

{\it TIC 419354107 = BW Vul:} The pulsation amplitudes of this famous ``stillstand" star are suppressed in our data due to its brightness.

{\it TIC 445619007 = HD 14645:} Possible binary or rotational variable with a period of 1.58293\,d.

{\it TIC 451932686 = HD 344880:} This is an eclipsing binary with an eccentric orbit and a long period; see main paper text for details.

{\it TIC 469221047 = HD 86214:} As for TIC 80897625, the amplitudes and phases of the two stellar oscillation modes are modulated in time. However, the present case is better described by a narrow frequency triplet ($\Delta f = 0.00327$\,\cd) and a frequency quadruplet ($\Delta f = 0.00492$\,\cd), respectively. We nevertheless prefer to list the two strongest signals only as we cannot rule out intrinsic amplitude/phase modulation.

{\it TIC 469223889 = HD 86162:} Possible \Bcep/SPB ``hybrid" pulsator.

\subsection{Candidate \Bcep stars}

{\it TIC 11158867 = TYC 4032-93-1:} Even though the available spectral classification for this star is B3, its dereddened $(B-V)$ color index and absolute magnitude rather point towards a \Dsct/\Gdor ``hybrid" pulsator. We keep this star as a doubtful candidate.

{\it TIC 12647534 = ALS 7011:} We detected signals in both the \Bcep and SPB domains. This star has a B-type neighbor separated by only 89" on the sky (TYC 4033-2268-1), and the frequency analysis of their data give consistent results. However, a blending analysis reveals that ALS 7011, and not TYC 4033-2268-1, is the source of this variability.

{\it TIC 13973539 = HD 229171:} The low frequency $F2$ we found for this Be star (cf. Table~\ref{tbl:cand-freqtable}) shows a slow amplitude and phase drift. However, as the time scales of the amplitude and phase change are different, we suspect a rotational modulation.

{\it TIC 28949811 = CD-44 4871:} The literature on this star suggests it is of early B type, but the single frequency we found is unusually high for a \Bcep star and its $S/N$ is low.

{\it TIC 29036690 = CD-46 4639:} This is an intermediate Helium star located in the \Bcep instability strip \citep{Groote1982}.

{\it TIC 29598925 = HD 76307:} The absolute magnitude from the star's Gaia DR2 parallax is lower than expected given its spectral class.

{\it TIC 30569481 = HD 76967:} The relative amplitudes and values of the combination frequencies are unusual for a \Bcep pulsator; the pulsation amplitudes may be variable as well. The Gaia DR2 parallax and Str\"omgren $H_{\beta}$ are consistent with the classification of a mid B-type star, leading to the (inconclusive) suspicion it may be a rapidly rotating SPB star.

{\it TIC 43301361 = HD 192003:} Taken at face value, the frequency spectrum and Gaia DR2 parallax point towards an unevolved \Bcep/SPB ``hybrid" pulsator. However, an F0 star in 43" distance may introduce some \Dsct/\Gdor-type variability due to blending.

{\it TIC 53709049 = TYC 4030-800-1:} The B1V star BD+60 175 is located 116" from this object and is some 2.5 mag brighter. Therefore we are not sure whether the reported variability indeed originates from TIC 53709049.

{\it TIC 53968977 = BD+60 192:} Our frequency analysis is based on the first two seasons of data only.

{\it TIC 74197071 = CD-45 4896:} The pulsation spectrum of this star is rather unusual for a \Bcep star and it is near (or even below) the low-luminosity end of the \Bcep instability strip, hence it may be a rapidly rotating SPB pulsator.

{\it TIC 78499882 = HD 249179:} This Be star is located in a high-mass X-ray binary system, and shows brightness variations consistent with disk variability that is typical of Be stars. It is not clear whether the high-frequency periodic variability detected by us is indeed caused by \Bcep pulsation.

{\it TIC 80814494 = HD 67980:} The Str\"omgren $H_\beta$ index for this star implies a spectral type around B3 \citep{Crawford1978} instead of B7II.

{\it TIC 94000461 = HD 74581:} This is a close visual double star, hence it is unclear from which source the variability originates.

{\it TIC 123828144 = HD 62894:} There is an A-type star of equal brightness 24" distant from this object, which could be a \Dsct pulsator.

{\it TIC 137489662 = HD 183535:} The Str\"omgren $H_\beta$ index for this star implies a spectral type around B1 \citep{Crawford1978} rather than B5.

{\it TIC 140309502 = CD-44 4484:} There is some weak evidence for shallow ($\approx 2$ mmag) eclipses with a 13.721(7)-d period, but more precise photometry is required to confirm or reject this suspicion.

{\it TIC 141903641 = CD-46 4437:} In the light curve for CD-46 4432 we identified signals consistent with a possible \Bcep/SPB ``hybrid" pulsator. However, the blending analysis shows this signal originating in the close visual double star CD-46 4437. Since the spectral type of the brighter of this pair is possibly consistent with \Bcep pulsation, we classify CD-46 4437 as a candidate.

{\it TIC 143530557 = HD 279639 and TIC 145672806 = CPD-52 1713:} Both stars are of later spectral type than usual \Bcep stars. We suspect they are rapidly rotating g-mode pulsators, but cannot prove this as of yet.

{\it TIC 190783321 = HD 76554:} All literature information consistently suggests this is an early B-type star, but its oscillation frequencies would be rather typical for a \Dsct star, and much higher than found among \Bcep pulsators.

{\it TIC 207045768 = HD 258853:} This star is probably not luminous enough to fall into the \Bcep strip, and each of the three variability frequencies is either the sum or the difference of the other two.

{\it TIC 231149746 = TYC 746-578-1:} The absolute magnitude of this star from its Gaia parallax is $2.6\pm0.1$. This suggests that either there is a problem with the original spectral classification \citep{Karlsson1972} or the parallax measurement.

{\it TIC 252864874 = TYC 3324-92-1:} The absolute magnitude of the star from Gaia DR2 and its pulsation frequencies suggest this is a \Dsct star rather than a \Bcep pulsator, and that perhaps there is a problem with the original spectral classification \citep{Heckmann1956}.

{\it TIC 290300414 = HD 78507:} The spectral classification and absolute magnitude of this star suggest it could be a rapidly rotating SPB star. Amplitude and frequency variations apparently occurred during the time span of our observations.

{\it TIC 307944768 = HD 260858:} This Helium-rich star's absolute magnitude from the Gaia DR2 parallax and effective temperature \citep{Netopil2008} suggest it has evolved off the main sequence. This is interesting because there is no post-main sequence \Bcep star known to date.

{\it TIC 319302209 = HD 220300:} The pulsation spectrum and basic parameters of this rapidly rotating Be star are more reminiscent of a rapidly rotating SPB star than of a \Bcep pulsator. It has been detected as variable by the HIPPARCOS mission (reported by \citealt{Kazarovets1999}), but only \citet{Labadie-Bartz2017} described its variability in detail.

{\it TIC 335484090 = V447 Cep:} Besides the known short-period variability of this star (it was also classified as a candidate \Bcep star by \citealt{Stankov2005}), we discovered that it is an eclipsing binary; see main paper text.

{\it TIC 347585038 = V352 Per:} This star was already listed as a \Bcep candidate by \citet{Stankov2005}.

{\it TIC 348609224 = BD+56 579:} This Be star has been identified as a visual double by Speckle Interferometry \citep{Hartkopf2009}, which may have had an effect on its Gaia parallax measurement.

{\it TIC 366108295 = ALS 6216:} This spectroscopic binary has a pulsation frequency quite high for a \Bcep star; a 1.5 mag brighter B0.5\,IV star is located 96" away and could have had an influence on the measured variability. Our blending analysis does not indicate a blend scenario, but this is so far not conclusive. 

{\it TIC 368237682 = BD+66 1651:} This Be and possible runaway star \citep{Tetzlaff2010} shows variability in both the \Bcep and SPB star frequency domains. Its luminosity computed from the Gaia DR2 parallax is much lower than expected for its B3Ve spectral class.

{\it TIC 370269139 = BD+55 334 and TIC 372115570 = BD+57 579:} These stars have several visual companions close enough to qualify as potential sources for the observed variability.

{\it TIC 384257658 = HD 78206:} Whereas the literature spectral classification is B7/8V, the star's absolute magnitude from the Gaia parallax corresponds to that of a B2V star.

{\it TIC 386693012 = HD 42896:} This rapidly rotating ($v \sin i = 318$\,\kms, \citealt{Huang2010}) luminous star shows quite high variability frequencies for a \Bcep star and several combinations thereof.

{\it TIC 387153140 = BD-09 4742:} The variability frequencies of this object are fairly high for a \Bcep star and each of these three frequencies is either the sum or the difference of the other two.

{\it TIC 419246605 = BD+61 77:} Whereas this star is repeatedly classified with an early B spectral type in the literature, its absolute magnitude from the Gaia DR2 parallax is only +2.8. We note that the error on this parallax is unusually high and that the star has a visual companion one magnitude fainter 0.3" apart, which may have affected the parallax measurement adversely.

{\it TIC 422533344 = BD+59 254:} In the light curve for TYC 4031-1770-1 we identified signals consistent with \Bcep pulsation. However, the blending analysis shows this signal originating in the neighboring star BD+59 254. The literature spectral type of BD+59 254 is A2, which is more consistent with \Dsct pulsation. The frequencies we detect are more consistent with \Bcep pulsation though, so it is unclear what the nature of this object is. 

{\it TIC 427396133 = HD 37115:} Because of some saturation issues (cf. \citealt{Labadie-Bartz2017}) with this data set and the removal of an apparent outburst of this Be star the results of our frequency analysis should be treated with caution.

{\it TIC 430625041 = CD-45 4501:} We identified signals consistent with \Bcep pulsation in the light curve of CPD-45 2977. Our blending analysis shows this signal instead comes from CD-45 4501, which we estimate to have a spectral type consistent with \Bcep pulsations. 

{\it TIC 447933173 = HD 77769:} Even though this star has three variability frequencies in the \Bcep domain and has been classified as such a pulsator in the past \citep{Pigulski2008}, it is cooler and less luminous than the pulsators of this class.

{\it TIC 458911894 = HD 19635:} Based on the available data it cannot be decided whether the detected triplet of frequencies is due to beating of multiple signals or due to amplitude/phase modulation of a single independent variation. In case of the interpretation as a triplet it should be noted that its frequency asymmetry is of the opposite sign than expected from the second-order effect of rotation \cite[e.g.,][]{Dziembowski1992}.

{\it TIC 459014234 = BD+60 770:} This is a visual double star with 8" separation with the primary classified with a B5 spectral type. The pulsation spectrum is reminiscent of that of a \Dsct star, and the Gaia DR2 parallax appears to corroborate that. However, given the visual binarity we are not sure how accurate the parallax measurement is and keep the star in our list of candidates.

\subsection{Rejected stars}

{\it TIC 12647620 = TYC 4033-2268-1:} The same signal exists in the light curves for both ALS 7011 and TYC 4033-2268-1, separated by 89''. Our blending analysis shows this signal as coming solely from ALS 7011, which we classify as a candidate. 

{\it TIC 26283875 = HD 221991/TYC 4000-2127-1:} This is a visual double star with 18" separation. The Gaia DR2 parallaxes of both objects show that they are A-type stars at very similar distance. We therefore conclude that at least one of them is a \Dsct pulsator, responsible for the detected variability, and that the B5 spectral classification is erroneous.

{\it TIC 28709025 = HD 75290:} \citet{Levato1975} give a spectral type of B9V, with is consistent with the measured Str\"omgren $H_\beta$ index of 2.765 that suggests B8/9 \citep{Crawford1978}, and with the star's absolute magnitude $M_v=-0.3$ from its Gaia DR2 parallax. We conclude that this cannot be a \Bcep star.

{\it TIC 40102236 = HD 190088:} As for the previous star, the Str\"omgren $H_\beta$ index (2.748) and its absolute magnitude $M_v=-0.1$ from its Gaia DR2 parallax suggest this is a late B-type and not a \Bcep star.

{\it TIC 60320306 = HD 59446:} Again, Str\"omgren photometry and the Gaia DR2 parallax of this star suggest a mid to late B spectral type. In addition, we are unsure that the detected variability is not an artifact of saturation in the images of this bright star.

{\it TIC 67985749 = HD 144941:} This is an extreme Helium star and therefore not of the \Bcep type; see Sect. \ref{sec:extreme-He}.

{\it TIC 90134626 = HD 72539:} Also in this case, Str\"omgren photometry and the Gaia DR2 parallax of this star consistently point towards a mid to late B spectral type. The variability is therefore likely due to SPB-type pulsation, with frequencies pushed towards the \Bcep domain by rapid rotation.

{\it TIC 93549165 = o Vel:} This is a very bright prototype SPB star \citep{DeCat2002A}. The variability detected in our light curves is certainly not due to the star, but an instrumental artifact.

{\it TIC 102161004 = HD 331621:} This is another case of a star that shows variability frequencies close to the \Bcep range, but its absolute magnitude is far below that of the known pulsators.

{\it TIC 118680798 = HD 29450:} Likewise, this star is too cool and not luminous enough to be even close to the \Bcep domain in the HR Diagram, as literature Str\"omgren photometry and the Gaia DR2 parallax demonstrate.

{\it TIC 123754451 = HD 62755:} Also in this case, literature Str\"omgren photometry and the Gaia DR2 parallax strongly argue against an interpretation of this star's variability as being caused by \Bcep pulsations due to its low effective temperature and luminosity.

{\it TIC 138905907 = HD 290564:} The absolute magnitude of this star and UBV photometry \citep{Guetter1979} suggest that this is an late A/early F star and hence the variability we detected is of the \Dsct type. Therefore our pre-selection of this star was based on an incorrect spectral classification.

{\it TIC 141903541 = CD-46 4432:} Initially identified as a possible \Bcep/SPB ``hybrid" pulsator. However, the blending analysis reveals that the signal originates in a source to the South-West, at the location of the close visual double CD-46 4437 (although we cannot resolve this visual double). Since it is possible that the brighter component is of the correct spectral type, we include CD-46 4437 as a candidate \Bcep star. 

{\it TIC 157535787 = HD 151654:} There are two vastly different spectral classifications for this star in the literature, B0.5\,V \citep{Garrison1977} and F0 \citep{Houk1999}. Str\"omgren photometry and the Gaia DR2 parallax corroborate the latter. Hence this is a \Dsct pulsator and not a \Bcep star.

{\it TIC 162012064 = HD 181124:} This case is similar to the former: \citet{Kelly1986} classified this star as B5, but \citet{Houk1999} as A3\,IV. The absolute magnitude of the star derived from Gaia DR2, its $(B-V)_0$ colour index and the high pulsation frequencies are clearly that of a \Dsct and not a \Bcep star.

{\it TIC 264485563 = HD 35612:} This instance is similar to HD 62755. Again, the effective temperature from Str\"omgren photometry and the absolute magnitude from the Gaia DR2 parallax are incompatible with a \Bcep classification. The weak variability signal is also not convincing.

{\it TIC 281803267 = TYC 4804-1086-1:} The signal in our light curve for this object is from a relatively bright W UMa type binary 3' to the South-East, V453 Mon. 

{\it TIC 291556636 = HD 59325:} Geneva photometry of the star \citep{Mermilliod1997} and its calibration \citep{Kunzli1997} suggest that this is a mid to late B star on the main sequence. The Gaia DR2 parallax implies an absolute magnitude corresponding to a post main sequence star, but even so the star would not fall in any instability domain related to \Bcep stars \citep{Daszynska2013}. We conclude that this is most likely an SPB star with observed frequencies modified towards the \Bcep domain by rapid rotation.

{\it TIC 352529679 = HD 190066:} The variability detected in the measurements of this star could be traced to the data quality being vastly different depending on whether the telescope was pointed East or West. Removing the poor observations, no variability remains.

{\it TIC 360661624 = ALS 10464:} Initially identified as a \Bcep pulsator, but our blending analysis shows the signal coming from the star LS II +23 36, which is $\sim$1 mag brighter and lies 85'' to the North-East. LS II +23 36 is included here as a candidate, as it may have a spectral type consistent with \Bcep pulsations. 

{\it TIC 377498419 = V652 Her:} This is another extreme Helium star that fulfilled our pre-selection criteria. The frequency solution in Table~\ref{tbl:rejected-freqtable} is only informal due to the large rate of change of the pulsation period; see Sect. \ref{sec:extreme-He}.

{\it TIC 384992041 = TYC 3315-1807-1:} A literature search readily shows that this is a close binary composed of a subdwarf star and a late-type main sequence companion \citep{Kawka2010}.

{\it TIC 399436782 = NGC 663 2:} Whereas our algorithms identified this variable with the open cluster member NGC 663 2, a search in the surroundings revealed that in reality we recovered the pulsations of the known \Bcep star NGC 663 4 which is here categorized as a \Bcep star.

{\it TIC 417438983 = HD 196035:} The variability detected for this bright star is an integer multiple of the sidereal daily alias, suggesting that there are saturation effects in the photometry. The effective temperature of the star from Str\"omgren and Geneva photometry and the absolute magnitude from the Gaia DR2 parallax are both too low to place the star at least close to the known \Bcep domain in the HR diagram.

{\it TIC 422533347 = TYC 4031-1770-1:} Initially identified as a \Bcep pulsator, but our blending analysis shows the signal coming from the neighboring star BD+59 254, which lies 77'' to the South-West. BD+59 254 is included as a candidate.

{\it TIC 424032547 = HD 350990 and TIC 424032634 = TYC 1624-299-1:} These two stars are separated by only 122" on the sky; they are located in a dense region in the open cluster Roslund 3 that contains several stars of similar brightness, leading to blending and saturation issues. The variability detected for both of these stars occurs at frequencies that are integer multiple of the sidereal daily alias. These stars are also of too late spectral type and too low absolute magnitude to fall into the \Bcep domain.

{\it TIC 426520557 = HD 254346:} The variability detected in this data set contains both pulsations and eclipses. Whereas the amplitude of the eclipses does not significantly change depending on whether the telescope is pointed East or West, the amplitudes of the pulsations do. Furthermore, the pulsating star HD 43385 that is more than one magnitude brighter than HD 254346 is located in 112" distance. Str\"omgren photometry places HD 43385 into the \Dsct instability strip. Hence we interpret the observed variability by \Dsct pulsations of HD 43385 in combination with eclipses of HD 254346, which itself shows no detectable pulsational variability; the eclipses are discussed in the main paper text.

{\it TIC 430625174 = CPD-45 2977:} The signal in our light curve comes from the $V=9$ neighbor 2' to the North-East, CD-45 4501, which is $\sim$2 mag brighter than CPD-45 2977 in the V-band. CD-45 4501 is categorized as a candidate. 

{\it TIC 434178307 = TYC 4269-482-1:} Str\"omgren photometry and the Gaia DR2 distance point towards a very late B or even an early A star. The detected variability signals have poor S/N. Thus there is no evidence that this could be a \Bcep pulsator.

{\it TIC 446041643 = HD 282433:} The effective temperature from Str\"omgren photometry and the absolute magnitude from the Gaia DR2 parallax are incompatible with a \Bcep classification. We allege that this is another case of a rapidly rotating SPB pulsator.

\clearpage 

\section{Frequency tables}

\twocolumngrid
All recovered frequencies and their amplitudes are reported in Tables 7, 8, and 9.

\subsection{\Bcep stars}

\startlongtable 
\begin{deluxetable}{lcr}  
\tablecaption{Frequencies and amplitudes determined for the \Bcep stars. Formal error estimates determined according to \citet{Montgomery1999} are given in braces in units of the last significant digit. \dag Amplitudes may be largely suppressed.}
\label{tbl:bcep-freqtable}
\tablenum{7}
\tablehead{ 
\colhead{ID} & \colhead{Freq.}    & \colhead{Amp.}\\
\colhead{ } & \colhead{(d$^{-1}$)} & \colhead{(mmag)}
}
\startdata
\hline \multicolumn{3}{c}{TIC 5076425} \\
\hline
F1&	5.63875(1)&37(1) \\
F2&	5.88842(1)&33(1) \\
F3&	5.81121(2)&	20(1) \\
F4&	5.45326(4)&10(1) \\
\hline 
\multicolumn{3}{c}{TIC 10891640} \\
\hline
F1&	6.799702(4)&13.8(2)\\
F2&	5.960461(6)&8.4(2) \\
F3&	0.162244(3)&4.5(2) \\
F4&	6.8001(2)&2.8(2) \\
\hline 
\multicolumn{3}{c}{TIC 11696250} \\
\hline
F1&	7.59638(1)&2.7(1) \\
F2&	6.81487(1)&2.5(1)\\
F3&	3.96790(2)&1.8(1) \\
F4&	5.28827(2)&1.7(1)\\
F5&	6.53336(3)&1.0(1)\\
F6&	5.55230(4)&0.9(1) \\
\hline 
\multicolumn{3}{c}{TIC 11698190} \\
\hline
F1&	5.872054(5)&9.7(2) \\
F2&	5.552150(6)&9.0(2) \\
F3&	5.56928(1)&3.5(2) \\
F4&	5.80301(3)&2.0(2) \\
\hline 
\multicolumn{3}{c}{TIC 13332837} \\
\hline
F1&	6.874397(7)&5.5(2) \\
F2&	6.48827(3)&1.5(2) \\
F3&	8.63307(3)&1.3(2) \\
\hline 
\multicolumn{3}{c}{TIC 13967727} \\
\hline
F1&	4.445459(3)&12.1(2) \\
F2&	4.378370(3)&11.9(2)\\
F3=F1+F2&	8.823830(4)&1.9(2)\\
F4&	0.18282(1)&3.5(2)\\
F5&	0.36563(2)&1.9(2)\\
F6&	0.14505(3)&1.3(2) \\
\hline 
\multicolumn{3}{c}{TIC 25070410} \\
\hline
F1&	7.447568(6)&12.2(2)\\
F2&	6.95940(1)&6.0(2) \\
F3&	6.95362(2)&3.8(2)\\
F4&	7.23907(2)&3.2(2) \\
F5&	6.63922(3)&2.6(2) \\
F6&	6.93872(3)&2.4(2) \\
\hline 
\multicolumn{3}{c}{TIC 29123576} \\
\hline
F1&	7.782614(8)&17.0(4)\\
F2&	7.32397(3)&i4.9(4) \\
\hline 
\multicolumn{3}{c}{TIC 29585482} \\
\hline
F1&	8.07168(2)&5.5(3)\\
\hline 
\multicolumn{3}{c}{TIC 31921921} \\
\hline
F1&	4.637292(4)&14.9(2)\\
F2=2F1&	9.274585(9)&4.8(2)\\
F3=3F1&	13.91188(1)& 1.2(2) \\
\hline 
\multicolumn{3}{c}{TIC 41327931} \\
\hline
F1&	8.97107(3)&1.2(2) \\
F2&	8.13509(3)&1.2(2)\\
\hline 
\multicolumn{3}{c}{TIC 42365645} \\
\hline
F1&	7.830516(2)&19.1(2)\\
F2&	7.794148(5)&8.0(2) \\
F3&	7.908111(9)&4.2(2) \\
F4&	7.94628(1)&4.1(2)\\
F5&	1.327791(6)&5.4(2)\\
F6&	7.831869(9)&4.6(2) \\
F7=2F5&	2.65558(1)&2.7(2)\\
F8&	7.79322(2)&2.4(2)\\
F9&	7.60063(2)&1.9(2)\\
F10=3F5&	3.98337(2)&1.8(2)\\
F11=4F5&	5.31117(2)&1.1(2) \\
F12&	7.70240(3)&1.4(2) \\
F13&	7.83856(3)&1.3(2)\\
\hline 
\multicolumn{3}{c}{TIC 42940133} \\
\hline
F1&	6.299788(5)&4.32(9) \\
F2&	6.921975(6)&3.99(9)\\
F3&	7.387334(6)&3.79(9)\\
F4&	7.164733(8)&2.66(9)\\
F5&	0.36111(2)&1.39(9)\\
F6&	6.72631(2)&1.19(9) \\
F7&	6.78095(2)&1.17(9) \\
F8&	0.32302(2)&0.93(9) \\
F9=2F10&	0.32628(2)&0.79(9) \\
F10&	0.16315(2)&0.70(9) \\
\hline 
\multicolumn{3}{c}{TIC 43881045} \\
\hline
F1&	7.122573(3)&7.5(1)\\
\hline 
\multicolumn{3}{c}{TIC 44980675} \\
\hline
F1&	5.158542(6)&8.4(1) \\
F2&	0.50941(2)&2.8(1) \\
F3&	0.32240(3)&1.6(1) \\
F4&	4.55869(3)&1.4(1) \\
\hline 
\multicolumn{3}{c}{TIC 45799839} \\
\hline
F1&	7.575293(7)&3.6(1)\\
F2&	9.29511(1)&1.8(1) \\
\hline 
\multicolumn{3}{c}{TIC 55702566} \\
\hline
F1&	6.54900(1)&6.5(3) \\
F2&	6.88965(2)&5.0(3)\\
F3&	0.65842(3)&2.8(3) \\
F4&	12.96233(4)&2.1(3) \\
F5&	7.41737(4)&2.0(3) \\
\hline 
\multicolumn{3}{c}{TIC 61516388} \\
\hline
F1&	7.14659(1)&7.5(2) \\
F2&	0.25205(2)&4.1(2) \\
F3&	0.28962(2)&3.3(2) \\
F4&	7.08337(3)&2.2(2) \\
\hline 
\multicolumn{3}{c}{TIC 65166720} \\
\hline
F1&	4.95583(1)&16.0(3)\\
F2&	5.04840(2)&10.8(3)\\
F3&	4.40346(4)&6.0(3) \\
F4&	4.50767(7)&3.4(3) \\
F5&	5.06972(8)&2.9(3) \\
\hline 
\multicolumn{3}{c}{TIC 75745359} \\
\hline
F1&	7.25905(1)&3.1(1)\\
F2&	6.97609(6)&0.8(1) \\
F3&	0.39610(6)&0.7(1) \\
\hline 
\multicolumn{3}{c}{TIC 80897625} \\
\hline
F1&	4.40603(1)&4.7(3) \\
F2&	4.33994(1)&4.4(3) \\
\hline 
\multicolumn{3}{c}{TIC 89756665} \\
\hline
F1&	8.41789(1)&3.7(2) \\
F2&	0.00819(2)&3.3(2) \\
F3&	9.85740(3)&1.8(2) \\
\hline 
\multicolumn{3}{c}{TIC 93723398} \\
\hline
F1&	5.220117(3)&29.2(3)\\
F2&	5.195248(6)&16.6(3)\\
F3&	5.23896(1)&9.5(3) \\
F4&	5.255634(9)&9.9(3)\\
F5&	5.20529(1)&8.0(3) \\
\hline 
\multicolumn{3}{c}{TIC 93730538} \\
\hline
F1&	5.445082(5)&6.8(1) \\
F2&	6.286168(7)&5.4(1) \\
F3&	5.74156(1)&3.6(1) \\
F4&	6.00031(2)&2.0(1)\\
F5&	0.29368(2)&2.1(1) \\
F6&	0.28095(2)&2.3(1)\\
F7&	0.25851(3)&1.3(1) \\
F8&	5.60304(4)&0.9(1) \\
\multicolumn{3}{c}{TIC 101423289} \\
\hline
F1&	7.48973(1)&6.2(2) \\
F2&	7.37917(1)&6.4(2) \\
F3&	7.25335(2)&4.9(2)\\
F4&	8.35985(4)&1.7(2) \\
F5&	6.96763(5)&1.4(2) \\
F6&	6.81801(6)&1.3(2) \\
\hline 
\multicolumn{3}{c}{TIC 105517114} \\
\hline
F1&	4.457141(2)&22.0(1)\\
F2&	4.666456(3)&11.6(1)\\
F3&	4.326864(3)&11.3(1) \\
F4&	4.190111(4)&7.6(1) \\
F5&	4.534389(6)&5.6(1) \\
F6&	4.809343(6)&5.3(1)\\
F7=F1+F5&	8.991530(6)&1.8(1)\\
F8=2F1&	8.914282(3)&1.3(1)\\
F9=F1+F2&	9.123597(3)&0.9(1)\\
F10=F2+F4&	8.856567(5)&0.8(1)\\
F11=F2+F3&	8.993320(4)&0.9(1) \\
F12=F4+F6&	8.999454(8)&0.9(1) \\
F13=F1+F3&	8.784005(3)&0.8(1) \\
F14=F2+F5&	9.200845(7)&0.4(1) \\
F15=2F3&	8.653729(4)&0.5(1) \\
\hline 
\multicolumn{3}{c}{TIC 123211175} \\
\hline
F1&	7.18926(4)&0.7(1) \\
\hline 
\multicolumn{3}{c}{TIC 128821888} \\
\hline
F1&	5.178946(6)&7.5(2) \\
F2&	5.06628(1)&3.4(2) \\
F3&	5.49008(2)&2.4(2) \\
F4&	5.33420(2)&2.0(2) \\
F5=F1+F4&	10.513145(6)&1.4(2) \\
\hline 
\multicolumn{3}{c}{TIC 129538133} \\
\hline
F1&	5.85028(2)&4.4(2) \\
F2&	5.50260(2)&3.6(2) \\
F3&	5.91130(3)&2.8(2) \\
\hline 
\multicolumn{3}{c}{TIC 141292310b} \\
\hline
F1&	4.54498(3)&4.7(1) \\
F2&	0.92020(6)&2.0(1) \\
F3&	4.2052(1)&1.3(1) \\
F4&	4.5765(1)&1.3(1) \\
F5&	1.35078(9)&1.3(1) \\
F6&	4.4974(1)&0.9(1) \\
F7&	1.1585(1)&0.9(1) \\
\hline 
\multicolumn{3}{c}{TIC 159932751} \\
\hline
F1&	6.966153(5)&8.5(1) \\
F2&	6.461838(5)&8.0(1) \\
F3&	6.449531(5)&8.2(1)\\
F4&	6.590981(9)&4.5(1) \\
F5&	6.97869(1)&3.2(1) \\
F6&	6.99092(2)&2.4(1)\\
F7&	6.43716(2)&2.4(1)\\
F8&	6.42472(4)&1.0(1) \\
F9&	0.68311(4)&0.9(1) \\
\hline 
\multicolumn{3}{c}{TIC 168996597} \\
\hline
F1&	5.82201(1)&2.9(1)\\
F2&	5.72972(1)&2.8(1)\\
F3&	0.39038(2)&1.6(1) \\
F4&	8.22823(3)&1.4(1) \\
F5&	0.36931(3)&1.4(1) \\
\hline 
\multicolumn{3}{c}{TIC 175760664} \\
\hline
F1&	5.486894(4)&28.4(3)\\
F2=2F1&	10.973787(7)&4.6(3)\\
F3=3F1&	16.46068(1)&3.1(3)\\
F4=4F1&	21.94757(1)&1.6(3) \\
\hline 
\multicolumn{3}{c}{TIC 180952390} \\
\hline
F1&	6.627465(9)&6.9(2) \\
F2&	6.23117(2)&3.0(2) \\
F3=F1+F2&12.85864(2)&1.3(2)\\
F4=2F1&	13.25493(2)&0.8(2) \\
\hline 
\multicolumn{3}{c}{TIC 182714198} \\
\hline
F1&	6.97786(2)&7.0(4) \\
F2=2F3&	0.76075(2)&6.9(4)\\
F3&	0.38037(1)&5.0(4)\\
F4&	6.41787(3)&3.7(4)\\
\hline 
\multicolumn{3}{c}{TIC 184874234} \\
\hline
F1&	5.877919(4)&36.1(4)\\
F2=2f1&	11.755838(7)&9.6(4)\\
F3=3f1&	17.63376(1)&5.4(4) \\
F4=4f1&	23.51168(1)&2.9(4)\\
F5=5f1&	29.38960(2)&2.1(4) \\
F6=6f1&	35.26752(2)&1.8(4) \\
F7&	0.00567(1)&9.7(4)\\
\hline 
\multicolumn{3}{c}{TIC 185476952} \\
\hline
F1&	8.29493(2)&2.1(2) \\
F2&	8.46231(3)&1.8(2) \\
F3&	9.02509(3)&1.5(2) \\
F4&	8.44342(5)&0.9(2) \\
F5&	8.12972(5)&0.9(2) \\
\hline 
\multicolumn{3}{c}{TIC 187483462} \\
\hline
F1&	6.73549(1)&0.90(7) \\
F2&	0.47329(2)&0.52(7) \\
\hline 
\multicolumn{3}{c}{TIC 197332002} \\
\hline
F1&	4.688242(9)&8.8(3) \\
F2&	1.89627(1)&8.1(3)\\
F3&	0.14873(2)&5.1(3) \\
F4&	5.89429(2)&4.2(3) \\
\hline 
\multicolumn{3}{c}{TIC 216976987} \\
\hline
F1&	6.777856(8)&2.65(9)\\
F2&	0.23294(1)&1.59(9) \\
F3&	1.76404(2)&0.93(9) \\
F4=2F2&	0.46587(3)&0.48(9) \\
\multicolumn{3}{c}{TIC 225687900} \\
\hline
F1&	5.443056(5)&24.2(4) \\
F2&	5.357764(9)&14.1(4) \\
\hline 
\multicolumn{3}{c}{TIC 232846315} \\
\hline
F1&	7.59044(1)&3.0(2) \\
F2=2f4&	1.47419(1)&3.0(2)\\
F3=4f4&	2.94838(2)&2.2(2) \\
F4&	0.73709(1)&0.7(2) \\
\hline 
\multicolumn{3}{c}{TIC 234068267} \\
\hline
F1&	5.598023(6)&9.6(2) \\
F2&	5.53356(1)&4.5(2)\\
F3&	5.48953(4)&1.5(2) \\
\hline 
\multicolumn{3}{c}{TIC 234230792} \\
\hline
F1&	10.85911(4)&1.7(2) \\
F2&	6.94930(5)&1.2(2) \\
\hline 
\multicolumn{3}{c}{TIC 245719692} \\
\hline
F1&	6.92500(5)&2.4(2) \\
F2&	0.45300(6)&1.9(2) \\
F3&	4.7635(1)&1.3(2) \\
\hline 
\multicolumn{3}{c}{TIC 251196433} \\
\hline
F1&	6.37298(2)&6.5(2) \\
F2&	4.12598(6)&2.0(2) \\
F3&	5.66421(8)&1.5(2) \\
\hline 
\multicolumn{3}{c}{TIC 251250184} \\
\hline
F1&	4.120389(9)&17.7(3)\\
F2&	4.03062(1)&14.2(3)\\
F3&	4.18986(2)&9.8(3) \\
F4&	4.27198(3)&5.9(3) \\
F5=2F3&	8.37973(3)&2.0(3)\\
F6=F2+F3&	8.22044(2)&1.2(3)\\
F7&	6.3903(1)&1.5(3)\\
F8=F1+F3&	8.31025(2)&1.1(3) \\
\hline 
\multicolumn{3}{c}{TIC 251250634} \\
\hline
F1&	12.20049(7)&1.9(2)\\
F2&	9.04715(9)&1.4(2) \\
\hline 
\multicolumn{3}{c}{TIC 251637508} \\
\hline
F1&	5.499909(4)&16.7(2)\\
F2&	5.461319(6)&12.7(2)\\
F3&	5.53836(2)&3.5(2) \\
F4&	5.38249(3)&2.5(2) \\
F5&	5.76330(4)&1.8(2)\\
F6=F1+F4&	10.88240(3)&1.3(2)\\
\hline 
\multicolumn{3}{c}{TIC 255974332} \\
\hline
F1&	4.82425(2)&11.0(4) \\
F2&	4.91960(3)&7.6(4) \\
F3&	0.18822(6)&3.6(4) \\
F4&	4.61661(9)&2.4(4) \\
F5&	4.3714(1)&2.2(4) \\
F6&	5.26261(1)&2.2(4) \\
F7=F1+F2&	9.74384(3)&1.5(4)\\
\hline 
\multicolumn{3}{c}{TIC 264612228} \\
\hline
F1&	6.27361(2)&5.8(2) \\
F2&	6.27212(5)&2.3(2) \\
F3&	5.78476(7)&1.7(2) \\
F4&	5.90656(7)&1.8(2) \\
\hline 
\multicolumn{3}{c}{TIC 264613043} \\
\hline
F1&	5.82463(2)&3.6(2) \\
F2&	1.50635(5)&1.9(2)\\
\hline 
\multicolumn{3}{c}{TIC 264613619} \\
\hline
F1&	4.07690(1)&8.3(2)\\
F2&	4.19978(2)&3.6(2) \\
F3&	4.04781(2)&3.8(2) \\
F4&	4.22886(3)&2.9(2)\\
F5&	4.21448(4)&2.1(2) \\
F6&	4.14531(7)&1.2(2)\\
F7&	4.03326(6)&1.4(2)\\
F8=F1+F2&	8.27668(3)&0.7(2) \\
\hline 
\multicolumn{3}{c}{TIC 264615882} \\
\hline
F1&	4.89703(1)&6.6(2) \\
F2&	4.74677(1)&5.6(2) \\
F3&	4.38948(2)&4.2(2) \\
F4&	4.62217(4)&2.0(2) \\
F5&	4.30801(4)&1.9(2) \\
F6&	4.20802(8)&1.1(2) \\
\hline 
\multicolumn{3}{c}{TIC 264731122} \\
\hline
F1&	4.127522(6)&20.0(2)\\
F2&	3.98217(2)&8.3(2) \\
F3&	4.00882(2)&7.2(2) \\
F4&	4.06179(2)&5.5(2) \\
F5&	3.95513(3)&4.6(2) \\
\hline 
\multicolumn{3}{c}{TIC 266338052} \\
\hline
F1&	5.74001(1)&5.5(1) \\
F2&	0.65881(3)&2.8(1) \\
F3&	6.91081(6)&1.1(1)\\
F4&	0.32941(2)&0.7(1)\\
F5&	6.92586(8)&0.8(1) \\
F6&	0.98822(4)&0.8(1)\\
\hline 
\multicolumn{3}{c}{TIC 269228628} \\
\hline
F1&	4.99191(6)&46(4)\\
F2&	4.83500(7)&38(4)\\
F3&	4.25270(9)&29(4)\\
\hline 
\multicolumn{3}{c}{TIC 272922818} \\
\hline
F1&	8.02788(4)&36(3)\\
\hline 
\multicolumn{3}{c}{TIC 279236955} \\
\hline
F1&	8.054088(2)&18.5(2)\\
F2&	7.731494(7)&5.9(2) \\
F3&	7.83966(1)&2.9(2) \\
F4&	8.23004(2)&2.4(2) \\
F5&	0.32259(3)&1.2(2)\\
F6&	15.78558(4)&1.0(2) \\
\hline 
\hline 
\multicolumn{3}{c}{TIC 279659875} \\
\hline
F1&	5.83009(2)&3.8(1)\\
F2&	0.61126(8)&0.9(1) \\
F3&	6.7508(1)&0.7(1) \\
\hline 
\multicolumn{3}{c}{TIC 282529703} \\
\hline
F1&	7.500674(5)&11.2(2)\\
F2&	7.67644(2)&3.7(2) \\
\hline 
\multicolumn{3}{c}{TIC 283136444} \\
\hline
F1&	7.95176(5)&14(1)\\
\hline 
\multicolumn{3}{c}{TIC 287467101} \\
\hline
F1&	5.60859(2)&2.0(1)\\
F2&	1.06253(4)&1.2(1) \\
\hline 
\multicolumn{3}{c}{TIC 287690192} \\
\hline
F1&	8.269534(4)&9.6(1) \\
F2&	7.662407(4)&8.6(1) \\
F3&	7.824390(4)&8.4(1) \\
F4&	0.63773(2)&2.0(1) \\
F5&	8.78432(2)&1.9(1) \\
F6&	1.12189(2)&1.4(1) \\
F7&	8.75905(3)&1.1(1) \\
\hline 
\multicolumn{3}{c}{TIC 293680998} \\
\hline
F1&	5.459807(3)&28.9(3) \\
F2&	5.363247(4)&24.4(3)\\
F3&	7.22067(3)&3.4(3)\\
F4&	5.41455(4)&2.5(3) \\
\hline 
\multicolumn{3}{c}{TIC 295435513} \\
\hline
F1&	7.70882(1)&10.8(2)\\
F2&	7.06709(5)&3.2(2) \\
\hline 
\multicolumn{3}{c}{TIC 296570221} \\
\hline
F1&	6.090863(8)&5.4(1) \\
F2&	6.04198(1)&3.7(1) \\
F3&	6.06932(3)&1.5(1)\\
F4&	5.50062(4)&1.2(1) \\
F5&	7.88333(4)&1.0(1) \\
\hline 
\multicolumn{3}{c}{TIC 297259536} \\
\hline
F1&	11.38469(1)&2.11(8) \\
F2&	9.94812(2)&1.02(8) \\
F3&	0.61498(3)&0.59(8) \\
F4&	9.44019(4)&0.47(8) \\
\hline 
\multicolumn{3}{c}{TIC 299821534} \\
\hline
F1&	4.543190(4)&30.7(4)\\
F2=2F1&	9.086379(9)&3.3(4)\\
F3=3F1&	13.62957(1)&2.2(4) \\
F4&	5.17349(4)&3.3(4)\\
F5&	5.62474(5)&2.8(4) \\
F6&	0.35691(5)&2.5(4) \\
F7&	6.05917(6)&2.4(4) \\
\hline 
\multicolumn{3}{c}{TIC 308954763} \\
\hline
F1&	5.37689(1)&7.6(2) \\
F2&	5.14309(2)&4.8(2) \\
F3&	5.12346(3)&3.7(2) \\
F4&	5.25366(5)&1.9(2) \\
F5&	5.15901(5)&1.8(2) \\
F6&	0.10893(8)&1.2(2) \\
F7&	5.06220(8)&1.2(2) \\
F8&	4.9591(1)&1.0(2) \\
F9&	1.14400(9)&1.2(2) \\
\hline 
\multicolumn{3}{c}{TIC 311943795} \\
\hline
F1&	8.28858(1)&4.8(3) \\
F2&	2.42342(2)&3.0(3)\\
F3&	7.20570(2)&2.7(3)\\
F4&	3.46824(3)&1.9(3) \\
F5&	7.09267(3)&1.8(3) \\
F6&	8.09053(4)&1.7(3) \\
\hline 
\multicolumn{3}{c}{TIC 312626970} \\
\hline
F1&	4.791218(4)&8.7(1) \\
F2&	4.577946(4)&8.4(1) \\
F3&	5.032253(5)&8.0(1)\\
F4&	5.420294(6)&6.6(1) \\
F5&	4.551897(7)&5.5(1) \\
F6&	4.86814(3)&1.2(1) \\
F7=2F5&	9.10379(1)&1.1(1)\\
\hline 
\multicolumn{3}{c}{TIC 312630206} \\
\hline
F1&	6.084691(2)&16.9(2)\\
F2&	6.260011(4)&11.8(2) \\
F3&	6.03265(1)&4.3(2) \\
F4&	6.11064(1)&3.0(2)\\
F5&	6.94350(2)&2.7(2) \\
F6&	6.59999(2)&1.9(2) \\
F7&	6.97220(2)&1.9(2) \\
F8&	6.68571(3)&1.5(2)\\
F9&	5.51236(3)&1.3(2) \\
F10&	7.48618(3)&1.3(2) \\
F11&	7.54980(3)&1.3(2) \\
F12&	6.42436(4)&1.2(2) \\
\hline 
\multicolumn{3}{c}{TIC 312637783} \\
\hline
F1&	8.182742(7)&5.5(2) \\
F2&	8.05626(1)&3.7(2) \\
F3&	7.62923(1)&3.6(2)\\
F4&	8.11316(2)&2.2(2) \\
F5&	7.98128(2)&2.1(2)\\
F6&	8.13077(2)&2.0(2) \\
\hline 
\multicolumn{3}{c}{TIC 312639883} \\
\hline
F1&	5.953170(6)&4.7(1) \\
F2&	7.12450(2)&1.7(1) \\
F3&	6.24850(2)&1.4(1) \\
F4&	8.49087(2)&1.2(1) \\
F5&	6.47017(3)&0.9(1)\\
\hline 
\multicolumn{3}{c}{TIC 314833456} \\
\hline
F1&	4.169867(9)&19.9(3) \\
F2&	4.54123(1)&15.8(3) \\
F3&	4.44952(2)&9.8(3) \\
F4&	4.51831(2)&8.2(3) \\
F5&	4.12392(3)&5.3(3) \\
F6&	4.19268(3)&5.1(3) \\
F7&	4.51793(2)&7.8(3) \\
F8=F2+F3&	8.99075(2)&3.1(3)\\
F9&	4.14687(7)&2.6(3)\\
F10&	4.10055(7)&2.5(3)\\
F11=2F2&	9.08246(2)&1.6(3)\\
\hline 
\multicolumn{3}{c}{TIC 331782797} \\
\hline
F1&	7.98732(5)&33(3)\\
\hline 
\multicolumn{3}{c}{TIC 337105253} \\
\hline
F1&	5.94609(4)&15(1)\\
F2&	5.7712(1)&7(1)	\\
\hline 
\multicolumn{3}{c}{TIC 338076483} \\
\hline
F1&	4.77848(3)&2.8(1)\\
F2&	1.24129(6)&1.1(1) \\
F3&	5.62774(9)&0.9(1) \\
\hline 
\multicolumn{3}{c}{TIC 347486043} \\
\hline
F1&	4.63603(1)&6.5(1) \\
F2=2F4&	1.14919(2)&3.2(1) \\
F3&	4.78138(5)&1.6(1) \\
F4&	0.57459(1)&1.6(1) \\
F5&	4.58469(6)&1.2(1) \\
F6&	4.61050(7)&1.1(1) \\
\hline 
\multicolumn{3}{c}{TIC 347684339} \\
\hline
F1&	5.89589(2)&3.5(2) \\
F2&	6.7620(1)&0.9(2) \\
\hline 
\multicolumn{3}{c}{TIC 348137274} \\
\hline
F1&	6.34881(3)&6.2(3) \\
F2&	6.23048(5)&3.7(3) \\
F3&	6.10761(9)&2.0(3) \\
\hline 
\multicolumn{3}{c}{TIC 348231245} \\
\hline
F1&	6.80616(5)&2.3(2) \\
\hline 
\multicolumn{3}{c}{TIC 348232246} \\
\hline
F1&	7.79032(5)&3.2(3) \\
\hline 
\multicolumn{3}{c}{TIC 348443241} \\
\hline
F1&	5.18990(3)&4.4(2) \\
F2&	2.20500(6)&1.9(2)\\
\hline 
\multicolumn{3}{c}{TIC 348506748} \\
\hline
F1&	4.60303(2)&2.8(1) \\
F2&	0.52494(3)&2.0(1) \\
F3&	4.48614(6)&1.1(1) \\
F4&	4.94911(6)&1.0(1) \\
F5&	4.77749(7)&0.9(1) \\
F6&	5.16733(8)&0.8(1) \\
F7&	4.56422(8)&0.8(1) \\
\hline 
\multicolumn{3}{c}{TIC 348607991} \\
\hline
F1&	4.58166(2)&3.9(2) \\
F2&	1.96881(8)&1.2(2) \\
F3&	5.01497(8)&1.2(2) \\
F4&	5.39336(9)&1.1(2)\\
\hline 
\multicolumn{3}{c}{TIC 348668924} \\
\hline
F1&	5.54861(3)&	 6.2(4) \\
F2&	5.05848(8)&	 2.4(4) \\
\hline 
\multicolumn{3}{c}{TIC 348671634} \\
\hline
F1&	4.41466(4)&	 3.9(3) \\
F2&	4.76344(9)&	 1.8(3)  \\
\hline 
\multicolumn{3}{c}{TIC 349332755} \\
\hline
F1&	7.64114(1)&	 4.1(2) \\
\hline 
\multicolumn{3}{c}{TIC 357853776} \\
\hline
F1&	10.257111(6)&	 5.5(2) \\
\hline 
\multicolumn{3}{c}{TIC 360063836} \\
\hline
F1&	5.139906(5)&	 8.0(2) \\
F2&	5.52154(2)&	 2.2(2) \\
F3&	6.14280(1)&	 3.6(2) \\
F4&	4.44844(2)&	 1.5(2) \\
F5&	5.71066(3)&	 1.5(2)  \\
F6&	5.55505(3)&	 1.1(2) \\
\hline 
\multicolumn{3}{c}{TIC 361324132} \\
\hline
F1&	6.568133(3)&10.1(1) \\
F2&	6.88566(1)&3.1(1) \\
F3&	6.19274(2)&1.9(1)\\
F4&	6.35290(2)&1.6(1)\\
F5&	6.56536(2)&1.8(1) \\
F6&	5.82671(3)&1.2(1)\\
F7&	7.05478(3)&0.9(1) \\
F8&	8.21035(3)&1.0(1) \\
\hline 
\multicolumn{3}{c}{TIC 370128780} \\
\hline
F1&	4.88959(7)&	 1.6(2)\\
F2&	0.10620(8)&1.5(2) \\
\hline 
\multicolumn{3}{c}{TIC 372724051} \\
\hline
F1&	5.263577(6)&	 19.9(2) \\
F2=2F4&	1.34570(1)&	 9.8(2) \\
F3&	5.28665(2)&	 6.3(2) \\
F4&	0.67285(1)&	 4.4(2) \\
F5&	6.90514(8)&	 1.5(2) \\
F6=2F1&	10.52715(1)&	 0.9(2) \\
\hline 
\multicolumn{3}{c}{TIC 374038418} \\
\hline
F1&	5.73406(6)&1.6(2) \\
\hline 
\multicolumn{3}{c}{TIC 377088966} \\
\hline
F1&	4.229863(4)&9.0(2) \\
F2&	4.216205(7)&5.9(2) \\
F3&	4.49732(2)&2.5(2)\\
\hline 
\multicolumn{3}{c}{TIC 378429048} \\
\hline
F1&	6.525448(5)&	6.4(1)\\
\hline 
\multicolumn{3}{c}{TIC 389532846} \\
\hline
F1&	4.32917(1)&8.7(2) \\
F2&	4.33116(4)&3.1(2) \\
\hline 
\multicolumn{3}{c}{TIC 391082033} \\
\hline
F1&	9.396502(8)&	 24.3(8) \\
\hline 
\multicolumn{3}{c}{TIC 391836737} \\
\hline
F1&	6.193273(1)&12.36(8)\\
F2&	6.09952(2)&1.15(8) \\
F3=2F1&	12.386546(2)&0.64(8)\\
\hline 
\multicolumn{3}{c}{TIC 392965683} \\
\hline
F1&	6.04430(2)&12.7(7)\\
F2&	6.12764(5)&4.2(7) \\
F3&	6.81849(5)&4.3(7) \\
\hline 
\multicolumn{3}{c}{TIC 393662110} \\
\hline
F1&	6.085935(2)&17.5(1) \\
F2&	5.914236(5)&6.5(1) \\
F3&	6.44171(1)&2.7(1) \\
F4&	7.23376(1)&2.4(1) \\
F5=2F1&	12.171870(4)&1.5(1) \\
F6&	6.26979(2)&1.2(1) \\
F7&	6.34677(3)&1.1(1) \\
\hline 
\multicolumn{3}{c}{TIC 399436828} \\
\hline
F1&	5.15360(5)&5.5(4)\dag \\
\hline 
\multicolumn{3}{c}{TIC 400852783} \\
\hline
F1&	6.88034(1)&8.9(3) \\
F2&	7.32561(1)&5.9(3) \\
F3&	7.15369(4)&2.0(3) \\
\hline 
\multicolumn{3}{c}{TIC 419354107} \\
\hline
F1&	4.974058(3)&29.5(3) \\
F2&	4.94025(2)&5.5(3) \\
F3=2F1&	9.948117(6)&4.1(3) \\
F4=3F1&	14.92218(1)&2.6(3) \\
F5=4F1&	19.89623(1)&1.8(3)\\
F6&	4.90630(2)&4.0(3) \\
\hline 
\multicolumn{3}{c}{TIC 421117635} \\
\hline
F1&	5.21831(3)&4.4(2) \\
F2&	4.73831(8)&1.5(2) \\
\hline 
\multicolumn{3}{c}{TIC 445619007} \\
\hline
F1&	5.21307(6)&1.7(2) \\
F2=2F4&	1.26348(8)&1.1(2)\\
F3=4F4&	2.5270(1)&1.2(2) \\
F4&	0.63174(4)&0.6(2) \\
\hline 
\multicolumn{3}{c}{TIC 451932686} \\
\hline
F1&	9.49288(2)&1.2(1) \\
\hline 
\multicolumn{3}{c}{TIC 452018494} \\
\hline
F1&	8.654173(7)&4.0(1) \\
\hline 
\multicolumn{3}{c}{TIC 461607866} \\
\hline
F1&	6.652979(6)&6.3(1) \\
F2&	6.279407(6)&6.0(1) \\
F3&	6.187924(8)&4.9(1) \\
F4&	6.58081(1)&3.5(1) \\
F5&	6.71918(1)&2.8(1)\\
F6&	7.77815(3)&1.3(1) \\
F7&	6.24561(4)&1.0(1) \\
\hline 
\multicolumn{3}{c}{TIC 469221047} \\
\hline
F1&	4.442633(7)&12.2(2) \\
F2&	4.404679(7)&12.2(2) \\
\hline 
\multicolumn{3}{c}{TIC 469223889} \\
\hline
F1&	7.84158(3)&1.1(1)\\
F2&	8.35141(5)&0.7(1)\\
F3&	6.79727(5)&0.7(1)\\
F4&	1.11389(5)&0.7(1)\\
\enddata
\end{deluxetable}
\newpage

\subsection{Candidate \Bcep stars}
\startlongtable 
\begin{deluxetable}{lcr}  
\tablecaption{Frequencies and amplitudes determined for the Candidate \Bcep stars. Formal error estimates determined according to \citet{Montgomery1999} are given in braces in units of the last significant digit. \dag Amplitudes may be largely suppressed.}
\label{tbl:cand-freqtable}
\tablenum{8}
\tablehead{ 
\colhead{ID} & \colhead{Freq.}    & \colhead{Amp.}\\
\colhead{ } & \colhead{(d$^{-1}$)} & \colhead{(mmag)}
}
\startdata
\hline
\multicolumn{3}{c}{TIC 92193} \\
\hline
F1&	5.97644(5)&1.6(2)\\
\hline 
\multicolumn{3}{c}{TIC 11158867} \\
\hline
F1&	12.91477(8)&2.0(3) \\
F2&	1.54085(9)&1.8(3) \\
\hline 
\multicolumn{3}{c}{TIC 12249117} \\
\hline
F1&	3.28913(1)&3.0(1) \\
\hline 
\multicolumn{3}{c}{TIC 12647534} \\
\hline
F1&	3.51179(2)&5.7(3) \\
F2&	1.24173(2)&5.5(3)\\
F3&	0.97933(7)&1.9(3) \\
F4&	1.25678(8)&1.8(3)\\
F5&	1.27325(9)&1.4(3) \\
\hline 
\multicolumn{3}{c}{TIC 13839931} \\
\hline
F1&	4.95417(6)&4.3(5) \\
\hline 
\multicolumn{3}{c}{TIC 13971092} \\
\hline
F1&	9.69457(4)&1.8(3) \\
\hline 
\multicolumn{3}{c}{TIC 13973539} \\
\hline
F1&	3.502964(3)&9.5(1)\\
F2&	1.38376(1)&2.7(1) \\
F3&	4.83592(2)&1.2(1)\\
F4=2F1&	7.005929(6)&0.7(1) \\
\hline 
\multicolumn{3}{c}{TIC 28949811} \\
\hline
F1&	13.12702(7)&1.2(2) \\
\hline 
\multicolumn{3}{c}{TIC 29036690} \\
\hline
F1&	7.92305(5)&0.7(1) \\
\hline 
\multicolumn{3}{c}{TIC 29598925} \\
\hline
F1&	4.01099(8)&4.0(3) \\
F2&	2.24307(7)&4.5(3) \\
F3&	0.7733(1)&3.1(3) \\
F4&	1.5457(1)&2.2(3)\\
F5&	1.9355(1)&2.5(3)\\
\hline 
\multicolumn{3}{c}{TIC 30569481} \\
\hline
F1&	5.578126(6)&5.29(9) \\
F2&	2.59672(2)&1.88(9)\\
F3=F2$-$F1&2.98140(2)&1.64(9) \\
F4&	5.08676(3)&1.05(9) \\
\hline 
\multicolumn{3}{c}{TIC 43301361} \\
\hline
F1&	8.849477(8)&3.02(9) \\
F2&	11.13493(2)&1.31(9)\\
F3&	0.85774(2)&1.32(9) \\
F4&	0.20085(3)&0.89(9) \\
F5&	0.30264(3)&0.75(9) \\
F6=2F3&	1.71548(3)&0.70(9) \\
\hline 
\multicolumn{3}{c}{TIC 49721269} \\
\hline
F1&	11.64652(1)&1.3(2) \\
F2&	5.82174(1)&1.2(2) \\
\hline 
\multicolumn{3}{c}{TIC 53709049} \\
\hline
F1&	6.61713(9)&6.9(9) \\
F2&	5.0123(1)&5.9(9) \\
\hline 
\multicolumn{3}{c}{TIC 53961302} \\
\hline
F1&	9.37902(1)&1.5(3) \\
F2&	1.03619(1)&1.4(3) \\
\hline 
\multicolumn{3}{c}{TIC 53968977} \\
\hline
F1&	0.76616(6)&8.1(5) \\
F2&	6.0136(1)&4.1(5)\\
\hline 
\multicolumn{3}{c}{TIC 60245596} \\
\hline
F1&	6.44274(3)&1.5(1)\\
F2&	3.18623(4)&1.1(1) \\
F3&	11.03042(6)&0.8(1) \\
\hline 
\multicolumn{3}{c}{TIC 65516748} \\
\hline
F1&	8.54807(7)&1.6(3) \\
\hline 
\multicolumn{3}{c}{TIC 72696935} \\
\hline
F1&	7.360422(8)&13.3(2) \\
F2&	11.04063(2)&1.1(2)\\
\hline 
\multicolumn{3}{c}{TIC 72944678} \\
\hline
F1&	3.32638(7)&10(1)\\
\hline 
\multicolumn{3}{c}{TIC 74197071} \\
\hline
F1&	2.96766(2)&4.7(2) \\
F2&	3.01456(5)&1.7(2) \\
F3&	5.76888(6)&1.3(2) \\
\hline 
\multicolumn{3}{c}{TIC 78499882} \\
\hline
F1&	3.72163(1)&2.7(1) \\
F2&	3.69110(2)&1.6(1) \\
\hline 
\multicolumn{3}{c}{TIC 80814494} \\
\hline
F1&	5.40452(3)&2.2(2) \\
\hline 
\multicolumn{3}{c}{TIC 93923487} \\
\hline
F1&	3.44624(3)&1.25(9) \\
F2&	1.90157(3)&1.23(9) \\
\hline 
\multicolumn{3}{c}{TIC 94000461} \\
\hline
F1&	4.545041(3)&15.1(1)\\
\hline 
\multicolumn{3}{c}{TIC 114249288} \\
\hline
F1&	3.133591(9)&4.0(2)\\
F2&	1.53094(1)&3.7(2) \\
F3&	1.52246(2)&2.2(2) \\
F4=F2+F3&	3.05339(2)&1.5(2)\\
\hline 
\multicolumn{3}{c}{TIC 123036723} \\
\hline
F1&	4.781420(8)&4.3(1) \\
F2&	2.365317(9)&3.7(1)\\
F3&	2.33445(2)&2.1(1) \\
F4&	2.44625(2)&1.5(1)\\
\hline 
\multicolumn{3}{c}{TIC 123828144} \\
\hline
F1&	5.52537(2)&1.02(7) \\
F2&	0.96250(3)&0.77(7) \\
\hline 
\multicolumn{3}{c}{TIC 137489662} \\
\hline
F1&	11.51964(1)&1.59(8) \\
\hline 
\multicolumn{3}{c}{TIC 140309502} \\
\hline
F1&	7.13915(1)&3.3(1) \\
F2&	0.07535(4)&0.8(1) \\
F3&	0.15069(8)&0.5(1) \\
F4&	0.2260(1)&0.5(1)\\
\hline 
\multicolumn{3}{c}{TIC 140429699} \\
\hline
F1&	6.17558(3)&1.6(1) \\
\hline 
\multicolumn{3}{c}{TIC 141190209} \\
\hline
F1&	4.74596(5)&1.5(2) \\
\hline 
\multicolumn{3}{c}{TIC 141903641 \dag} \\ 
\hline
F1&	5.29110(1)&5.9(2) \\ 
F2&	0.23796(2)&3.2(2) \\ 
F3&	0.14353(4)&1.3(2) \\ 
F4&	6.13873(5)&1.1(2) \\ 
\hline 
\multicolumn{3}{c}{TIC 143530557} \\
\hline
F1&	3.23172(1)&4.1(2) \\
F2&	1.57798(3)&1.4(2) \\
F3&	3.24230(4)&1.2(2) \\
F4&	3.48728(4)&1.1(2) \\
\hline 
\multicolumn{3}{c}{TIC 145672806} \\
\hline
F1&	3.16275(4)&1.8(2) \\
\hline 
\multicolumn{3}{c}{TIC 155977294} \\
\hline
F1&	4.44906(1)&1.56(9) \\
F2&	2.05883(2)&1.31(9) \\
F3&	2.57197(3)&0.73(9) \\
\hline 
\multicolumn{3}{c}{TIC 172560369} \\
\hline
F1&	8.90489(3)&2.5(3)\\
\hline 
\multicolumn{3}{c}{TIC 184226481} \\
\hline
F1&	3.70079(2)&4.5(3) \\
\hline 
\multicolumn{3}{c}{TIC 190783321} \\
\hline
F1&	15.59057(4)&1.0(1) \\
F2&	21.16694(5)&0.8(1) \\
F3&	14.28699(5)&0.8(1) \\
\hline 
\multicolumn{3}{c}{TIC 193573033} \\
\hline
F1&	3.55629(2)&2.7(2) \\
\hline 
\multicolumn{3}{c}{TIC 193608395} \\
\hline
F1&	6.93971(3)&2.5(3) \\
F2&	3.39643(3)&2.3(3) \\
\hline 
\multicolumn{3}{c}{TIC 207045768} \\
\hline
F1&	6.13987(2)&2.5(2)\\
F2&	2.77901(2)&2.5(2)\\
F3=F1-F2&	3.36086(3)&1.9(2)\\
\hline 
\multicolumn{3}{c}{TIC 220238856} \\
\hline
F1&	3.71927(2)&10.9(8)\\
F2&	1.69782(3)&8.9(8) \\
F3&	4.41429(3)&8.6(8) \\
F4&	3.29601(3)&7.2(8) \\
\hline 
\multicolumn{3}{c}{TIC 220298408} \\
\hline
F1&	6.08901(2)&5.1(3) \\
\hline 
\multicolumn{3}{c}{TIC 226144867} \\
\hline
F1&	12.67603(3)&6.4(7) \\
\hline 
\multicolumn{3}{c}{TIC 231149746} \\
\hline
F1&	9.05289(5)&2.9(5) \\
F2&	12.23358(6)&2.5(5) \\
\hline 
\multicolumn{3}{c}{TIC 231177417} \\
\hline
F1&	5.90535(6)&6(1) \\
\hline 
\multicolumn{3}{c}{TIC 231236273} \\
\hline
F1&	9.66445(4)&2.4(3) \\
\hline 
\multicolumn{3}{c}{TIC 244327575} \\
\hline
F1&	3.724081(2)&18.0(1)\\
F2=2F1&	7.448162(3)&2.7(1)\\
F3=3F1&	11.17224(5)&1.3(1)\\
F4&	0.61728(1)&2.1(1)\\
F5&	0.78466(2)&1.3(1)\\
F6=4F1&	14.896324(6)&0.6(1) \\
\hline 
\multicolumn{3}{c}{TIC 252864874} \\
\hline
F1&	5.38477(4)&6.9(5) \\
F2&	12.96472(8)&3.7(5) \\
F3&	4.91754(9)&3.1(5) \\
\hline 
\multicolumn{3}{c}{TIC 272625912} \\
\hline
F1&	3.30518(1)&11.6(2)\\
F2&	1.38055(2)&5.0(2)\\
\hline 
\multicolumn{3}{c}{TIC 281533292} \\
\hline
F1&	5.53994(3)&7.9(8) \\
\hline 
\multicolumn{3}{c}{TIC 282737511} \\
\hline
F1&	3.215232(5)&6.9(1) \\
F2&	0.72957(2)&1.7(1) \\
F3&	0.27360(2)&1.5(1)\\
\hline 
\multicolumn{3}{c}{TIC 285122116} \\
\hline
F1&	7.9203(1)&1.1(2) \\
\hline 
\multicolumn{3}{c}{TIC 285684866} \\
\hline
F1&	3.07295(2)&7.3(3) \\
F2&	0.08417(2)&7.2(3) \\
\hline 
\multicolumn{3}{c}{TIC 286352720} \\
\hline
F1&	6.06353(1)&3.6(1) \\
F2&	5.99286(5)&0.7(1) \\
\hline 
\multicolumn{3}{c}{TIC 290300414} \\
\hline
F1&	4.36798(3)&0.95(8) \\
F2&	2.43643(4)&0.64(8)\\
F3&	5.41055(5)&0.61(8) \\
\hline 
\multicolumn{3}{c}{TIC 294306184} \\
\hline
F1&	7.22638(6)&5(1) \\
\hline 
\multicolumn{3}{c}{TIC 296449179} \\
\hline
F1=2F3&	5.967590(6)&9.2(2) \\
F2&	1.95179(2)&2.5(2) \\
F3&	2.983795(3)&2.6(2) \\
\hline 
\multicolumn{3}{c}{TIC 297264447} \\
\hline
F1&	8.78504(4)&1.4(2) \\
\hline 
\multicolumn{3}{c}{TIC 307944768} \\
\hline
F1&	5.89021(5)&1.2(2)\\
\hline 
\multicolumn{3}{c}{TIC 319302209} \\
\hline
F1&	3.383001(3)&8.5(4) \\
F2&	2.687471(6)&4.6(4) \\
F3&	1.880402(5)&5.1(4)\\
F4&	2.721681(6)&4.1(4) \\
\hline 
\multicolumn{3}{c}{TIC 335484090} \\
\hline
F1&	3.18454(1)&18.1(3)\\
F2=2F9&	0.266751(8)&6.3(3) \\
F3=6F9&	0.80025(2)&6.1(3) \\
F4=4F9&	0.53350(2)&5.6(3) \\
F5=8F9&	1.06700(3)&5.4(3) \\
F6=10F9&	1.33375(4)&4.7(3) \\
F7=12F9&	1.60050(5)&3.7(3) \\
F8=14F9&	1.86725(6)&3.6(3) \\
F9&	0.133375(4)&2.8(3) \\
F10=16F9&	2.13400(6)&3.0(3) \\
F11=5F9&	0.66688(2)&2.6(3) \\
F12=18F9&	2.40076(7)&2.2(3) \\
\hline 
\multicolumn{3}{c}{TIC 345652701} \\
\hline
F1&	3.18286(3)&2.4(2) \\
\hline 
\multicolumn{3}{c}{TIC 347585038} \\
\hline
F1&	3.64383(3)&3.1(2) \\
F2&	0.49554(5)&2.0(2)\\
F3&	1.41879(5)&1.8(2) \\
\hline 
\multicolumn{3}{c}{TIC 348609224} \\
\hline
F1&	3.95599(8)&1.5(2) \\
\hline 
\multicolumn{3}{c}{TIC 360749347 \dag} \\ 
\hline
F1&	4.88831(2)&	 12(1) \\ 
\hline 
\multicolumn{3}{c}{TIC 364146227} \\
\hline
F1&	3.09012(3)&5.6(4) \\
F2&	1.68288(5)&2.9(4) \\
\hline 
\multicolumn{3}{c}{TIC 366108295} \\
\hline
F1&	12.95685(9)&	 2.0(3) \\
\hline 
\multicolumn{3}{c}{TIC 368237682} \\
\hline
F1&	7.28839(5)&1.9(2)\\
F2&	0.37825(7)&1.4(2) \\
F3&	1.83304(8)&1.1(2)\\
F4&	7.54582(8)&1.2(2) \\
F5&	1.63789(8)&1.1(2) \\
F6&	1.7997(1)&0.8(2) \\
\hline 
\multicolumn{3}{c}{TIC 369708912} \\
\hline
F1&	3.25928(3)&2.5(2)\\
F2&	2.30628(3)&2.0(2) \\
\hline 
\multicolumn{3}{c}{TIC 370269139} \\
\hline
F1&	5.91020(9)&1.5(3)\\
\hline 
\multicolumn{3}{c}{TIC 372115570} \\
\hline
F1&	3.96205(2)&4.0(2) \\
\hline 
\multicolumn{3}{c}{TIC 374696382} \\
\hline
F1&	5.77910(4)&3.6(2) \\
F2&	0.30931(8)&1.7(2)\\
F3&	0.28025(8)&1.6(2) \\
\hline 
\multicolumn{3}{c}{TIC 379932247} \\
\hline
F1&	8.30337(4)&2.4(4)\\
\hline 
\multicolumn{3}{c}{TIC 384257658} \\
\hline
F1&	3.69970(2)&1.86(9) \\
F2&	3.65490(4)&0.90(9) \\
\hline 
\multicolumn{3}{c}{TIC 384540878} \\
\hline
F1&	3.55128(4)&3.6(4) \\
\hline 
\multicolumn{3}{c}{TIC 386513392} \\
\hline
F1&	3.12176(3)&1.6(1) \\
\hline 
\multicolumn{3}{c}{TIC 386693012} \\
\hline
F1& 13.597032(2)&7.2(1) \\
F2& 16.249506(8)&2.3(1) \\
F3=F2-F1& 2.652474(8)&2.2(1) \\
F4=2F1& 27.194064(5)& 0.9(1) \\
F5=F1+F2& 29.846538(8)&0.6(1) \\
F6& 13.60250(2)&0.9(1) \\
\hline 
\multicolumn{3}{c}{TIC 387153140} \\
\hline
F1&	11.32157(3)&2.1(2) \\
F2&	1.88019(4)&1.5(2) \\
F3=F1+F2&	13.20176(4)&1.2(2) \\
\hline 
\multicolumn{3}{c}{TIC 400080677} \\
\hline
F1&	9.02070(6)&0.6(1) \\
\hline 
\multicolumn{3}{c}{TIC 400533469} \\
\hline
F1&	11.1849(1)&4.5(8)\\
\hline 
\multicolumn{3}{c}{TIC 401847213} \\
\hline
F1&	5.1564(1)&4.8(8) \\
\hline 
\multicolumn{3}{c}{TIC 408941991} \\
\hline
F1&	4.61682(1) & 1.5(1)\\
F2&	2.29288(2) & 1.1(1)\\
\hline 
\multicolumn{3}{c}{TIC 408991617} \\
\hline
F1&	5.77126(2)&1.4(1) \\
\hline 
\multicolumn{3}{c}{TIC 419246605} \\
\hline
F1&	8.02047(9)&1.3(2) \\
\hline 
\multicolumn{3}{c}{TIC 420545430} \\
\hline
F1&	3.31306(7)&2.5(3) \\
F2&	1.4800(1)&1.7(3) \\
\hline 
\multicolumn{3}{c}{TIC 421691939} \\
\hline
F1&	6.01647(7)&3.3(4) \\
F2&	5.73585(9)&2.7(4) \\
\hline 
\multicolumn{3}{c}{TIC 422533344 \dag} \\ 
\hline
F1&	5.39483(7)&3.9(5) \\
F2&	6.15953(9)&2.9(5) \\
\hline 
\multicolumn{3}{c}{TIC 427396133} \\
\hline
F1&	5.90994(2)&5.8(3)\\
F2&	6.10060(3)&3.8(2)\\
\hline 
\multicolumn{3}{c}{TIC 430625041 \dag} \\ 
\hline
F1&	5.7728(1) &	 2.3(3) \\ 
\hline 
\multicolumn{3}{c}{TIC 440971753} \\
\hline
F1&	3.17120(7)&4.1(3)\\
F2&	1.7627(1)&2.4(3) \\
F3&	3.1851(1)&2.4(3) \\
F4&	0.35458(1)&2.7(3) \\
\hline 
\multicolumn{3}{c}{TIC 447933173} \\
\hline
F1&	5.46811(1)&3.9(1) \\
F2&	5.82001(2)&2.8(1) \\
F3&	2.70908(2)&2.3(1)\\
F4&	5.80719(3)&1.7(1) \\
\hline 
\multicolumn{3}{c}{TIC 458911894} \\
\hline
F1&	3.87623(3)&2.5(2) \\
F2&	3.85813(5)&1.6(2) \\
F3&	3.89609(5)&1.5(2) \\
\hline
\multicolumn{3}{c}{TIC 459014234} \\
\hline
F1&	11.78938(1)&10.5(2)\\
F2&	13.58245(3)&3.7(2) \\
F3&	11.01875(4)&3.0(2)\\
F4&	10.78396(5)&2.4(2) \\
F5&	13.79750(9)&1.4(2) \\ 
\hline 
\multicolumn{3}{c}{TIC 459105076} \\
\hline
F1&	10.92534(4)&2.3(2)\\
\hline 
\multicolumn{3}{c}{TIC 469095496} \\
\hline
F1&	8.00798(3)&4.2(3) \\
\hline 
\enddata
\end{deluxetable}
\newpage

\subsection{Rejected stars}

\startlongtable 
\begin{deluxetable}{lcr}  
\tablecaption{Frequencies and amplitudes determined for the rejected stars. Formal error estimates determined according to \citet{Montgomery1999} are given in braces in units of the last significant digit.}
\label{tbl:rejected-freqtable}
\tablenum{9}
\tablehead{ 
\colhead{ID} & \colhead{Freq.}    & \colhead{Amp.}\\
\colhead{ } & \colhead{(d$^{-1}$)} & \colhead{(mmag)}
}
\startdata
\hline 
\multicolumn{3}{c}{TIC 12647620} \\ 
\hline
F1&	3.51173(5)&4.8(4) \\
F2&	1.24177(8)&3.0(4) \\
\hline \multicolumn{3}{c}{TIC 26283875} \\ \hline
F1&	14.24270(9)&0.8(1) \\ 
\hline \multicolumn{3}{c}{TIC 28709025} \\ \hline
F1&	5.48688(6)&1.9(1) \\ 
F2&	5.5341(1)&0.7(1) \\ 
F3&	4.9785(2)&0.7(1) \\ 
\hline \multicolumn{3}{c}{TIC 40102236} \\ \hline
F1&	5.239879(8)&3.0(1) \\ 
F2&	2.76522(3)&0.9(1) \\ 
\hline \multicolumn{3}{c}{TIC 60320306} \\ \hline
F1&	7.01910(4)&0.80(9) \\ 
\hline \multicolumn{3}{c}{TIC 67985749} \\ \hline
F1&	6.34398(5)&1.5(2) \\ 
\hline \multicolumn{3}{c}{TIC 90134626} \\ \hline
F1&	2.98520(2)&1.5(1) \\ 
F2&	3.65605(3)&1.1(1)\\ 
\hline \multicolumn{3}{c}{TIC 93549165} \\ \hline
F1&	3.97459(7)&0.69(6) \\ 
\hline \multicolumn{3}{c}{TIC 102161004} \\ \hline
F1&	3.34360(3)&1.2(1) \\ 
F2&	4.19498(5)&0.6(1) \\ 
\hline \multicolumn{3}{c}{TIC 118680798} \\ \hline
F1&	4.61715(3)&0.65(7) \\ 
\hline \multicolumn{3}{c}{TIC 123754451} \\ \hline
F1&	3.43280(3)&0.80(7) \\ 
F2&	2.72317(3)&0.73(7) \\ 
F3&	4.24874(3)&0.74(7) \\ 
F4&	5.62140(4)&0.58(7) \\ 
F5&	4.49144(5)&0.48(7) \\ 
\hline \multicolumn{3}{c}{TIC 138905907} \\ \hline
F1&	8.65749(1)&11.9(5)\\ 
F2&	12.46790(2)&5.5(5)\\ 
F3&	7.77313(3)&4.9(5) \\ 
F4&	13.55768(4)&3.5(5) \\ 
\hline \multicolumn{3}{c}{TIC 141903541} \\ 
\hline
F1&	5.29110(1)&5.9(2) \\
F2&	0.23796(2)&3.2(2) \\
F3&	0.14353(4)&1.3(2) \\
F4&	6.13873(5)&1.1(2) \\
\hline \multicolumn{3}{c}{TIC 157535787} \\ \hline
F1&	12.58931(2)&6.4(2)\\ 
F2&	1.27797(3)&3.5(2) \\ 
\hline \multicolumn{3}{c}{TIC 162012064} \\ \hline
F1&	22.18628(3)&2.2(2) \\ 
F2&	19.71143(3)&1.8(2) \\ 
F3&	24.07994(4)&1.5(2) \\ 
\hline \multicolumn{3}{c}{TIC 264485563} \\ \hline
F1&	9.08388(6)&0.8(1) \\ 
\hline 
\multicolumn{3}{c}{TIC 281803267} \\
\hline
F1&	3.91390(2)&19(1)\\
\hline \multicolumn{3}{c}{TIC 291556636} \\ \hline
F1&	3.71521(2)&4.0(2) \\ 
F2&	1.84478(2)&3.5(2) \\ 
F3&	3.67380(3)&2.6(2)\\ 
F4&	2.98491(5)&1.5(2) \\ 
\hline \multicolumn{3}{c}{TIC 352529679} \\ \hline
n/a \\ 
\hline \multicolumn{3}{c}{TIC 360661624} \\ 
\hline
F1&	4.88831(2)&	 12(1) \\
\hline \multicolumn{3}{c}{TIC 377498419} \\ \hline
F1&	18.69049(3)&7.0(5) \\ 
F2&	9.34525(2)&2.8(5)\\ 
F3&	28.03574(4)&1.4(5) \\ 
F4&	37.38098(6)&1.1(5)\\ 
\hline \multicolumn{3}{c}{TIC 384992041} \\ \hline
F1&	3.76144(1)&44.5(8)\\ 
F2=2F1&	7.52289(2)&6.4(8) \\ 
\hline \multicolumn{3}{c}{TIC 417438983} \\ \hline
F1&	4.00826(2)&3.3(2) \\ 
\hline 
\multicolumn{3}{c}{TIC 422533347} \\ 
\hline
F1&	5.39483(7)&3.9(5) \\
F2&	6.15953(9)&2.9(5) \\
\hline \multicolumn{3}{c}{TIC 424032547} \\ \hline
F1&	5.02512(2)&3.5(3) \\ 
\hline \multicolumn{3}{c}{TIC 424032634} \\ \hline
F1&	7.03050(2)&1.4(1) \\ 
F2&	2.40547(3)&1.2(1) \\ 
\hline \multicolumn{3}{c}{TIC 426520557} \\ \hline
F1&	17.123158(6)&2.07(7) \\ 
F2&	12.96614(1)&1.36(7)\\ 
F3&	15.14342(1)&1.20(7) \\ 
F4&	14.64399(2)&0.64(7) \\ 
\hline 
\multicolumn{3}{c}{TIC 430625174} \\
\hline
F1&	5.7728(1)&2.3(3) \\
\hline \multicolumn{3}{c}{TIC 434178307} \\ \hline
F1&	5.9868(1)&1.9(4) \\ 
F2&	16.7695(1)&1.9(4) \\ 
\hline \multicolumn{3}{c}{TIC 446041643} \\ \hline
F1&	4.483797(5)&5.3(1)\\ 
F2&	4.164086(8)&3.3(1) \\ 
F3&	4.18027(2)&1.4(1) \\ 
\hline
\enddata
\end{deluxetable}


\bibliography{Bib}{}
\bibliographystyle{aasjournal}

\end{document}